\documentclass[lettersize,onecolumn]{IEEEtran}
\usepackage{braket}
\usepackage{bm}
\usepackage{graphicx}
\usepackage{tikz}
\usepackage{marvosym}
\usepackage{ifthen}
\usepackage{tikzsymbols}
\usetikzlibrary{calc}
\usetikzlibrary{matrix}
\usetikzlibrary{tikzmark}
\usetikzlibrary{fit}
\usetikzlibrary{shadows.blur}
\usetikzlibrary{shapes.symbols}
\usetikzlibrary{shapes.geometric}
\usetikzlibrary{patterns.meta}
\usepackage[normalem]{ulem}

\usepackage{hyperref}

\usepackage{algorithm}
\usepackage[noend]{algpseudocode}

\usepackage{mdframed}

\usetikzlibrary{positioning}
\usepackage{cancel}
\usepackage{amssymb, amsmath}
\usepackage{amsthm}
\usepackage{thmtools}
\usepackage{epstopdf}
\usepackage{pgfplots}
\usepackage{tikz}
\usepackage{tikzpeople}
\usepackage{circuitikz}
\usetikzlibrary{tikzmark, shapes, fit,backgrounds}
\usetikzlibrary{matrix, decorations.pathreplacing, angles, quotes}
\usetikzlibrary{patterns, calc, positioning}
\usetikzlibrary{arrows, arrows.meta}
\usetikzlibrary{shapes.multipart}
\usetikzlibrary{fit}
\usetikzlibrary{shapes.geometric, positioning}
\usetikzlibrary{decorations.pathmorphing}
\tikzstyle{block} = [rectangle, draw, 
    minimum width=4em, text centered, rounded corners, minimum height=1em]
\tikzstyle{line} = [draw, -latex]
\tikzset{meter/.append style={draw, inner sep=10, rectangle, font=\vphantom{A}, minimum width=30, scale=.7, path picture={\draw[black] ([shift={(.1,.3)}]path picture bounding box.south west) to[bend left=50] ([shift={(-.1,.3)}]path picture bounding box.south east);\draw[black,-{Latex[scale=.5]}] ([shift={(0,.1)}]path picture bounding box.south) -- ([shift={(.3,-.1)}]path picture bounding box.north);}}}
\tikzset{snake it/.style={decorate, decoration=snake}}

\newtheorem{lemma}{Lemma}
\newtheorem{theorem}{Theorem}

\newtheorem{remark}{Remark}
\declaretheorem[style=definition,numbered=yes]{definition}

\let\oldsqcap\sqcap
\renewcommand{\sqcap}{\scalebox{1}[1.25]{$\oldsqcap$}}

\newcommand\blfootnote[1]{%
  \begingroup
  \renewcommand\thefootnote{}\footnote{#1}%
  \addtocounter{footnote}{-1}%
  \endgroup
}

\begin{document}

\title{
On the Capacity of Erasure-prone Quantum Storage with Erasure-prone Entanglement Assistance}
\author{Hua Sun, Syed A. Jafar}
\date{}                                           
\maketitle

\blfootnote{Hua Sun (email: hua.sun@unt.edu) is with the Department of Electrical Engineering at the University of North Texas. Syed A. Jafar (email: syed@uci.edu) is with the  Department of Electrical Engineering and Computer Science (EECS) at the University of California Irvine. }

\begin{abstract}
A quantum message is encoded into $N$ quantum storage systems ($Q_1\dots Q_N$) with assistance from $N_B$ maximally entangled bi-partite quantum systems $A_1B_1, \dots, A_{N_B}B_{N_B}$, that are prepared in advance such that $B_1\dots B_{N_B}$ are stored separately as entanglement assistance (EA) systems, while $A_1\dots A_{N_B}$ are made available to the encoder. Both the storage systems and EA systems are erasure-prone. The quantum message must be recoverable given any $K$ of the $N$ storage systems along with any $K_B$ of the $N_B$ EA systems. The capacity for this setting is the maximum size of the quantum message, given that the size of each EA system is $\lambda_B$. All system sizes are relative to the size of a storage system, which is normalized to unity. The exact capacity is characterized as a function of $N,K,N_B,K_B, \lambda_B$ in all cases, with one exception. The capacity remains open for an intermediate range of $\lambda_B$ values when a strict majority of the $N$ storage systems, and a strict non-zero minority of the $N_B$ EA systems, are erased. As a key stepping stone, an analogous classical storage (with shared-randomness assistance) problem is introduced. A set of constraints is identified for the classical problem, such that classical linear code constructions translate to quantum storage codes, and the converse bounds for the two settings utilize similar insights. In particular, the capacity characterizations for the classical and quantum settings are shown to be identical in all cases where the capacity is settled. 
\end{abstract}

\newpage

\allowdisplaybreaks

\section{Introduction}
As a cornerstone of the theory of error-correcting codes, coding for \emph{erasures} plays a critical role in ensuring the reliable storage of information. In quantum systems, where both the information and the storage medium are quantum in nature, even the act of error detection can  introduce additional errors. Erasures --- errors at \emph{known} locations --- are less susceptible to this vulnerability and are consequently easier to correct. This underpins the concept of erasure conversion \cite{Erasure_conversion, levine2024demonstrating}, which utilizes physical principles to transform other types of quantum errors into erasures. Erasures may also emerge as the predominant error mode in future quantum distributed storage systems, particularly if such systems can be scaled to the size of modern classical data centers, where each storage system functions as a fault-tolerant memory unit \cite{QLRC,Sun_Jafar_QuStorage}. Furthermore, from an information theoretic perspective, erasures are among the most tractable of all quantum error paradigms \cite{Sun_Jafar_QuStorage, Grassl_Huber_Winter}, creating opportunities for synergies between the studies of fundamental limits on one hand and practical code designs on the other \cite{CQE}. Let us denote `\emph{erasure-prone quantum storage}' as $\widetilde{\mbox{QS}}$, with the $\sim$ indicating  susceptibility to erasures.

The capabilities of $\widetilde{\mbox{QS}}$ are significantly enhanced by \emph{entanglement assistance} (EA), i.e., when quantum entanglement is available in advance as an additional shared resource between the encoder and decoder. For example, it is known that quantum erasure codes do not exist in the parameter regime where the locality is less than the code distance \cite{QLRC}. This limitation arises primarily due to the \emph{no-cloning theorem} \cite{Nielsen_Chuang}, which prohibits the recovery of quantum information from a set of storage systems whose erasure (i.e., absence) must be tolerated. Otherwise, \emph{two} copies of the stored quantum information could be reconstructed -- one by a legitimate receiver without access to the erased/missing systems, and another by a hypothetical party with access only to the `missing' systems. However, such limitations can be circumvented by utilizing EA. Remarkably, EA is also associated with improved analytical tractability from an information theoretic standpoint. A textbook example is the  capacity of a quantum channel, which remains unresolved in general but is settled if EA is available \cite{EAQC}.

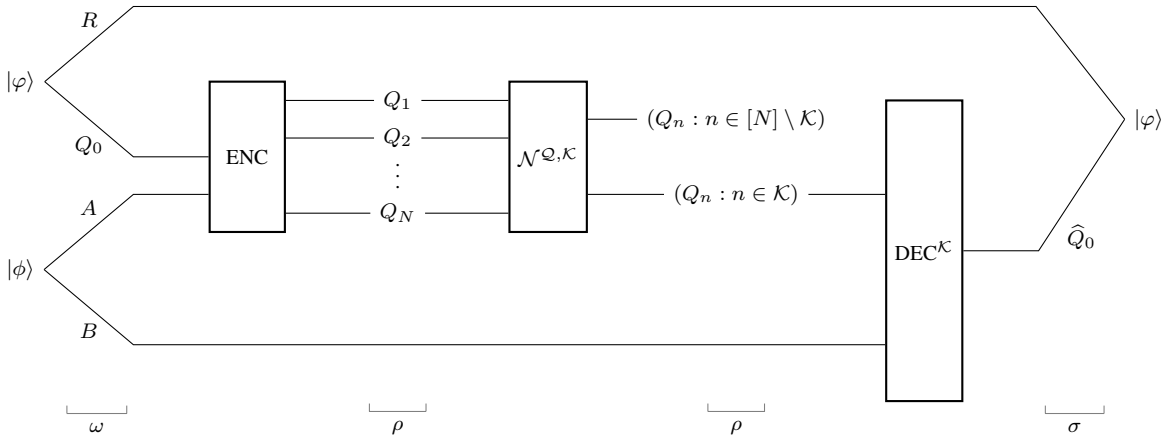
\begin{figure}[h]
\begin{tikzpicture}
\node (ENC) at (0,0) [draw, thick, rectangle, minimum width=1cm, minimum height=2cm]{\footnotesize ENC};
\node (QR) at ($(-2.5cm,1cm)+(ENC.west)$) {\footnotesize $\ket{\varphi}$};
\node (QRhat) at ($(12.5cm,0.5cm)+(ENC.west)$) {\footnotesize $\ket{\varphi}$};
\draw [-] (QR.east)--($(-1cm,0cm)+(ENC.west)$)node[midway,below=0.1cm]{\footnotesize $Q_0$}-- (ENC.west);

\node (AB) at ($(-2.5cm,-1.5cm)+(ENC.west)$) {\footnotesize $\ket{\phi}$};
\draw [-] (AB.east)--($(-1cm,-0.5cm)+(ENC.west)$)node[midway,above=0.1cm]{\footnotesize $A$}-- ($(0cm,-0.5cm)+(ENC.west)$); 

\draw [-] (QR.east)--($(-1cm,2cm)+(ENC.west)$)node[midway,above=0.1cm]{\footnotesize $R$}-- ($(11cm,2cm)+(ENC.west)$)--(QRhat.west);
\node (Ne) at (4,0) [draw, thick, rectangle, minimum width=1cm, minimum height=2cm, align=center]{\footnotesize $\mathcal{N}^{\mathcal{Q},\mathcal{K}}$};

\node (NBe) at (4,-2.5) {};

\node (De) at (9,-1.25) [draw, thick, rectangle, minimum width=1cm, minimum height=4cm]{\footnotesize $\mbox{DEC}^{\mathcal{K}}$};
\draw [-] (QRhat.west)--($(1cm,0cm)+(De.east)$)node[midway,below=0.4cm]{\footnotesize $\widehat{Q}_0$}-- (De.east);

\node (omega) at (-2,-3.6) {\footnotesize $\omega$};
\draw[black!50] ([xshift=-5pt,yshift=0pt]omega.north west) -- ([xshift=5pt,yshift=0pt]omega.north east)
      ++(0,0) -- ++(0,3pt) 
      ([xshift=-5pt,yshift=0pt]omega.north west) -- ++(0,3pt);
\node (rho) at (2,-3.6) {\footnotesize $\rho$};
\draw[black!50] ([xshift=-5pt,yshift=0pt]rho.north west) -- ([xshift=5pt,yshift=0pt]rho.north east)
      ++(0,0) -- ++(0,3pt) 
      ([xshift=-5pt,yshift=0pt]rho.north west) -- ++(0,3pt);
\node (rho2) at (6.5,-3.6) {\footnotesize $\rho$};
\draw[black!50] ([xshift=-5pt,yshift=0pt]rho2.north west) -- ([xshift=5pt,yshift=0pt]rho2.north east)
      ++(0,0) -- ++(0,3pt) 
      ([xshift=-5pt,yshift=0pt]rho2.north west) -- ++(0,3pt);
\node (omega2) at (11,-3.6) {\footnotesize $\sigma$};
\draw[black!50] ([xshift=-5pt,yshift=0pt]omega2.north west) -- ([xshift=5pt,yshift=0pt]omega2.north east)
      ++(0,0) -- ++(0,3pt) 
      ([xshift=-5pt,yshift=0pt]omega2.north west) -- ++(0,3pt);

\node (Q1) at (2,0.75) {\footnotesize $Q_1$};
\node (Q2) at (2,0.25) {\footnotesize $Q_2$};
\node (vdots) at (2,-0.15) {\footnotesize $\vdots$};
\node (QN) at (2,-0.75) {\footnotesize $Q_N$};
\draw (ENC.east|-Q1) -- (Q1)--(Q1-|Ne.west);
\draw (ENC.east|-Q2) -- (Q2)--(Q2-|Ne.west);
\draw (ENC.east|-QN) -- (QN)--(QN-|Ne.west);

\node (Qerased) at (6.5,0.5) {\footnotesize $(Q_n: n\in [N]\setminus \mathcal{K})$};
\draw (Ne.east|-Qerased)-- (Qerased);
\node (Qunerased) at (6.5,-0.5) {\footnotesize $(Q_n: n\in \mathcal{K})$};
\draw (Ne.east|-Qunerased) -- (Qunerased)--(Qunerased-|De.west);

\coordinate (x) at ($(-1cm,-2.5cm)+(ENC.west)$);
\draw [-] (AB.east)--($(-1cm,-2.5cm)+(ENC.west)$)node[midway,below=0.1cm]{\footnotesize $B$}--(x.east-|De.west);
\end{tikzpicture}
\caption{A $\widetilde{\mbox{QS}}\mbox{EA}$ setting studied in  \cite{Grassl_Huber_Winter} is illustrated. A quantum message $Q_0$ (possibly entangled with a purifying reference system $R$) is encoded into $N$ quantum storage systems $(Q_1\dots Q_N)$ by an encoder with entanglement assistance $A$. Any $N-K$ of these $N$ systems are erased by a channel. A decoder must recover the quantum information from the remaining (unerased) systems and the entanglement assistance $B$. Note that the entanglement assistance is perfect (not prone to erasures). A feasible coding scheme specifies an encoder ENC, and an array of decoders $\mbox{DEC}^{\mathcal{K}}$, guaranteeing recovery for every $\mathcal{K}\in\binom{[N]}{K}$.}
\label{fig:QSperfectEA}
\end{figure}

Aside from a few notable exceptions \cite{lai2012entanglement, fujiwara2013quantum}, $\widetilde{\mbox{QS}}$ with EA has been studied primarily under the assumption that the EA is perfect, a paradigm that we may refer to as $\widetilde{\mbox{QS}}\mbox{EA}$ \cite{Grassl_Huber_Winter, brun2006correcting, brun2014catalytic, lai2017linear, guenda2018constructions, luo2018mds, grassl2021entanglement}. Such a setting is illustrated in Figure \ref{fig:QSperfectEA}. While entanglement distribution is seldom perfect, the assumption of perfect EA may be  justified if the encoder and decoder can perform entanglement distillation \cite{bennett1996purification, bennett1996mixed, devetak2005distillation} prior to the encoding of the quantum information, and the coded systems are immediately sent to the receiver so the decoding can take place prior to any new corruption in the EA resource. This may not be possible in some settings, e.g., distributed storage over an extended period. The classical communication required in advance for entanglement distillation may not be feasible, or there may be significant risk of corruption of the EA resource after the quantum information is encoded and before it can be decoded.  It is therefore important to study settings with imperfect EA. References \cite{lai2012entanglement, fujiwara2013quantum} have found various coding schemes for error-prone quantum storage with error prone EA, to show that even imperfect EA can be quite useful. Motivated by these observations, in this work we explore the information theoretic capacity of $\widetilde{\mbox{QS}}\widetilde{\mbox{EA}}$, i.e., erasure-prone quantum storage with erasure-prone entanglement assistance.

\begin{figure}[h]
\begin{tikzpicture}
\node (ENC) at (0,0) [draw, thick, rectangle, minimum width=1cm, minimum height=2cm]{\footnotesize ENC};
\node (QR) at ($(-2.5cm,1cm)+(ENC.west)$) {\footnotesize $\ket{\varphi}$};
\node (QRhat) at ($(12.5cm,0.5cm)+(ENC.west)$) {\footnotesize $\ket{\varphi}$};
\draw [-] (QR.east)--($(-1cm,0cm)+(ENC.west)$)node[midway,below=0.1cm]{\footnotesize $Q_0$}-- (ENC.west);

\node (AB) at ($(-2.5cm,-1.5cm)+(ENC.west)$) {\footnotesize $\ket{\phi}$};
\draw [-] (AB.east)--($(-1cm,-0.5cm)+(ENC.west)$)node[midway,above=0.1cm]{\footnotesize $A$}-- ($(0cm,-0.5cm)+(ENC.west)$); 

\draw [-] (QR.east)--($(-1cm,2cm)+(ENC.west)$)node[midway,above=0.1cm]{\footnotesize $R$}-- ($(11cm,2cm)+(ENC.west)$)--(QRhat.west);
\node (Ne) at (4,0) [draw, thick, rectangle, minimum width=1cm, minimum height=2cm, align=center]{\footnotesize $\mathcal{N}^{\mathcal{Q},\mathcal{K}}$};

\node (NBe) at (4,-2.5) [draw, thick, rectangle, minimum width=1cm, minimum height=2cm, align=center]{\footnotesize $\mathcal{N}^{\mathcal{B},\mathcal{K}_B}$};
\draw [-] (AB.east)--($(-1cm,-2.5cm)+(ENC.west)$)node[midway,below=0.1cm]{\footnotesize $B$}-- ($(0cm,-2.5cm)+(ENC.west)$) --(NBe.west);
\node (Bn) at ($(-2cm,-0.3cm)+(NBe.west)$) {\footnotesize $B = B_1 \dots B_{N_B}$};

\node (De) at (9,-1.25) [draw, thick, rectangle, minimum width=1cm, minimum height=4.5cm]{\footnotesize $\mbox{DEC}^{\mathcal{K}, \mathcal{K}_B}$};
\draw [-] (QRhat.west)--($(1cm,0cm)+(De.east)$)node[midway,below=0.4cm]{\footnotesize $\widehat{Q}_0$}-- (De.east);

\node (omega) at (-2,-4) {\footnotesize $\omega$};
\draw[black!50] ([xshift=-5pt,yshift=0pt]omega.north west) -- ([xshift=5pt,yshift=0pt]omega.north east)
      ++(0,0) -- ++(0,3pt) 
      ([xshift=-5pt,yshift=0pt]omega.north west) -- ++(0,3pt);
\node (rho) at (2,-4) {\footnotesize $\rho$};
\draw[black!50] ([xshift=-5pt,yshift=0pt]rho.north west) -- ([xshift=5pt,yshift=0pt]rho.north east)
      ++(0,0) -- ++(0,3pt) 
      ([xshift=-5pt,yshift=0pt]rho.north west) -- ++(0,3pt);
\node (rho2) at (6.5,-4) {\footnotesize $\rho$};
\draw[black!50] ([xshift=-5pt,yshift=0pt]rho2.north west) -- ([xshift=5pt,yshift=0pt]rho2.north east)
      ++(0,0) -- ++(0,3pt) 
      ([xshift=-5pt,yshift=0pt]rho2.north west) -- ++(0,3pt);
\node (omega2) at (11,-4) {\footnotesize $\sigma$};
\draw[black!50] ([xshift=-5pt,yshift=0pt]omega2.north west) -- ([xshift=5pt,yshift=0pt]omega2.north east)
      ++(0,0) -- ++(0,3pt) 
      ([xshift=-5pt,yshift=0pt]omega2.north west) -- ++(0,3pt);

\node (Q1) at (2,0.75) {\footnotesize $Q_1$};
\node (Q2) at (2,0.25) {\footnotesize $Q_2$};
\node (vdots) at (2,-0.15) {\footnotesize $\vdots$};
\node (QN) at (2,-0.75) {\footnotesize $Q_N$};
\draw (ENC.east|-Q1) -- (Q1)--(Q1-|Ne.west);
\draw (ENC.east|-Q2) -- (Q2)--(Q2-|Ne.west);
\draw (ENC.east|-QN) -- (QN)--(QN-|Ne.west);

\node (Qerased) at (6.5,0.5) {\footnotesize $(Q_n: n\in [N]\setminus \mathcal{K})$};
\draw (Ne.east|-Qerased)-- (Qerased);
\node (Qunerased) at (6.5,-0.5) {\footnotesize $(Q_n: n\in \mathcal{K})$};
\draw (Ne.east|-Qunerased) -- (Qunerased)--(Qunerased-|De.west);

\node (Berased) at (6.5,-2) {\footnotesize $(B_i: i\in [N_B]\setminus \mathcal{K}_B)$};
\draw (NBe.east|-Berased)-- (Berased);
\node (Bunerased) at (6.5,-3) {\footnotesize $(B_i: i\in \mathcal{K}_B)$};
\draw (NBe.east|-Bunerased) -- (Bunerased)--(Bunerased-|De.west);

\end{tikzpicture}
\caption{The $\widetilde{\mbox{QS}}\widetilde{\mbox{EA}}$ setting studied in this work. Note that the entanglement assistance systems $B$ are also subject to erasures.}
\label{fig:channel}
\end{figure}
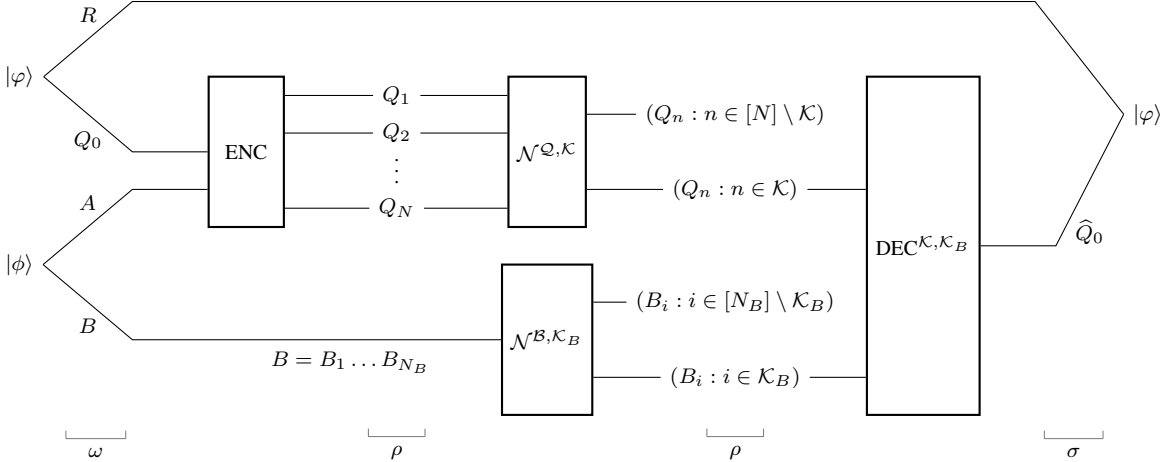

Building upon the foundation laid in \cite{Grassl_Huber_Winter} for a Shannon theoretic study of  $\widetilde{\mbox{QS}}\mbox{EA}$, we incorporate EA erasures and define a $\widetilde{\mbox{QS}}\widetilde{\mbox{EA}}(N,K,N_B,K_B)$ problem. The schematic for $\widetilde{\mbox{QS}}\mbox{EA}$ appears in Figure \ref{fig:channel}. The formal problem description appears in Section \ref{sec:defqsea}. An informal description is provided here. In the $\widetilde{\mbox{QS}}\widetilde{\mbox{EA}}(N,K,N_B,K_B)$ problem, a quantum message is encoded into $N$ storage systems (quantum systems $Q_1\dots Q_N$) with assistance from $N_B$ maximally entangled bi-partite quantum systems $A_1B_1, \dots, A_{N_B}B_{N_B}$, that are prepared in advance such that $B_1\dots B_{N_B}$ are stored separately as EA systems, while $A_1\dots A_{N_B}$ are made available to the encoder. Both the storage systems and EA systems are erasure-prone. The quantum message must be recoverable given any $K$ of the $N$ storage systems along with any $K_B$ of the $N_B$ EA systems. The information theoretic capacity for this setting is the maximum size of the quantum message, given that the size of each EA system is $\lambda_B$. All system sizes are relative to the size of a storage system, which is normalized to unity. The goal of this work is to characterize the information theoretic capacity of $\widetilde{\mbox{QS}}\widetilde{\mbox{EA}}(N,K,N_B,K_B)$. Note that setting $N_B=K_B$ would mean no EA systems are erased,  reducing our $\widetilde{\mbox{QS}}\widetilde{\mbox{EA}}$ setting to the $\widetilde{\mbox{QS}}\mbox{EA}$ setting. It is noteworthy that there are certain cases identified in \cite{Grassl_Huber_Winter} for which the capacity of $\widetilde{\mbox{QS}}\mbox{EA}$ is open, which will be resolved (see Observation 6 following Theorem \ref{thm:region}) as a byproduct of our capacity results for $\widetilde{\mbox{QS}}\widetilde{\mbox{EA}}$. A brief summary of our results is presented next.

As the main result of this work, the exact capacity of  $\widetilde{\mbox{QS}}\widetilde{\mbox{EA}}$ is characterized as a function of $N,K,N_B,K_B, \lambda_B$ in all cases, with one exception. The capacity remains open for an intermediate range of $\lambda_B$ values, specifically $K/(2K_B) <\lambda_B< (N-2K)/N_B+K/K_B$,  when a strict majority of the $N$ storage systems, and a strict non-zero minority of the $N_B$ EA systems, are erased, i.e., $K<N/2$ and $N_B>K_B> N_B/2$. For the setting that remains open, we present a lower bound (a feasible coding scheme) that we conjecture (Observation $1$ following Theorem \ref{thm:region}) to be information theoretically optimal. As a key stepping stone for $\widetilde{\mbox{QS}}\widetilde{\mbox{EA}}(N,K,N_B,K_B)$, an analogous classical storage (with shared-randomness assistance (SRA)) problem, called $\widetilde{\mbox{CS}}\widetilde{\mbox{SRA}}(N,K,N_B,K_B)$, is introduced, where both the classical storage and the SRA are erasure-prone. A set of constraints is identified for the classical problem, such that classical linear code constructions translate to quantum storage codes via a CSS construction.  It is worth noting that even after the corresponding classical problem $\widetilde{\mbox{CS}}\widetilde{\mbox{SRA}}$ is identified, challenges remain on both achievability and converse fronts. On the achievability side, since $\widetilde{\mbox{CS}}\widetilde{\mbox{SRA}}$ has not been previously studied, new optimal codes have to be designed for this problem. As it turns out these optimal codes involve rather non-trivial alignment structures. On the converse side, first a classical converse is needed to test the optimality of the codes for $\widetilde{\mbox{CS}}\widetilde{\mbox{SRA}}$, and then after the codes are translated into quantum codes via the CSS construction, a tight quantum information-theoretic converse is still needed to establish the information-theoretic optimality of the corresponding quantum code. Yet, it is remarkable that the capacity characterizations for the classical and quantum settings are found to be identical in all cases where the capacity is settled.

\begin{figure}[thbp]
\begin{center}
\begin{tikzpicture}
\node (Box1) at (0,0) [draw, rectangle, anchor=west]{\scalebox{0.9}{\begin{tikzpicture}[xscale=0.45]
\node (Q0) at (-1,2) [draw, circle, inner sep=0.1cm, thick, fill=blue!10] {\footnotesize ${Q}_0$};
\node (A1) at (1,2) [draw, circle, inner sep=0.1cm, thick, fill=red!10] {\footnotesize ${A}_1$};
\node (A2) at (3,2) [draw, circle, inner sep=0.1cm, thick, fill=red!10] {\footnotesize ${A}_2$};
\node (A3) at (5,2) [draw, circle, inner sep=0.1cm, thick, fill=red!10] {\footnotesize ${A}_3$};

\node (name) [right=1.51cm of A3] {\footnotesize $\underline{\widetilde{\mbox{QS}}\widetilde{\mbox{EA}}(3,1,3,2)}$};

\node (Q1) at (0,0) [draw, circle, inner sep=0.1cm, thick] {\footnotesize ${Q}_1$};
\node (Q2) at (2,0) [draw, circle, inner sep=0.1cm, thick] {\footnotesize ${Q}_2$};
\node (Q3) at (4,0) [draw, circle, inner sep=0.1cm, thick] {\footnotesize ${Q}_3$};
\node (B1) at (8,0) [draw, circle, inner sep=0.1cm, thick, fill=red!10] {\footnotesize ${B}_1$};
\node (B2) at (10,0) [draw, circle, inner sep=0.1cm, thick, fill=red!10] {\footnotesize ${B}_2$};
\node (B3) at (12,0) [draw, circle, inner sep=0.1cm, thick, fill=red!10] {\footnotesize ${B}_3$};

\node (ENC)   [draw, thick, rectangle, minimum width = 3.6cm, minimum height=0.5cm, above=0.34cm of Q2, fill=black!10] {\footnotesize ENC};

\node (DEC1)   at (-1,-1.5) [draw, rectangle, minimum width = 1cm, minimum height=0.5cm, black!50] {\footnotesize $\mbox{DEC}^{\{1\},\{1,2\}}$};

\node (DEC3)   at (5.2,-1.5) [draw, thick, rectangle, minimum width = 1cm, minimum height=0.5cm, fill=black!10] {\footnotesize $\mbox{DEC}^{\{2\},\{1,3\}}$};

\node [left=0cm of DEC3]{$\cdots$};
\node [right=0cm of DEC3]{$\cdots$};

\node (DEC5)   at (11.3,-1.5) [draw, rectangle, minimum width = 1cm, minimum height=0.5cm, black!50] {\footnotesize $\mbox{DEC}^{\{3\},\{2,3\}}$ };

\draw[-latex, dashed, black!50](Q1.south)--(DEC1.north);
\draw[-latex, dashed, black!50](B1.south)--(DEC1.north);
\draw[-latex, dashed, black!50](B2.south)--(DEC1.north);

\draw[-latex, thick](Q2.south)--(DEC3.north);
\draw[-latex, thick](B1.south)--(DEC3.north);
\draw[-latex, thick](B3.south)--(DEC3.north);

\draw[-latex, dashed, black!50](Q3.south)--(DEC5.north);
\draw[-latex, dashed, black!50](B2.south)--(DEC5.north);
\draw[-latex, dashed, black!50](B3.south)--(DEC5.north);

\node (Q0hat) at (-1,2) [draw, circle, inner sep=0.1cm, thick, fill=blue!10, below=0.5cm of DEC3] {\footnotesize $\widehat{Q}_0$};
\draw [-latex] (DEC3)--(Q0hat);
\draw [-latex] (Q0.south) -- (Q0.south |-, |-ENC.north);
\draw [-latex] (A1.south) -- (A1.south |-, |-ENC.north);
\draw [-latex] (A2.south) -- (A2.south |-, |-ENC.north);
\draw [-latex] (A3.south) -- (A3.south |-, |-ENC.north);
\draw [-latex] (Q1.north |-, |-ENC.south)--(Q1.north);
\draw [-latex] (Q2.north |-, |-ENC.south)--(Q2.north);
\draw [-latex] (Q3.north |-, |-ENC.south)--(Q3.north);
\end{tikzpicture}}};

\def\gapa{0.1cm}

\node (Q2C) at ($(Box1.east|-,|-Box1.north)+(1,0)$) [draw,rectangle, anchor=north west]{\scalebox{0.81}{\small 
$\begin{array}{l}
\underline{\widetilde{\mbox{QS}}\widetilde{\mbox{EA}}(3,1,3,2)\rightarrow\widetilde{\mbox{CS}}\widetilde{\mbox{SRA}}(3,1,3,2)}\\[0.2cm]
\mbox{Message: }Q_0\rightarrow Y_{0}\sim\mbox{Unif.}(\mathbb{F}_q^{\kappa\lambda_0})\\[\gapa]
\mbox{Storage: }Q_n\rightarrow Y_{n}\in\mathbb{F}_q^{\kappa}, n\in[3]\\[\gapa]
\underbrace{A_nB_n}_{(EA)}\rightarrow \underbrace{A_n=B_n}_{(SRA)}\stackrel{i.i.d.}{\sim}\mbox{Unif.}(\mathbb{F}_q^{\kappa\lambda_B}), n\in[3]\\[\gapa]
\mbox{EA: Entanglement Assistance}\\[\gapa]
\mbox{SRA: Shared-Randomness Assistance}\\[0.45cm]
\underline{\mbox{Entropic Statements of Constraints}}\\[\gapa]
H(Y_0\mid Y_i,B_j,B_k)=0,~~ i\in[3],~j,k\in[3], j\neq k\\[\gapa]
I(Y_0;Y_i,Y_j,B_k)=0,~~ i,j\in[3], i\neq j, k\in[3]\\[\gapa]
{\color{blue}H(B_1,B_2,B_3\mid Y_1,Y_2,Y_3)=0}
\end{array}
$}
};
\def\gapa{0.2cm}
\node (Qcode) at ($(Box1.west|-,|-Box1.south)+(0,-0.25)$) [draw,rectangle, anchor=north west]{\scalebox{0.9}{\small 
$\begin{array}{l}
\mbox{\underline{$\widetilde{\mbox{QS}}\widetilde{\mbox{EA}}(3,1,3,2)$ code}: For all $\bm{a}\in\mathbb{F}_q^3$, map  $\ket{\bm{a}}_{Q_0}\sum_{\bm{b}}\ket{\bm{b}}_A\ket{\bm{b}}_B$ to } \sum_{\bm{b}}\ket{Y_1Y_2Y_3}_{Q_1Q_2Q_3}\ket{\bm{b}}_B\\[\gapa]
\bm{a}\triangleq (a_1,a_2,a_3), ~~~\bm{b}\triangleq (\bm{b}^{(1)},\bm{b}^{(2)},\bm{b}^{(3)}), ~~~~~~\bm{b}^{(n)}\triangleq (b_0^{(n)}, b_{11}^{(n)},b_{12}^{(n)},b_{21}^{(n)},b_{22}^{(n)})\in\mathbb{F}_q^5, n\in[3]\\[\gapa]
\mbox{Quantum Systems: }A\triangleq A_1A_2A_3, ~~~B\triangleq B_1B_2B_3\\[\gapa]
\end{array}
$}
};
\def\gapa{0.5cm}
\node (Sol) at ($(Q2C.east|-,|-Qcode.south)+(0,-0.25)$)[draw,rectangle, anchor=north east]{\scalebox{0.84}{\small 
$
\def\gapa{0.2cm}
\begin{array}{l}
\underline{\widetilde{\mbox{CS}}\widetilde{\mbox{SRA}}(3,1,3,2) \mbox{ code}}\\[0.2cm]
\mbox{$\mathbb{F}_q$-linear, ~~~$q:$ any prime power}\\[\gapa]
\kappa=6, \lambda_0=1/2, \lambda_B=5/6\\[\gapa]
Y_0=(a_1,a_2,a_3), ~~~~~a_4\triangleq a_1+a_2\\[\gapa]
B_n=(b_0^{(n)}, b_{11}^{(n)},b_{12}^{(n)},b_{21}^{(n)},b_{22}^{(n)}), n\in[3]\\[\gapa]
b_{3i}^{(n)}\triangleq b_{1i}^{(n)}+b_{2i}^{(n)}, ~~~i\in[2], n\in[3]
\end{array}
\hspace{1cm}
\def\gapa{0.7cm}
\begin{array}{l}
Y_1=\begin{bmatrix}
a_1+b_0^{(2)}+b_{11}^{(1)}&a_3+b_0^{(3)}+b_{11}^{(2)}&a_3+b_0^{(1)}+b_{11}^{(3)} \\
a_2+b_0^{(3)}+b_{12}^{(1)}&a_4+b_0^{(1)}+b_{12}^{(2)}&a_2+b_0^{(2)}+b_{12}^{(3)} 
\end{bmatrix}\\[\gapa]
Y_2=\begin{bmatrix}
a_1+b_0^{(2)}+b_{21}^{(1)}&a_3+b_0^{(3)}+b_{21}^{(2)}&a_3+b_0^{(1)}+b_{21}^{(3)} \\
a_2+b_0^{(3)}+b_{22}^{(1)}&a_4+b_0^{(1)}+b_{22}^{(2)}&a_2+b_0^{(2)}+b_{22}^{(3)} \\
\end{bmatrix}\\[\gapa]
Y_3=\begin{bmatrix}
a_1+b_0^{(2)}+b_{31}^{(1)}&a_3+b_0^{(3)}+b_{31}^{(2)}&a_3+b_0^{(1)}+b_{31}^{(3)} \\
a_2+b_0^{(3)}+b_{32}^{(1)}&a_4+b_0^{(1)}+b_{32}^{(2)}&a_2+b_0^{(2)}+b_{32}^{(3)}
\end{bmatrix}
\end{array}
$}
};

\draw[thick, -latex] ($(Box1.east|-,|-Q2C.east)+(0,0)$)--(Q2C);
\draw[thick, -latex] ($(Q2C.south)+(3,0)$)--($(Q2C.south|-,|-Sol.north)+(3,0)$);
\draw[thick, -latex] ($(Q2C.south|-,|-Sol.north)+(2,0)$)|-(Qcode.east);
\end{tikzpicture}
\end{center}
\caption{Constructing a $\widetilde{\mbox{QS}}\widetilde{\mbox{EA}}$ code from a $\widetilde{\mbox{CS}}\widetilde{\mbox{SRA}}$ code for $(N,K,N_B,K_B)=(3,1,3,2)$ and $(\lambda_0,\lambda_B)=(1/2,5/6)$. First, quantum systems are mapped to classical systems. In addition to the decodability and security constraints identified in prior works \cite{Sun_Jafar_QuStorage, Smith, Hayashi_Song}, here the constraints for the classical setting also include (shown in blue) recovery of the shared-randomness from the storage systems. Next, a  classical code is constructed by optimally aligning shared-randomness, local randomness (not needed in this example) and message terms to satisfy all constraints. Finally, the classical code is mapped to a quantum code via the CSS construction \cite{Calderbank_Shor, Steane}.}\label{fig:flowchart}
\end{figure}
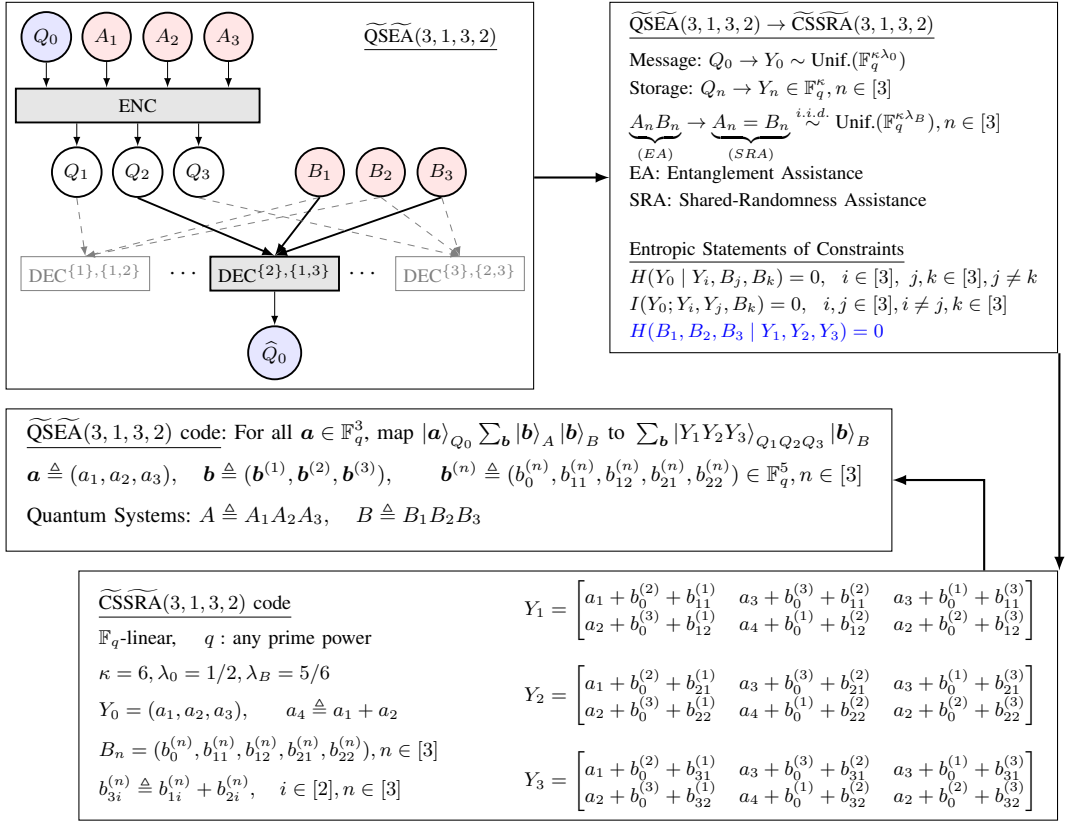

Figure \ref{fig:flowchart} previews briefly via an example how a classical code  for $\widetilde{\mbox{CS}}\widetilde{\mbox{SRA}}$ is translated into a quantum code for $\widetilde{\mbox{QS}}\widetilde{\mbox{EA}}$.  The setting illustrated in the figure is $\widetilde{\mbox{QS}}\widetilde{\mbox{EA}}(3,1,3,2)$, where any $N-K=2$ out of $N=3$ storage systems and any $N_B-K_B=1$ out of $N_B=3$ EA systems can be erased. A decoder is specified for each possible erasure scenario, and the goal is to find the maximum size $\lambda_0$ of the quantum message that can be stored and perfectly recovered in every allowed erasure scenario, when all storage systems have size normalized to unity and all EA systems $(A_1,A_2,A_3,B_1,B_2,B_3)$ have size $\lambda_B$. When the $\widetilde{\mbox{QS}}\widetilde{\mbox{EA}}$ setting is mapped to a  $\widetilde{\mbox{CS}}\widetilde{\mbox{SRA}}$ setting, all quantum systems become classical systems, and we are left with a classical problem to solve, subject to certain entropic constraints. The constraints are listed in the figure. Detailed explanations are relegated to subsequent sections. The optimal classical code (construction of $Y_1, Y_2, Y_3$) is also shown in the figure, and it is not difficult to verify that the construction satisfies all entropic constraints, mainly because of the careful alignments of message and SR terms. The $\widetilde{\mbox{CS}}\widetilde{\mbox{SRA}}$ code is then mapped to a $\widetilde{\mbox{QS}}\widetilde{\mbox{EA}}$ code via the CSS construction \cite{Calderbank_Shor, Steane}. The alignments reflect the non-trivial nature of code design for $\widetilde{\mbox{CS}}\widetilde{\mbox{SRA}}(N,K,N_B,K_B)$, which is inherited by the optimal quantum codes for $\widetilde{\mbox{QS}}\widetilde{\mbox{EA}}(N,K,N_B,K_B)$. 

The construction shown in Figure \ref{fig:flowchart} achieves storage size $\lambda_0=1/2$ with $\lambda_B=5/6$. Information theoretic converse arguments, not represented in Figure \ref{fig:flowchart}, but established later in the paper (Theorem \ref{thm:region} in Section \ref{sec:capresult}), imply that the message size $\lambda_0=1/2$ is optimal not only for $\lambda_B=5/6$ but also for any value of $\lambda_B\geq 5/6$, i.e., larger sizes of EA systems do not increase the storage capacity for this setting. The open problem in the context of  Figure \ref{fig:flowchart} corresponds to the question: is the message size $\lambda_0=1/2$ also achievable with $\lambda_B$ strictly smaller than $5/6$?  If our conjecture holds, then applied to this example it would imply that $\lambda_0=1/2$ cannot be achieved by any coding scheme with EA system sizes strictly smaller than $5/6$. 

Lastly, note that the coding scheme shown  in Figure \ref{fig:flowchart} does not require particularly large alphabet. Indeed, it works over any finite field $\mathbb{F}_q$, including the binary field $\mathbb{F}_2$. Over $\mathbb{F}_2$ it corresponds to the setting where each quantum storage system has size $6$ qubits, each EA system has size $5$ qubits, and the quantum message has size $3$ qubits. Generally in this work, since we pursue a Shannon theoretic formulation of storage capacity, which is not sensitive to field size, we will not attempt to minimize the field size for our coding schemes. Indeed, in Section \ref{ex:very2} we present the same example from a slightly different perspective that is easier to generalize, but requires a larger field size.

\vspace{0.1in}

\subsection{Notation} \label{sec:notation}
$\mathbb{Z}, \mathbb{R}, \mathbb{C}$ are the sets of integers, reals, and complex numbers respectively. The notations  $\mathbb{Z}_{\geq a}$  $(\mathbb{R}_{\geq a})$ represent the sets of integers (reals) greater than or equal to $a$, respectively. $[N]$ denotes the set $\{1,2,\cdots, N\}, N \in \mathbb{Z}_{\geq 1}$. For $i,j \in \mathbb{Z}$, the notation $[i:j]$ denotes the set $\{i, i+1, \cdots, j\}$ when $i \leq j$ and the empty set otherwise.  
$|X|$ denotes the cardinality of $X$ if $X$ is a set, the support of $X$ if $X$ is a classical random variable, and the dimension of the Hilbert space associated with $X$ if $X$ is a quantum system. 
For a set $\mathcal{X}$, the set of its cardinality-$k$ subsets is denoted as $\binom{\mathcal{X}}{k} \triangleq \{ \mathcal{Y} : \mathcal{Y} \subset \mathcal{X}, |\mathcal{Y}| = k \}$, and the notation $Q_{\mathcal{X}}$ denotes  $(Q_i: i \in \mathcal{X})$.
For two sets $\mathcal{X}, \mathcal{Y}$, the notation $\mathcal{X} \setminus \mathcal{Y}$ denotes the set of elements that are in $\mathcal{X}$ but not in $\mathcal{Y}$. We will frequently use the notation $\mathcal{K}\in\binom{[N]}{K}$ and $\mathcal{K}_B\in\binom{[N_B]}{K_B}$ to represent subsets of $[N]$ and $[N_B]$ of size $K$ and $K_B$, respectively. It will be convenient to denote their complements as $\mathcal{K}^c=[N]\setminus\mathcal{K}$ and $\mathcal{K}_B^c=[N_B]\setminus\mathcal{K}_B$, respectively.
For matrices ${\bf M}_1, {\bf M}_2$ of compatible dimensions, $({\bf M}_1, {\bf M}_2)$ and $({\bf M}_1; {\bf M}_2)$ represent their horizontal and vertical concatenations, respectively.  ${\bf M}^T$ represents the transpose of the matrix ${\bf M}$.  The notation ${\bf M}(\mathcal{I}, \mathcal{J})$ represents the sub-matrix of ${\bf M}$ formed by retaining only the rows from the index tuple $\mathcal{I}$ and the columns from the index tuple $\mathcal{J}$. For example, ${\bf M}((1,2),(3,4))$ represents the $2\times 2$ sub-matrix comprising rows $1,2$ and columns $3,4$ of ${\bf M}$. 
For quantum systems $X, Y$, the notation $X \rightsquigarrow Y$ denotes that a unitary (isometric) map is applied to $X$ so that it is transformed to $Y$. 
When $X$ is a set of classical random variables or quantum systems, $H(X)$ denotes the joint Shannon or von Neumann (quantum) entropy of the elements in $X$. $\mathbb{F}_q$ denotes the finite field with $q$ elements where $q$ is a power of a prime.
For non-zero vectors $u,v$, we write $u\equiv v$ to mean $u = v/||v||$, i.e., $u$ is the unit vector obtained by normalizing $v$. 

\section{Problem Statement}
\subsection{Preliminaries}
The dimension $|Q|$ of a quantum system $Q$ corresponds to the dimension of its associated Hilbert space $\mathcal{H}_Q$. Only finite dimensional Hilbert spaces are considered in this work.  Define $S(\mathcal{H}_Q)$  as the set of all states of $Q$, represented as density matrices (unit-trace positive semidefinite matrices) over $\mathcal{H}_Q$. 
The set of unit rank density matrices, $S_1(\mathcal{H}_Q)\triangleq\{\ket{x}\bra{x} : \ket{x}\in\mathcal{H}_Q, \braket{x|x}=1\}$ corresponds to pure states. Pure states can also be represented as unit vectors $\ket{x}\in\mathcal{H}_Q, \braket{x|x}=1$.
A composite system $Q_1Q_2\dots Q_N$ is associated with the Hilbert space $\mathcal{H}_{Q_1}\otimes \mathcal{H}_{Q_2}\otimes\cdots\otimes\mathcal{H}_{Q_N}$, where each $Q_n$ is associated with the Hilbert space $\mathcal{H}_{Q_n}$, $n\in[N]$.
A \emph{qudit}, representing a $q$-dimensional quantum system, is a unit for expressing the size of a quantum system. 
The size of a quantum system $Q$ is $\log_q|Q|$ qudits.  
The composite system $Q_1Q_2\dots Q_N$ has size $\log_q|Q_1Q_2\dots Q_N|=\log_q|Q_1|+\log_q|Q_2|+\cdots+\log_q|Q_N|$ qudits.

\subsection{Problem Formulation: $\widetilde{\mbox{QS}}\widetilde{\mbox{EA}}(N,K,N_B,K_B)$}\label{sec:defqsea}

The goal of $\widetilde{\mbox{QS}}\widetilde{\mbox{EA}}$  is to encode an arbitrary quantum message $Q_0$, into $N$ storage systems (quantum systems), $Q_1Q_2\dots Q_N$, with assistance from a quantum-entanglement resource $AB$, $A=A_1\dots A_{N_B}$, $B=B_1\dots B_{N_B}$ where all of $A$ is available to the encoder, such that given any $K$ of the $N$ storage systems $Q_1Q_2\dots Q_N$ and any $K_B$ of the $N_B$ EA systems $B=B_1\dots B_{N_B}$, there exists a decoder that can recover $Q_0$.  Note that $N,K, N_B,K_B\in\mathbb{Z}_{\geq 0}$ and $K\leq N$, $K_B\leq N_B$. To highlight key parameters we  refer to an instance of $\widetilde{\mbox{QS}}\widetilde{\mbox{EA}}$ as $\widetilde{\mbox{QS}}\widetilde{\mbox{EA}}(N,K,N_B,K_B)$. The problem is formalized by the following description of a coding scheme.
\subsubsection{Encoding and Decoding for $\widetilde{\mbox{QS}}\widetilde{\mbox{EA}}$}
A diagram of the encoding and decoding procedure is shown in Figure \ref{fig:channel}.

A  coding scheme for $\widetilde{\mbox{QS}}\widetilde{\mbox{EA}}(N,K,N_B,K_B)$ is specified by the parameters 
\begin{align}
\textstyle\left(\lambda_0,\lambda_B,\kappa,q,\mbox{ENC},\left(\mbox{DEC}^{\mathcal{K},\mathcal{K}_B}:\mathcal{K}\in\binom{[N]}{K}, \mathcal{K}_B\in\binom{[N_B]}{K_B}\right)\right),
\end{align}
where $\kappa\in\mathbb{Z}_{\geq 1}$, $q\in\mathbb{Z}_{\geq 2}$, and $q^{\kappa \lambda_0}, q^{\kappa \lambda_B} \in \mathbb{Z}_{\geq 1}$.
In order to emphasize the dependence of parameters on the coding scheme, given a  coding scheme identified by a label $\mathfrak{S}$, we may refer to its parameters as $\lambda_0(\mathfrak{S}),\lambda_B(\mathfrak{S}),\kappa(\mathfrak{S}),q(\mathfrak{S}),\mbox{ENC}(\mathfrak{S}),\mbox{DEC}^{\mathcal{K},\mathcal{K}_B}(\mathfrak{S})$. Usually however, as in the following discussion, the label for the coding scheme will be suppressed in order to simplify the notation if there is no potential for ambiguity.

 The scheme encodes an arbitrary quantum \emph{message}, $Q_0$, of size $\log_q|Q_0|=\kappa\lambda_0$ qudits. Associated with $Q_0$, define $R$, without loss of generality also of the same size $\log_q|R|=\kappa\lambda_0$ qudits, as a {\em reference} quantum system such that $RQ_0$ is an arbitrary pure state $\ket{\varphi}\bra{\varphi}_{RQ_0}\in S_1(\mathcal{H}_R\otimes \mathcal{H}_{Q_0})$.

The scheme utilizes EA, represented by a maximally entangled quantum system $AB$, where  $A = A_1\dots A_{N_B}$, and $B = B_1 \dots B_{N_B}$ are $N_B$-partite systems, with sizes  $\log_q|A_n| = \log_q|B_n|=\kappa\lambda_B$ qudits, $\forall n \in [N_B]$. The system $AB$ is in a pure state,
\begin{align}
&\ket{\phi}_{A B} =  \ket{\psi}_{A_1 B_1} \otimes \cdots \otimes \ket{\psi}_{A_{N_B} B_{N_B}}, \notag \\
&\ket{\psi}_{A_n B_n} \equiv  \sum_{\bm{j}\in q^{\kappa{\lambda_B}}} \ket{\bm{j}, \bm{j}}_{A_n B_n}, \hspace{1cm} n\in[N_B]. \label{ent}
\end{align}
$R Q_0$ and $AB$ are in a product state.
\begin{eqnarray}
\omega_{R Q_0 A B} = \ket{\varphi}\bra{\varphi}_{RQ_0} \otimes  \ket{\phi}\bra{\phi}_{A B}. \label{product}
\end{eqnarray}
The encoding operation applies the CPTP (completely positive trace preserving)  map `$\mbox{ENC}$', 
\begin{align}
& \mbox{ENC}: S(\mathcal{H}_{Q_0} \otimes \mathcal{H}_{A})\rightarrow S(\mathcal{H}_{Q_1}\otimes\cdots\otimes\mathcal{H}_{Q_N}),
\end{align}
 to the collective input comprised of the quantum message $Q_0$ and the entanglement resource $A$, and produces the quantum storage systems $Q_1Q_2\dots Q_N$. The size of each storage system, $\log_q|Q_n|=\kappa$ qudits, for all $n\in[N]$.
Following the encoding, the resulting state of $RQ_1Q_2\dots Q_N B$,
\begin{align}
& \rho_{R Q_1 \dots Q_N B} = {I}_{R B} \otimes \mbox{ENC}_{Q_0 A \rightarrow Q_1 \dots Q_N} (\omega_{R Q_0 A B}).\label{eq:rho}
\end{align}
Next we describe the decoding operation. Let $\mathcal{K}, \mathcal{K}_B$ denote the sets of indices of storage and EA systems that remain available to the decoder, respectively, after the rest of the storage and EA systems are erased (see Figure  \ref{fig:channel}). For each $\mathcal{K} \in \binom{[N]}{K}, \mathcal{K}_B \in \binom{[N_B]}{K_B}$, the coding scheme specifies a CPTP decoding map,
\begin{align}
& \mbox{DEC}^{\mathcal{K}, \mathcal{K}_B}: S(\otimes_{i\in \mathcal{K}} \mathcal{H}_{Q_i} \otimes_{j \in \mathcal{K}_B} \mathcal{H}_{B_{j}} )\rightarrow S(\mathcal{H}_{Q_0}), \\
\intertext{which acts on the `unerased systems' to output a quantum system $\widehat{{Q}}_0$. The resulting state $\sigma$ following the decoding operation is expressed as follows.}
& \sigma_{R \widehat{Q}_0 Q_{\mathcal{K}^c} B_{\mathcal{K}_B^c} } = {I}_{R Q_{\mathcal{K}^c} B_{\mathcal{K}_B^c}} \otimes \mbox{DEC}^{\mathcal{K}, \mathcal{K}_B}_{Q_{\mathcal{K}} B_{\mathcal{K}_B} \rightarrow \widehat{Q}_0} (\rho_{R Q_1 \dots Q_N B_1 \dots B_{N_B}}).
\end{align}
Recall from Section \ref{sec:notation} that $\mathcal{K}^c\triangleq [N]\setminus\mathcal{K}$ and $\mathcal{K}_B^c\triangleq [N_B]\setminus\mathcal{K}_B$.
A coding scheme is feasible if it satisfies the recovery condition,
\begin{align}
\sigma_{R \widehat{Q}_0} &= \ket{\varphi}\bra{\varphi}_{RQ_0}, \label{q:dec}
\end{align}
for all $\ket{\varphi}\bra{\varphi}\in S_1(\mathcal{H}_R\otimes \mathcal{H}_{Q_0})=S_1(\mathbb{C}^{q^{\kappa\lambda_0}}\otimes \mathbb{C}^{q^{\kappa\lambda_0}})$ and for all $\mathcal{K} \in \binom{[N]}{K}, \mathcal{K}_B \in \binom{[N_B]}{K_B}$. 

\subsubsection{Storage Capacity Region and Storage Capacity for $\widetilde{\mbox{QS}}\widetilde{\mbox{EA}}(N,K,N_B,K_B)$}
Taking a Shannon theoretic perspective as in \cite{Sun_Jafar_QuStorage}, the dependence on $\kappa, q$ is relaxed by allowing coding schemes to freely optimize over these parameters. Accordingly,  a tuple $({\lambda}_0,{\lambda}_B)\in\mathbb{R}_{\geq 0}^2$ is  said to be \emph{achievable} for $\widetilde{\mbox{QS}}\widetilde{\mbox{EA}}(N,K,N_B,K_B)$ if there exists a feasible coding scheme $\mathfrak{S}$ with $\lambda_0(\mathfrak{S})\geq {\lambda}_0$ and $\lambda_B(\mathfrak{S})\leq {\lambda}_B$. The closure of the set of achievable $({\lambda}_0,{\lambda}_B)$ tuples is defined to be the storage capacity region $\mathcal{C}_Q(N,K,N_B,K_B)$. For each ${\lambda}_B\in\mathbb{R}_{\geq 0}$,  the storage capacity of $\widetilde{\mbox{QS}}\widetilde{\mbox{EA}}(N,K,N_B,K_B)$ is defined as
\begin{align}
C_Q(N,K,N_B,K_B,{\lambda}_B) &\triangleq \max\{{\lambda}_0: ({\lambda}_0,{\lambda}_B)\in \mathcal{C}_Q(N,K,N_B,K_B)\}.
\end{align}
The storage capacity region and the storage capacity may be abbreviated as $\mathcal{C}_Q, C_Q(\lambda_B)$, respectively, when the values of the remaining parameters are obvious from the context.

\subsection{Problem Formulation: $\widetilde{\mbox{CS}}\widetilde{\mbox{SRA}}(N,K,N_B,K_B)$}\label{sec:defcssra}
Closely related to $\widetilde{\mbox{QS}}\widetilde{\mbox{EA}}(N,K,N_B,K_B)$, here we introduce the problem of \emph{classical storage with shared-randomness assistance}, i.e., $\widetilde{\mbox{CS}}\widetilde{\mbox{SRA}}(N,K,N_B,K_B)$. The latter is a conceptually simpler setting that will turn out to be quite insightful for exploring $\widetilde{\mbox{QS}}\widetilde{\mbox{EA}}(N,K,N_B,K_B)$. To emphasize the relationship between the two problems, we use similar notation for corresponding quantities.

The goal of $\widetilde{\mbox{CS}}\widetilde{\mbox{SRA}}$  is to encode an arbitrary classical message $Y_0$, into $N$ classical storage systems, $Y_1,Y_2, \dots, Y_N$, with assistance from shared-randomness (SR) resource $AB$, $A=A_1\dots A_{N_B}$, $B=B_1\dots B_{N_B}$ where all of $A$ is available to the encoder, such that given any $K$ of the $N$ storage systems $Y_1, Y_2, \dots, Y_N$ and any $K_B$ of the $N_B$ SRA systems $B_1, \dots, B_{N_B}$, there exists a decoder that can recover $Y_0$.  Since classical information can be copied, in this case it suffices to set $A_n=B_n$ for all $n\in[N_B]$. Note that $N,K, N_B,K_B\in\mathbb{Z}_{\geq 0}$ and $K\leq N$, $K_B\leq N_B$.  The problem is formalized by the following description of a coding scheme.
\subsubsection{Encoding and Decoding for $\widetilde{\mbox{CS}}\widetilde{\mbox{SRA}}$}\label{sec:enccssra}
A  coding scheme  for $\widetilde{\mbox{CS}}\widetilde{\mbox{SRA}}(N,K,N_B,K_B)$ is specified by the parameters 
\begin{align}
\textstyle\left(\lambda_0,\lambda_B,\kappa,q,\mbox{ENC},\left(\mbox{DEC}^{\mathcal{K},\mathcal{K}_B}:\mathcal{K}\in\binom{[N]}{K}, \mathcal{K}_B\in\binom{[N_B]}{K_B}\right)\right),
\end{align}
where $\kappa\in\mathbb{Z}_{\geq 1}$, $q\in\mathbb{Z}_{\geq 2}$, and $q^{\kappa \lambda_0}, q^{\kappa \lambda_B} \in \mathbb{Z}_{\geq 1}$. 
For  a  coding scheme labeled $\mathfrak{S}$, we may refer to its parameters as $\lambda_0(\mathfrak{S}),\lambda_B(\mathfrak{S}),\kappa(\mathfrak{S}),q(\mathfrak{S}),\mbox{ENC}(\mathfrak{S}),\mbox{DEC}^{\mathcal{K},\mathcal{K}_B}(\mathfrak{S})$. The label identifying the coding scheme will be suppressed  if there is no potential for ambiguity.

 The scheme encodes a classical message, $Y_0$, drawn uniformly from its alphabet $\mathcal{Y}_0$, of size $\log_q|\mathcal{Y}_0|=\kappa\lambda_0$ \emph{dits}, where we define the term \emph{dit} to mean a $q$-ary classical symbol.
 \begin{align}
\Pr(Y_0=y_0) = \frac{1}{q^{\kappa\lambda_0}}, &&\forall  y_0\in\mathcal{Y}_0. \label{message}
\end{align}
 
The scheme utilizes SRA, represented by $B$, where  $B = (B_1, \dots, B_{N_B})\in\mathcal{B}^{N_B}$, with size  $\log_q|\mathcal{B}|=\kappa\lambda_B$ dits. We refer to each $B_i$ as an SR-system. The SR is uniformly distributed. 
\begin{align}
\Pr(B=\bm{b})=\frac{1}{q^{\kappa\lambda_B}},&&\forall \bm{b}\in \mathcal{B}^{N_B}.
\end{align}
The scheme also utilizes local randomness, represented by $Z\in\mathcal{Z}$, generated locally by the encoder.
The message, SR and local randomness are independent.
\begin{align}
\Pr(B=\bm{b},Z=\bm{z},Y_0=y_0)= \Pr(B=\bm{b})\Pr(Z=\bm{z})\Pr(Y_0=y_0),&&\forall \bm{b}\in \mathcal{B}^{N_B}, \bm{z}\in\mathcal{Z},y_0\in\mathcal{Y}_0. \label{ind}
\end{align}
The encoding operation applies the map $$\mbox{ENC}: \mathcal{Y}_0 \times \mathcal{B}^{N_B}\times \mathcal{Z}\rightarrow \mathcal{Y}^N,$$  to the collective input comprised of the classical message $Y_0$, the SR-resource $B$, and the local randomness-resource $Z$, and produces the storage systems $Y_1,Y_2,\dots,Y_N\in\mathcal{Y}$. 
 \begin{align}
(Y_1, \dots, Y_N) = \mbox{ENC}(Y_0, B,Z).
\end{align}
 The size of each storage system, $\log_q|\mathcal{Y}|=\kappa$ dits.

Next consider the decoding operation. Let $\mathcal{K}, \mathcal{K}_B$ denote the sets of indices of storage and SR systems available to the decoder, respectively, after the remaining systems are erased. For each $\mathcal{K} \in \binom{[N]}{K}, \mathcal{K}_B \in \binom{[N_B]}{K_B}$, the coding scheme specifies a  decoder $$\mbox{DEC}^{\mathcal{K}, \mathcal{K}_B}: \mathcal{Y}^K \times \mathcal{B}^{K_B}\rightarrow \mathcal{Y}_0,$$ which maps the unerased systems  to a decoded message,
\begin{align}
 \widehat{Y}_0 = \mbox{DEC}^{\mathcal{K}, \mathcal{K}_B}(Y_{\mathcal{K}} , B_{\mathcal{K}_B}),
\end{align}
where we define  $Y_{\mathcal{K}}=(Y_n, n \in \mathcal{K})$ and $B_{\mathcal{K}_B}=(B_i, i \in \mathcal{K}_B)$.
The coding scheme is feasible if it guarantees perfect decodability, i.e.,
\begin{align}
 \Pr(\widehat{Y}_0=Y_0)=1, &&\textstyle\forall\mathcal{K} \in \binom{[N]}{K}, \mathcal{K}_B \in \binom{[N_B]}{K_B}.\label{c:dec}
 \end{align}
The  decodability constraint can be expressed equivalently as the following entropic condition,
\begin{align}
H( Y_0 \mid Y_{\mathcal{K}} , B_{\mathcal{K}_B} ) = 0, &&\textstyle\forall\mathcal{K} \in \binom{[N]}{K}, \mathcal{K}_B \in \binom{[N_B]}{K_B}. \label{c:edec}
\end{align}
As a key distinctive feature, the $\widetilde{\mbox{CS}}\widetilde{\mbox{SRA}}(N,K,N_B,K_B)$ problem contains  two additional constraints, stated in entropic terms in \eqref{c:sec} and \eqref{c:decb} as follows.
\begin{enumerate}
\item Security Constraint:
The erased systems must not contain any information about the classical message.
\begin{align}
&I(Y_0 ; Y_{\mathcal{K}^c} , B_{\mathcal{K}^c_B}) = 0,&& \textstyle\forall\mathcal{K} \in \binom{[N]}{K}, \mathcal{K}_B \in \binom{[N_B]}{K_B}.  \label{c:sec}
\intertext{\item SR Recovery Constraint:
Collectively the storage systems $Y_1, \dots, Y_N$ must fully determine the SR $B$.}
&H(B \mid Y_1, \dots, Y_N) = 0. \label{c:decb}
\end{align}
\end{enumerate}
These constraints are imposed to strengthen the connection between $\widetilde{\mbox{CS}}\widetilde{\mbox{SRA}}(N,K,N_B,K_B)$ and $\widetilde{\mbox{QS}}\widetilde{\mbox{EA}}(N,K,N_B,K_B)$, so that insights from the former may transfer to the latter. The motivation for the security constraint  (\ref{c:sec})  is apparent as it corresponds to the no-cloning condition inherent in  $\widetilde{\mbox{QS}}\widetilde{\mbox{EA}}(N,K,N_B,K_B)$.  The motivation for the SR recovery constraint (\ref{c:decb}) is perhaps less obvious. As it turns out (Theorem \ref{thm:region} in Section \ref{sec:results}), imposing these constraints allows the capacity results for $\widetilde{\mbox{CS}}\widetilde{\mbox{SRA}}(N,K,N_B,K_B)$ to parallel those for $\widetilde{\mbox{QS}}\widetilde{\mbox{EA}}(N,K,N_B,K_B)$.

\subsubsection{Storage Capacity Region and Storage Capacity for $\widetilde{\mbox{CS}}\widetilde{\mbox{SRA}}(N,K,N_B,K_B)$}
A tuple $({\lambda}_0,{\lambda}_B)\in\mathbb{R}_{\geq 0}^2$ is  said to be \emph{achievable} for $\widetilde{\mbox{CS}}\widetilde{\mbox{SRA}}(N,K,N_B,K_B)$ if there exists a feasible coding scheme $\mathfrak{S}$ with $\lambda_0(\mathfrak{S})\geq \lambda_0$ and $\lambda_B(\mathfrak{S})\leq \lambda_B$. The closure of the set of achievable $(\lambda_0,\lambda_B)$ tuples is defined to be the storage capacity region $\mathcal{C}(N,K,N_B,K_B)$. For each $\lambda_B\in\mathbb{R}_{\geq 0}$,  the storage capacity of $\widetilde{\mbox{CS}}\widetilde{\mbox{SRA}}(N,K,N_B,K_B)$ is defined as
\begin{align}
C(N,K,N_B,K_B,{\lambda}_B) &\triangleq \max\{{\lambda}_0: ({\lambda}_0,{\lambda}_B)\in \mathcal{C}(N,K,N_B,K_B)\}.
\end{align}
The storage capacity region and the storage capacity may be abbreviated as $\mathcal{C}, C(\lambda_B)$, respectively, when the values of the remaining parameters are obvious from the context.

\section{Results}\label{sec:results}
\subsection{Capacity Characterizations for  $\widetilde{\mbox{QS}}\widetilde{\mbox{EA}}$ and $\widetilde{\mbox{CS}}\widetilde{\mbox{SRA}}$}\label{sec:capresult}

Define the following sets.
\begin{eqnarray}
&& \mathcal{R}_{cut} \triangleq \left\{ (\lambda_0, \lambda_B)\in\mathbb{R}^2_{\geq 0}: \lambda_0 \leq \max(2K-N,0) + \lambda_B\cdot\max(2K_B - N_B, 0)   \right\}, \\
&& \mathcal{R}_\infty \triangleq \left\{ (\lambda_0, \lambda_B)\in\mathbb{R}^2_{\geq 0}:  \lambda_0 \leq \min(N, 2K) - \min(N-K, K) \frac{N_B}{K_B} \right\},\\
&& \mathcal{R}_{1} \triangleq
\left\{  
(\lambda_0, \lambda_B)\in\mathbb{R}^2_{\geq 0}: 
\begin{array}{l}
\lambda_0 \leq K(2K_B-N_B) \frac{(N-2K)+N_B\lambda_B}{(N-2K)2K_B + KN_B}\end{array}
 \right\}.
\end{eqnarray}
Our main result  appears in the following theorem.
\begin{theorem} \label{thm:region}
For the  $\widetilde{\mbox{QS}}\widetilde{\mbox{EA}}(N,K,N_B,K_B)$ and $\widetilde{\mbox{CS}}\widetilde{\mbox{SRA}}(N,K,N_B,K_B)$ problems  defined in Sections \ref{sec:defqsea} and \ref{sec:defcssra}, the capacity regions $\mathcal{C}_Q$ and $\mathcal{C}$, respectively, are characterized as follows.
\begin{enumerate}
\item[] {\bf Case 1:} $K_B/N_B \leq 1/2$,
\begin{align}
 \forall X\in\{\mathcal{C}_Q, \mathcal{C}\},  &&X&=  \mathcal{R}_{cut}.
\intertext{\item[] {\bf Case 2:} $K_B/N_B > 1/2, K/N \geq 1/2$,}
\forall X\in\{\mathcal{C}_Q, \mathcal{C}\}, &&X&  =  \mathcal{R}_{cut} \cap \mathcal{R}_\infty.
\intertext{\item[] {\bf Case 3:} $K_B/N_B > 1/2, K/N < 1/2$,}
\forall X\in\{\mathcal{C}_Q, \mathcal{C}\},&&X& \subset \mathcal{R}_{cut} \cap \mathcal{R}_\infty,\\
&& X&\supset  \mathcal{R}_{cut} \cap \mathcal{R}_\infty \cap \mathcal{R}_{1}.
\end{align} 
\end{enumerate}
\end{theorem}

The proof of Theorem \ref{thm:region} is presented in Section \ref{sec:region}. In particular,  Section \ref{sec:qregion} considers the quantum case and Sections \ref{sec:cregion} and \ref{sec:c} consider the converse and achievability of the classical case, respectively.

\begin{figure}[h]
\center
\begin{tikzpicture}[scale=1.2]
\begin{scope}[shift={(0,0)},xscale=0.9]
\draw [-{Latex[length=1.5mm]}, thick] (0,0)--(3.7,0) node [above] {$\lambda_B$};
\draw [-{Latex[length=1.5mm]}, thick] (0,0)--(0,3) node [right] { $\lambda_0$};
\draw [thick] (0,1)--(1,2);
\draw [thick, black] (1,2)--(3.5,2);
\draw[thick, dashed] (0,2)--(1,2);
\draw[thick, dashed] (1,0)--(1,2);
\node at (0,-0.3) {\small $0$};
\node at (1,-0.3) {\small $\frac{N-K}{K_B}$};
\node at (-0.6,1) {\footnotesize $2K-N$};
\node at (-1.2, 2) {\footnotesize $N - (N-K)\frac{N_B}{K_B}$};
\end{scope}
\begin{scope}[shift={(6.3,0)},xscale=0.9]
\draw [-{Latex[length=1.5mm]}, thick] (0,0)--(5.1,0) node [above] {$\lambda_B$};
\draw [-{Latex[length=1.5mm]}, thick] (0,0)--(0,3) node [right] {$\lambda_0$};
\draw [thick] (0,0)--(1,1);
\draw [thick, blue] (1,1)--(3,2.3);
\draw [thick, black] (3,2.3)--(4.8,2.3);
\draw[thick, dashed] (2.3,2.3)--(3,2.3);
\draw[thick, dashed] (0,2.3)--(2.3,2.3);
\draw[thick, dashed] (0,1)--(1,1);
\draw[thick, dashed] (1,0)--(1,1);
\draw[thick, dashed] (3,0)--(3,2.3);
\draw[thick, dashed] (1,1)--(2.3,2.3);
\node at (0,-0.3) {\small $0$};
\node at (1,-0.3) {\small $\frac{K}{2} \frac{1}{K_B}$};
\node at (2.9,-0.3) {\small $\frac{N-2K}{N_B} + \frac{K}{K_B}$};
\node at (-1,1) {\footnotesize $\frac{K}{2}\left( \frac{2K_B - N_B}{K_B}\right)$};
\node at (-1.1, 2.3) {\footnotesize $2K\left(1 - \frac{N_B}{2K_B}\right)$};

\fill[pattern={Dots[distance=8pt]}] (0,0) -- (1,1) -- (3,2.3) -- (4.8,2.3) -- (5,0) -- (0,0);
\end{scope}
\end{tikzpicture}
\vspace{-0.1in} 
\caption{Left: Capacity region for Case $2$, i.e., when $K_B/N_B>1/2,K/N\geq 1/2$. Note that the upper boundary of the capacity region represents the capacity $C_Q(\lambda_B)=C(\lambda_B)$. Right: Inner and outer bounds for capacity region in Case $3$, i.e., when $K/N < 1/2, K_B/N_B > 1/2$. Note that the bounds match if $\lambda_B\leq K/(2K_B)$ or if $\lambda_B\geq (N-2K)/N_B+K/K_B$, providing an exact capacity characterization in both cases.}
\label{fig:2}
\end{figure}
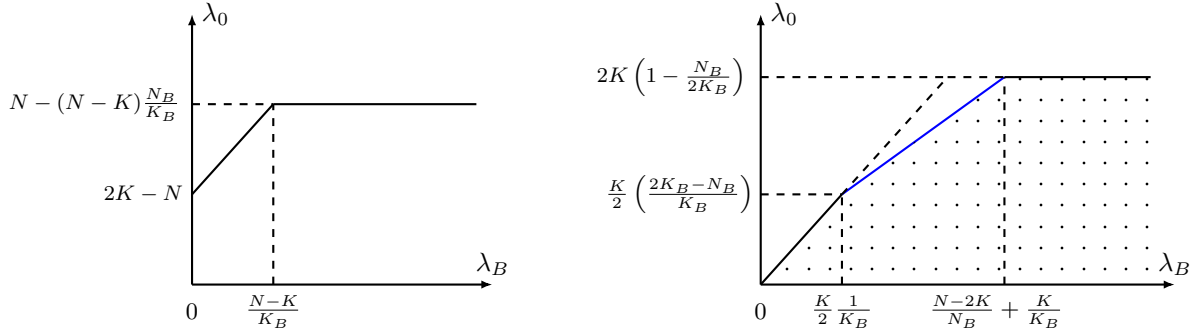

Theorem \ref{thm:region} implies that in Case 1, the capacity is simply $C_Q(\lambda_B)=C(\lambda_B)=  \max(2K-N,0)$, same as the capacity without assistance from entanglement or SR. Figure \ref{fig:2} illustrates the capacity regions for Case $2$ and Case $3$.

The following observations are in order.
\begin{enumerate}
\item $\mathcal{C}_{Q}$ vs $\mathcal{C}$: Our results on $\mathcal{C}_{Q}$ and $\mathcal{C}$ are identical in all cases, highlighting the relationship between the two problems. The exact capacities ${C}_{Q}$ and ${C}$ are characterized in all cases except if $K/N<1/2, K_B/N_B>1/2$ and $K/(2K_B)< \lambda_B<(N-2K)/N_B+K/K_B$. We conjecture that even in the case which remains open, the capacity regions are identical, so that $\mathcal{C}_{Q}=\mathcal{C}$ in all cases. Furthermore, we conjecture that in the case that remains open, it is our \emph{inner bound} that is tight, i.e., $\mathcal{C}_{Q}=\mathcal{C}=\mathcal{R}_{cut} \cap \mathcal{R}_\infty \cap \mathcal{R}_{1}$ in Case 3.

\item Achievability: Our coding schemes for $\widetilde{\mbox{QS}}\widetilde{\mbox{EA}}(N,K,N_B,K_B)$ are obtained by first designing coding schemes for classical setting of $\widetilde{\mbox{CS}}\widetilde{\mbox{SRA}}(N,K,N_B,K_B)$, and then translating them to the quantum setting via CSS codes 
to achieve the same rate tuple. The connection is developed in Section \ref{sec:CS2QS} and formalized in Theorem \ref{thm:ab}. Remarkably, all the extreme points (see Figure \ref{fig:2}) of our inner bound for $\mathcal{C}_{Q}$ are achieved by \emph{pure} coded states $\rho_{RQ_1\dots Q_NB_1\dots B_{N_B}}$ (see Figure \ref{fig:channel}).

\item Converse: Theorem \ref{thm:region} contains two converse bounds, i.e., the two hyperplanes in $\mathcal{R}_{cut}$ and $\mathcal{R}_{\infty}$. The converse bound in $\mathcal{R}_{cut}$ is
\begin{eqnarray}
\lambda_0 \leq \max(2K-N,0) + \lambda_B\cdot\max(2K_B - N_B, 0). \label{cut}
\end{eqnarray}
This bound  has a cut-set interpretation via the classical storage problem $\widetilde{\mbox{CS}}\widetilde{\mbox{SRA}}$ as follows. In (\ref{cut}), the LHS corresponds to the message size $H(Y_0)$. The first term on the RHS corresponds to the total storage size of the $K$ unerased storage systems after subtracting from those $K$ systems another $N-K$ systems which need to be kept secure, and therefore cannot provide information about the message. Intuitively, the first term on the RHS corresponds to $H(Y_{[K] \setminus [N-K]})$.  Similarly,  the second term on the RHS is the total storage size available in the $K_B$ unerased SR-systems, after removing $N_B-K_B$ of those systems, which need to be kept secure, i.e., $H(B_{[K_B] \setminus [N_B-K_B]})$. The insight extends to the quantum storage problem $\widetilde{\mbox{QS}}\widetilde{\mbox{EA}}$ by the connection between the two problems.

The converse bound in $\mathcal{R}_{\infty}$ is
\begin{eqnarray}
\lambda_0 \leq \min(N, 2K) - \min(N-K, K) \frac{N_B}{K_B}. \label{infinity}
\end{eqnarray}
This can be interpreted  as the unlimited EA (or unlimited SR) bound, corresponding to $\lambda_B\rightarrow\infty$. Note that the RHS of (\ref{infinity}) is equal to $K$, the number of surviving storage systems out of $Q_1\dots Q_N$, if $N_B=K_B$, i.e., EA is \emph{not} erasure-prone. On the other hand, if $K_B < N_B$, i.e., entanglement \emph{is} erasure-prone, then the RHS is strictly smaller than $K$. Thus, erasure-prone EA imposes a non-zero penalty, even if the size $\lambda_B$ of the entanglement systems is arbitrarily large, which limits the storage capacity to a value strictly  less than $K$. The exact magnitude of the penalty, which can be deduced from (\ref{infinity}), is perhaps not intuitively obvious, and depends on the ratio $N_B/K_B$. An example of (\ref{infinity}) when $(N,K,N_B, K_B) = (3,2,4,3)$ is given in Section \ref{sec:infinity} for the classical problem; the proof also generalizes to the quantum setting.

\item In Case 1, i.e., when $K_B/N_B \leq 1/2$, evidently EA is useless.

\item In Case 2, i.e., when $K/N \geq 1/2$ and $K_B/N_B > 1/2$, we have only one non-trivial extreme point (see Figure \ref{fig:2}). In the classical problem $\widetilde{\mbox{CS}}\widetilde{\mbox{SRA}}$, the SR $B$ is sufficient for the coding scheme and no local randomness $Z$ is used; correspondingly, for the quantum problem $\widetilde{\mbox{QS}}\widetilde{\mbox{EA}}$, no ancilla qudits are needed for the encoding map $\mbox{ENC}$, i.e., $\mbox{ENC}$ is unitary. An example of $(N,K,N_B,K_B) = (2,1,3,2)$ is given in Section \ref{ex:less} for the classical problem.

\item Case 3, where $K/N < 1/2, K_B/N_B > 1/2$, turns out to be the most challenging, and also the most interesting regime. Here more storage systems are erased than unerased $(N-K>K)$, so if there is no EA, then no message can be stored, i.e., capacity is zero. EA, even though it is itself erasure-prone, still improves the capacity (see Figure \ref{fig:2}). Remarkably, our inner bound (conjectured optimal) contains two non-trivial extreme points. Highlights of the codes are discussed below from the $\widetilde{\mbox{CS}}\widetilde{\mbox{SRA}}$ perspective.
\begin{enumerate}
\item Extreme point $(\lambda_0, \lambda_B) = \left(\frac{K}{2K_B}, \frac{K(2K_B-N_B)}{2K_B} \right)$: In addition to SR $B$, local randomness $Z$ is used by the encoder; correspondingly, the quantum  scheme has to carefully design ancilla qudits for the encoding. The classical (and quantum) codes involve rather delicate structures. An example is given in Section \ref{ex:very1} for the classical problem. For example, in (\ref{eq:ex1y}), each storage system $Y_n$ contains two parts $f, h$, where $f$ is some generic combination of $(Y_0,B)$, while $h$ is some generic combination of $(Y_0, Z, f_0)$, where $f_0$ is a part of $f$. Thus $h$ contains a mixture of the message $Y_0$, the same $f_0$ (across all $Y_1,\dots,Y_N$) which already contains the SR $B$, and the local randomness $Z$. Further, $f$ is repetition coded (MDS coded in general) across $Y_1\dots Y_N$ (see (\ref{eq:ex1y})).

\item Extreme point $(\lambda_0, \lambda_B) = \left(2K\left(1- \frac{N_B}{2K_B}\right), \frac{N-2K}{N_B}+\frac{K}{K_B} \right)$: The classical and quantum codes in this case again involve rather interesting alignment structure. Figure \ref{fig:flowchart} illustrates the structure via a small example, which  is presented again in Section \ref{ex:very2} from a more conceptual, and therefore more generalizable perspective, albeit for larger field sizes. 

\item Consider $N_B = K_B = 1$, so EA is not erasure-prone. The above two extreme points reduce to $(\lambda_0, \lambda_B) = (K/2, K/2)$ and $(\lambda_0, \lambda_B) = (K, N-K)$; space sharing of these two rate tuples achieves the blue line in Figure \ref{fig:2} (see Appendix \ref{app:convex}). 
This resolves an open problem\footnote{{\color{black} Remarkably,  the same space-sharing argument also proves the tightness of the entropic quantum Singleton bounds, which was left open in \cite{Mani_Winter} and \cite{Prasad_Grassl}, for settings that include both classical and quantum messages \cite{Mani_Winter} or only classical messages \cite{Prasad_Grassl}, respectively.}} in \cite{Grassl_Huber_Winter}, which left the achievability of the blue line segment open, while providing a proof that it represents a valid converse bound. Evidently, our inner bound for $\mathcal{C}_Q$ is indeed optimal when $N_B = K_B = 1$, lending support to our conjecture that our inner bound  is  \emph{always} optimal in Case $3$. 

\end{enumerate}
\end{enumerate}


\subsection{Translating a Linear Coding Scheme from $\widetilde{\mbox{CS}}\widetilde{\mbox{SRA}}$ to $\widetilde{\mbox{QS}}\widetilde{\mbox{EA}}$}\label{sec:CS2QS}
Recall from Section \ref{sec:enccssra} that a  $\widetilde{\mbox{CS}}\widetilde{\mbox{SRA}}(N,K,N_B,K_B)$ coding scheme   is specified by the parameters 
$\Big(\lambda_0,\lambda_B,\kappa,q,\mbox{ENC},$ $\left(\mbox{DEC}^{\mathcal{K},\mathcal{K}_B}:\mathcal{K}\in\binom{[N]}{K}, \mathcal{K}_B\in\binom{[N_B]}{K_B}\right)\Big)$. 

\begin{definition}[Linear Scheme] \label{def:linear}
A $\widetilde{\mbox{CS}}\widetilde{\mbox{SRA}}(N,K,N_B,K_B)$ coding scheme is \emph{linear} if, 
\begin{enumerate}
\item the parameter $q$ is a power of a prime, so that the finite field $\mathbb{F}_q$ exists, and 
\item the encoding map ENC is $\mathbb{F}_q$-linear, with the message, storage, SR and local randomness  represented as vectors over $\mathbb{F}_q$, i.e., message $Y_0\in\mathbb{F}_q^{1\times \kappa\lambda_0}$,  storage $Y_n\in\mathbb{F}_q^{1\times \kappa}$ for all $n\in[N]$, i.i.d. uniform SR $B_i\in\mathbb{F}_q^{1\times\kappa\lambda_B}$ for all $i\in[N_B]$, and local randomness $Z=(Z_1,\dots,Z_L)$ with i.i.d. uniform $Z_l \in \mathbb{F}_q$ for all $l\in[L]$.
\end{enumerate}
\end{definition}

Since the encoder for a linear scheme is $\mathbb{F}_q$-linear, it follows that it can be specified by matrices ${\bf A}_n\in\mathbb{F}_q^{\kappa\lambda_0\times\kappa}$, ${\bf B}_n\in\mathbb{F}_q^{N_B\kappa\lambda_B\times\kappa}$ and ${\bf Z}_n\in\mathbb{F}_q^{L\times\kappa}$ for all $n\in[N]$, such that
\begin{align}
(Y_1,\dots,Y_N)&=\mbox{ENC}(Y_0,B,Z)\\
&=Y_0\begin{bmatrix}{\bf A}_1&\dots&{\bf A}_N\end{bmatrix}+B\begin{bmatrix}{\bf B}_1&\dots&{\bf B}_N\end{bmatrix}
+Z\begin{bmatrix}{\bf Z}_1&\dots&{\bf Z}_N\end{bmatrix}.\label{eq:linenc}
\end{align}
Given the parameters $\lambda_0,\lambda_B,\kappa,q$ and matrices ${\bf A}_n,{\bf B}_n,{\bf Z}_n,n\in[N]$ that specify the encoder $\mbox{ENC}$ as in \eqref{eq:linenc},  note that the feasibility of the scheme is guaranteed (and the existence of suitable linear decoders is  implied) if the constraints (\ref{c:edec}), (\ref{c:sec}), (\ref{c:decb}) are satisfied.

A rate tuple $(\lambda_0,\lambda_B)$ is said to be linearly achievable if there exists a feasible linear scheme $\mathfrak{S}$ such that $\lambda_0(\mathfrak{S})\geq \lambda_0$ and $\lambda_B(\mathfrak{S})\leq\lambda_B$. The following theorem  formalizes the relationship between $\widetilde{\mbox{CS}}\widetilde{\mbox{SRA}}$ and $\widetilde{\mbox{QS}}\widetilde{\mbox{EA}}$ via linear achievability.

\begin{theorem}\label{thm:ab}
If a rate tuple $(\lambda_0, \lambda_B)$ is linearly achievable for $\widetilde{\mbox{CS}}\widetilde{\mbox{SRA}}(N,K,N_B,K_B)$, then the same rate tuple is achievable for $\widetilde{\mbox{QS}}\widetilde{\mbox{EA}}(N,K,N_B,K_B)$.
\end{theorem}
The proof of Theorem \ref{thm:ab} is presented in Section \ref{sec:ab}.
The proof  is a generalization of the corresponding proof of \cite[Theorem 2]{Sun_Jafar_QuStorage}. The latter setting assumes no entanglement (or SR) assistance, and the classical-quantum connection in such cases has been noted in the prior literature \cite{Sun_Jafar_QuStorage, Smith, Hayashi_Song}. To the best of our knowledge, the generalization in Theorem \ref{thm:ab} to include EA (and the corresponding SRA in the classical setting) is new, i.e., (erasure-prone) entanglement-assisted quantum storage codes have not been previously connected to (erasure-prone) SR-assisted classical storage codes.
Apparently, the main difference due to EA is that in order to ensure that the classical code can be translated to a quantum code where the quantum encoding map is unitary (or an isometry), we end up including the additional constraint (\ref{c:decb}) to the classical problem (i.e., we require recovery of  $B$ from $Y_{[N]}$).

\section{Examples}
In this section, we provide some examples that illustrate the key ideas in settings with smaller values of the parameters $N, K, N_B, K_B$, so that the notation is simpler. Sections \ref{ex:less}, \ref{ex:very1}, and \ref{ex:very2} present the achievable schemes for various extreme points of $\mathcal{C}$, and Section \ref{sec:infinity} presents the proof of the $\mathcal{R}_\infty$ (unlimited SR)  bound.

\subsection{$(N,K,N_B,K_B) = (2,1,3,2)$: Achievability of $(\lambda_0, \lambda_B) = (1/2, 1/2)$}\label{ex:less}
Consider the $\widetilde{\mbox{CS}}\widetilde{\mbox{SRA}}(2, 1, 3, 2)$ setting as our next example. We present a linear scheme that achieves the rate tuple $(\lambda_0, \lambda_B) = (1/2, 1/2)$. The parameters of the linear scheme are  as follows. 
\begin{enumerate}
\item Field size $q\geq 8$ is a prime power.
\item Scaling factor $\kappa = 2$.
\item Precoding matrices are chosen such that
\begin{eqnarray}
\begin{bmatrix} {\bf A}_1^{1\times 2} & {\bf A}_2^{1\times 2}\\[0.2cm]
{\bf B}_1^{3 \times 2} & {\bf B}_2^{3\times 2}\end{bmatrix} = {\bf H}_{4 \times 4} = \left[ \frac{1}{\alpha_i - \beta_j} \right]_{ij}.
\end{eqnarray}
$\alpha_i, \beta_j, i, j \in [4]$ are distinct elements in $\mathbb{F}_q$. Since $q\geq 8$, these distinct elements exist. Note that by construction ${\bf H}$ is a $4 \times 4$ Cauchy matrix with $\frac{1}{\alpha_i - \beta_j}$ being its element in the $i^{th}$ row and $j^{th}$ column. 
\end{enumerate}
Local randomness $Z$ is not needed for this scheme.  The storage systems are set as
\begin{eqnarray}
\left( Y_1, Y_2 \right)_{1 \times 4} = \left(Y_0, B_1, B_2, B_3 \right)_{1\times 4} \times {\bf H}, \label{eq:y12}
\end{eqnarray}
where $Y_0\in\mathbb{F}_q^{1\times\kappa\lambda_0}=\mathbb{F}_q$, the SR  $B_1, B_2, B_3\in\mathbb{F}_q^{1\times\kappa\lambda_B}=\mathbb{F}_q$, and the storage systems $Y_1, Y_2\in\mathbb{F}_q^{1\times\kappa}=\mathbb{F}_q^{1\times 2}$.

To show that the scheme works it suffices to verify that the constraints (\ref{c:edec}), (\ref{c:sec}),  (\ref{c:decb}) are satisfied. Let us consider the case $\mathcal{K} = \{1\}, \mathcal{K}_B = \{1,3\}$. Other cases follow similarly.

\begin{enumerate}
\item  (\ref{c:edec}): We need to show that $Y_0$ can be recovered from the unerased systems $Y_1, B_1, B_3$. Proceed as follows. Subtract the contributions of $B_1,B_3$ from $Y_1$ to obtain $(Y_0, B_2) \times {\bf H}_{sub}$ where ${\bf H}_{sub}={\bf H}((1,3),(1,2))$ is the $2\times 2$ sub-matrix of ${\bf H}$ corresponding to row-indices $1,3$, and column-indices $1,2$, according to \eqref{eq:y12}. As a square sub-matrix of a Cauchy matrix, ${\bf H}_{sub}$ is an invertible matrix. Invert ${\bf H}_{sub}$ to recover $Y_0$.
\item 
(\ref{c:sec}): We need to show that $Y_0$ is independent of $(Y_2, B_2)$. Proceed as follows.
\begin{eqnarray}
I\left( Y_0; Y_2, B_2 \right) &=& I\left( Y_0; Y_2 \mid B_2 \right) \label{eq:t1} \\
&\leq& H \left( Y_2  \right) - H\left( Y_2 \mid Y_0, B_2 \right) \label{eq:t1+}\\
&\leq& 2 - H\left(  \left(B_1, B_3 \right) {\bf H}_{sub}^{2\times 2}\mid Y_0,B_2 \right) \label{eq:t20}  \\
&=& 2 - H\left(  \left(B_1, B_3 \right) {\bf H}_{sub}^{2\times 2} \right) \label{eq:t2}  \\
&=& 2 - H\left(B_1, B_3 \right) \label{eq:t2+}  \\
&=& 2 - 2 = 0. 
\end{eqnarray}
(\ref{eq:t1}) follows from the independence of $Y_0$ and $B_2$ (refer to (\ref{ind})). \eqref{eq:t1+} follows from the fact that $H(Y_2|B_2)\leq H(Y_2)$ (conditioning reduces entropy). In (\ref{eq:t20}), the first term (expressed in $q$-ary units) follows from the facts that $Y_2$ contains $2$ symbols from $\mathbb{F}_q$,  and the uniform distribution maximizes Shannon entropy; the second term follows from subtracting the contribution of $Y_0, B_2$ from $Y_2$, to obtain $(B_1,B_3){\bf H}_{sub}$ where ${\bf H}_{sub}={\bf H}((2,4),(3,4))$. Step \eqref{eq:t2} uses the fact that $Y_0, B_1, B_2, B_3$ are independent. Step \eqref{eq:t2+} uses the fact that as a $2\times 2$ sub-matrix of a Cauchy matrix, ${\bf H}_{sub}$ is invertible. The last step uses the fact that $B_1,B_3$ are i.i.d. uniform symbols from $\mathbb{F}_q$, so that $H(B_1,B_3)=2$ in $q$-ary units.

\item  (\ref{c:decb}): We need to show that $B=(B_1,B_2,B_3)$ can be recovered from $(Y_1,Y_2)$. This follows immediately from (\ref{eq:y12}), because as a Cauchy matrix by construction, ${\bf H}$ is  invertible.
\end{enumerate}

\subsection{$(N,K,N_B,K_B) = (3,1,3,2)$: Achievability of $(\lambda_0, \lambda_B) = (1/4, 1/4)$} \label{ex:very1}
For this example, we present a linear scheme for $\widetilde{\mbox{CS}}\widetilde{\mbox{SRA}}(3, 1, 3, 2)$ that achieves the rate tuple $(\lambda_0, \lambda_B) = (1/4, 1/4)$. Local randomness $Z$ will be useful in this example. The parameters of the linear scheme are  as follows.

\begin{enumerate}
\item Field size $q \geq 13$ is a prime power.
\item Scaling factor $\kappa = 4$.
\item Local randomness $Z = (Z_1, Z_2, Z_3, Z_4) \in \mathbb{F}_q^{1\times 4}$ is uniformly distributed.
\item Precoding matrices are specified through the following steps. Note that $Y_0$ contains $\kappa \lambda_0 = 1$ element in $\mathbb{F}_q$ and $B_1, B_2, B_3$ each contains $\kappa \lambda_B = 1$ element in $\mathbb{F}_q$. 
First, choose a $4 \times 3$ Cauchy matrix ${\bf F} = \left[\frac{1}{\alpha_i - \beta_j}\right]_{ij}$ where the element in the $i^{th}$ row and $j^{th}$ column is $\frac{1}{\alpha_i - \beta_j}$ and $\alpha_i, \beta_j, i \in [4], j \in [3]$ are distinct elements in $\mathbb{F}_q$. Define
\begin{eqnarray}
\left(f_0, f_1, f_2\right)_{1\times 3} \triangleq \left(Y_0, B_1, B_2, B_3 \right)_{1 \times 4} \times {\bf F}_{4 \times 3}.\label{eq:deff012}
\end{eqnarray}
Second, choose a $6 \times 6$ Cauchy matrix ${\bf H} = \left[\frac{1}{\gamma_i - \delta_j}\right]_{ij}$ where the element in the $i^{th}$ row and $j^{th}$ column is $\frac{1}{\gamma_i - \delta_j}$ and $\gamma_i, \delta_j, i, j \in [6]$ are distinct elements in $\mathbb{F}_q$. Define
\begin{eqnarray}
\left(h_1, h_2, h_3, h_4, h_5, h_6\right)_{1\times 6} \triangleq \left(f_0, Y_0, Z_1, Z_2, Z_3, Z_4 \right)_{1 \times 6} \times {\bf H}_{6 \times 6}.
\end{eqnarray}
We are now ready to specify the storage systems. Note that $Y_1, Y_2, Y_3$ each contains $\kappa = 4$ elements in $\mathbb{F}_q$.
\begin{eqnarray}
Y_1 &=& \left( f_1, f_2, h_1, h_2 \right), \notag \\
Y_2 &=& \left( f_1, f_2, h_3, h_4 \right), \label{eq:ex1y}\\
Y_3 &=& \left( f_1, f_2, h_5, h_6 \right). \notag
\end{eqnarray}
Note that the first two elements in $Y_1, Y_2, Y_3$ are identical, i.e., $f_1, f_2$. The above encoding procedure is linear and thus the precoding matrices ${\bf A}_n, {\bf B}_n, {\bf Z}_n, n \in [3]$ are implied.
\end{enumerate}

To show that the scheme works we need to verify the constraints (\ref{c:edec}), (\ref{c:sec}),  (\ref{c:decb}). Consider the case $\mathcal{K} = \{1\}, \mathcal{K}_B = \{1,3\}$. Other cases follow similarly.
\begin{enumerate}
\item (\ref{c:edec}): 
We need to show that $Y_0$ is determined by $(Y_1, B_1, B_3)$. Proceed as follows. Subtracting the contribution of $B_1, B_3$ from $f_1, f_2$ (the first two elements of $Y_1$) in \eqref{eq:deff012}, we obtain $\left( Y_0, B_2 \right)_{1\times 2} \times {\bf F}_{sub}$. 
As a $2\times 2$ sub-matrix of a Cauchy matrix, ${\bf F}_{sub}={\bf F}((1,3), (2,3))_{2\times 2}$ is invertible. Invert ${\bf F}_{sub}$ to obtain $Y_0$.

\item (\ref{c:sec}): We need to show that $Y_0$ is independent of $Y_2, Y_3, B_2$. Proceed as follows.
\begin{align}
& I\left(Y_0; Y_2, Y_3, B_2 \right) \notag\\
&= I\left( Y_0; f_1, f_2, h_3, h_4, h_5, h_6 \mid B_2 \right) \label{eq:ex11} \\
&\leq  H\left(f_1, f_2, h_3, h_4, h_5, h_6\right)  -  H \left( f_1, f_2, h_3, h_4, h_5, h_6 \mid Y_0, B_2 \right) \label{eq:ex11+}\\
&\leq 6 -  H \left( f_1, f_2 \mid Y_0, B_2 \right) -  H \left( h_3, h_4, h_5, h_6 \mid f_1, f_2, Y_0, B_1, B_2, B_3 \right) \label{eq:ex12} \\
&\leq 6 -  H \left( (B_1, B_3) {\bf F}((2,4), (2,3)) \mid Y_0, B_2 \right) \notag\\
&\hspace{1cm}-  H \left( (Z_1, Z_2, Z_3, Z_4) {\bf H}\left((3,4,5,6), (3,4,5,6)\right) \mid Y_0, B_1, B_2, B_3 \right) \label{eq:ex13}\\
&= 6 - H(B_1, B_3 \mid Y_0, B_2 ) - H(Z_1, Z_2,Z_3, Z_4 \mid Y_0, B_1, B_2, B_3) \label{eq:ex14} \\ 
&= 6 - 2 - 4 = 0. \label{eq:ex15}
\end{align}
In (\ref{eq:ex11}), we plug in the design of $Y_2, Y_3$ and use the fact that message $Y_0$ is independent of SR $B_2$ (refer to (\ref{ind})). Step \eqref{eq:ex11+} follows because dropping conditioning cannot reduce entropy. In (\ref{eq:ex12}), the first term follows from the fact that $f_1, f_2, h_3, h_4, h_5, h_6$ are $6$ symbols in $\mathbb{F}_q$ and the uniform distribution maximizes  entropy; the third term follows from the property that adding conditioning (on $B_1, B_3$) cannot increase entropy. In (\ref{eq:ex13}), the second term is obtained by subtracting the contribution of $Y_0, B_2$ from $f_1, f_2$ and we are left with $2$ linear combinations in $B_1, B_3$; to obtain the third term, we note that $f_1, f_2$ are deterministic functions of $Y_0, B_1, B_2, B_3$ so that they can be dropped from the conditioning term, and subtracting the contribution of $Y_0, B_1, B_2, B_3$ from $h_3, h_4, h_5, h_6$ we are left with $4$ linear combinations in $Z_1, Z_2, Z_3, Z_4$. (\ref{eq:ex14}) follows because ${\bf F}$ and ${\bf H}$ are Cauchy matrices by construction, so their sub-matrices have full rank. The last step (\ref{eq:ex15}) is due to the fact that  $Z_1, Z_2, Z_3, Z_4, B_1, B_2, B_3$ comprise i.i.d. uniform symbols in $\mathbb{F}_q$ and are independent of the message $Y_0$.

\item (\ref{c:decb}): We need to show that $(B_1, B_2, B_3)$ is determined by $(Y_1, Y_2, Y_3)$. Proceed as follows. First, as Cauchy ${\bf H}_{6\times 6}$ is invertible, from $h_1, h_2, h_3, h_4, h_5, h_6$ (the last two elements of $Y_1, Y_2, Y_3$) we obtain $f_0, Y_0, Z_1, Z_2, Z_3, Z_4$. Second, we  subtract the contribution of $Y_0$ from $f_0, f_1, f_2$ ($f_0, Y_0$ have been recovered and $f_1, f_2$ are the first two elements of $Y_1, Y_2, Y_3$) and obtain $(B_1, B_2, B_3) \times {\bf F}((2,3,4), (1,2,3))$, from which $B_1, B_2, B_3$ is recovered as ${\bf F}$ is Cauchy. Thus $(B_1, B_2, B_3)$ is determined by $(Y_1, Y_2, Y_3)$.
\end{enumerate}

\subsection{$(N,K,N_B,K_B) = (3,1,3,2)$: Achievability of $(\lambda_0, \lambda_B) = (1/2, 5/6)$} \label{ex:very2}
Continuing with the same setting of $\widetilde{\mbox{CS}}\widetilde{\mbox{SRA}}(3, 1, 3, 2)$, here we present the linear scheme for a different rate tuple $(\lambda_0, \lambda_B) = (1/2, 5/6)$. Local randomness $Z$ will not be used in this scheme. The parameters of the linear scheme are  as follows. 

\begin{enumerate}
\item Field size $q$ is any prime power such that $q \geq 13$.
\item Scaling factor $\kappa = 6$.
\item Precoding matrices are specified through the following steps. Note that $Y_0$ contains $\kappa \lambda_0 = 3$ elements in $\mathbb{F}_q$, denoted as $Y_0 = (a_1, a_2, a_3)$ and $B_1, B_2, B_3$ each contains $\kappa \lambda_B = 5$ elements in $\mathbb{F}_q$. We divide the $5$ elements into $2$ parts, labelled by the superscript $x, y$.
\begin{eqnarray}
B_1 = \left( \underbrace{ \bm{b}^{1,x}_1}_{1\times 2} , \underbrace{ \bm{b}^{2,x}_1}_{1\times 2}, \underbrace{b_1^y}_{1\times 1} \right), B_2 = \left( \underbrace{ \bm{b}^{1,x}_2}_{1\times 2} , \underbrace{ \bm{b}^{2,x}_2}_{1\times 2}, \underbrace{b_2^y}_{1\times 1} \right), B_3 = \left( \underbrace{ \bm{b}^{1,x}_3}_{1\times 2} , \underbrace{ \bm{b}^{2,x}_3}_{1\times 2}, \underbrace{b_3^y}_{1\times 1} \right).
\end{eqnarray}

Choose a $6 \times 6$ Cauchy matrix ${\bf H} = \left[\frac{1}{\alpha_i - \beta_j}\right]_{ij}$ where the element in the $i^{th}$ row and $j^{th}$ column is $\frac{1}{\alpha_i - \beta_j}$ and $\alpha_i, \beta_j, i \in [6], j \in [6]$ are distinct elements in $\mathbb{F}_q$. Then define
\begin{eqnarray}
\left( \underbrace{ \bm{h}_1}_{1 \times 2}, \underbrace{\bm{h}_2}_{1\times 2}, \underbrace{\bm{h}_3}_{1 \times 2} \right)_{1\times 6} \triangleq \left(a_1, a_2, a_3, b_1^y, b_2^y, b_3^y \right)_{1 \times 6} \times {\bf H}_{6 \times 6}. \label{very2:h}
\end{eqnarray}

We are now ready to specify the storage systems. Note that $Y_1, Y_2, Y_3$ each contains $\kappa = 6$ elements in $\mathbb{F}_q$.
\begin{eqnarray}
Y_1 &=& \left( \bm{h}_1 + \bm{b}_1^{1,x}, \bm{h}_2 + \bm{b}_2^{1,x}, \bm{h}_3 + \bm{b}_3^{1,x} \right)_{1\times 6}, \notag \\
Y_2 &=& \left( \bm{h}_1 + \bm{b}_1^{2,x}, \bm{h}_2 + \bm{b}_2^{2,x}, \bm{h}_3 + \bm{b}_3^{2,x} \right)_{1\times 6},  \label{very2:y}\\
Y_3 &=& \left( \bm{h}_1 + \bm{b}_1^{1,x} + \bm{b}_1^{2,x}, \bm{h}_2 + \bm{b}_2^{1,x} + \bm{b}_2^{2,x}, \bm{h}_3 + \bm{b}_3^{1,x} + \bm{b}_3^{2,x}\right)_{1\times 6}.\notag
\end{eqnarray}
Observe that across $Y_1, Y_2, Y_3$, the pattern of $\bm{h}$ terms is repeated, whereas $\bm{b}^{1,x}, \bm{b}^{2,x}$ go through a $(3,2)$ MDS code $(x,y) \rightarrow (x, y, x+y)$. 
\end{enumerate}

To show that the scheme works we verify the constraints (\ref{c:edec}), (\ref{c:sec}),  (\ref{c:decb}). Consider the case $\mathcal{K} = \{1\}, \mathcal{K}_B = \{1,3\}$. Other cases follow similarly.
\begin{enumerate}
\item (\ref{c:edec}):
We need to show that $Y_0$ is determined by $(Y_1, B_1, B_3)$. Let us proceed as follows. $\bm{b}_1^{1,x}, \bm{b}_3^{1,x}$ are known from $B_1, B_3$. Subtract the contributions of $\bm{b}_1^{1,x}, \bm{b}_3^{1,x}$ from $Y_1$ (consider the first and third terms) and obtain
\begin{eqnarray}
\left( \bm{h}_1, \bm{h}_3 \right)_{1\times 4} = \left(a_1, a_2, a_3, b_1^y, b_2^y, b_3^y \right)_{1 \times 6} \times {\bf H}(:, (1,2,5,6) )_{6 \times 4}.
\end{eqnarray}
Observe that $b_1^y, b_3^y$ are known from $B_1, B_3$. Subtract the contribution of $b_1^y, b_3^y$ from $\bm{h}_1, \bm{h}_3$ and obtain
\begin{eqnarray}
 \left(a_1, a_2, a_3, b_2^y \right)_{1 \times 4} \times {\bf H}((1,2,3,5), (1,2,5,6))_{4 \times 4}.
\end{eqnarray}
As a sub-matrix of the Cauchy matrix ${\bf H}$, the $4\times 4$ matrix ${\bf H}((1,2,3,5), (1,2,5,6))$ is invertible. Invert to recover $Y_0 = (a_1, a_2, a_3)$.

\item (\ref{c:sec}): We need to show that $Y_0$ is independent of $(Y_2, Y_3, B_2)$. Proceed as follows.
\begin{align}
& I\left(Y_0; Y_2, Y_3, B_2 \right) \notag\\
&\overset{(\ref{ind}), (\ref{very2:y})}{=} I\left( Y_0; \bm{h}_1 + \bm{b}_1^{2,x}, \bm{h}_2, \bm{h}_3 + \bm{b}_3^{2,x},  \bm{b}_1^{1,x},  \bm{b}_3^{1,x} \mid B_2 \right) \label{eq:ex21}\\
&=  H\left( \underbrace{ \bm{h}_1 + \bm{b}_1^{2,x}, \bm{h}_2, \bm{h}_3 + \bm{b}_3^{2,x},  \bm{b}_1^{1,x},  \bm{b}_3^{1,x}}_{1 \times 10}  \mid B_2 \right)  \notag\\
&\hspace{1cm}-  H \left( \bm{h}_1 + \bm{b}_1^{2,x}, \bm{h}_2, \bm{h}_3 + \bm{b}_3^{2,x},  \bm{b}_1^{1,x},  \bm{b}_3^{1,x} \mid Y_0, B_2 \right) \\
&\leq 10 -  H \left(\bm{h}_2 \mid Y_0, B_2 \right) -  H \left( \bm{h}_1 + \bm{b}_1^{2,x}, \bm{h}_3 + \bm{b}_3^{2,x},  \bm{b}_1^{1,x},  \bm{b}_3^{1,x}  \mid Y_0, B_2, \bm{h}_2, b_1^y, b_2^y, b_3^y \right) \notag\\
\\
&\overset{(\ref{very2:h})}{\leq} 10 -  H \left( (b_1^y, b^y_3) \times {\bf H}((4,6), (3,4)) \mid Y_0, B_2 \right) \notag\\
&\hspace{1cm}-  H \left( \bm{b}_1^{2,x}, \bm{b}_3^{2,x},  \bm{b}_1^{1,x},  \bm{b}_3^{1,x}  \mid Y_0, B_2, b_1^y, b_2^y, b_3^y \right) \label{eq:ex23}\\
&= 10 - 2 - 8 = 0. \label{eq:ex24} 
\end{align}
In (\ref{eq:ex21}), we plug in the design of $Y_2, Y_3$, subtract the contribution of $B_2$, i.e., $\bm{b}_2^x$, and perform invertible transformations to simplify the terms. 
In (\ref{eq:ex23}), the third term follows from the fact that $\bm{h}$ are known given $Y_0, b^y$.
(\ref{eq:ex24}) follows from the fact that ${\bf H}$ is Cauchy and the SR $B$ is i.i.d. uniform and independent of $Y_0$.

\item (\ref{c:decb}):
We need to show that $B =(B_1, B_2, B_3)$ is determined by $(Y_1, Y_2, Y_3)$. Proceed as follows.
\begin{eqnarray}
&& Y_3 - Y_1 = \left(  \bm{b}_1^{2,x},  \bm{b}_2^{2,x}, \bm{b}_3^{2,x} \right)_{1\times 6}, \\
&& Y_3 - Y_2 = \left(  \bm{b}_1^{1,x},  \bm{b}_2^{1,x}, \bm{b}_3^{1,x} \right)_{1\times 6}.
\end{eqnarray}
Subtract the contribution of $\bm{b}^{x}$ terms from $Y_1, Y_2, Y_3$ and obtain $\bm{h}_1,  \bm{h}_2,  \bm{h}_3$, from which $a_1$, $a_2$, $a_3$, $b_1^y$, $b_2^y$, $b_3^y$ are determined because ${\bf H}$ is invertible (refer to (\ref{very2:h})). Thus  $B$ ($\bm{b}^x, b^y$ terms) is recovered.
\end{enumerate}

\subsection{$(N,K,N_B,K_B) = (3,2,4,3)$: Proof of Converse $\lambda_0 \leq 5/3$}\label{sec:infinity}
For the setting of $\widetilde{\mbox{CS}}\widetilde{\mbox{SRA}}(3, 2, 4, 3)$, here we present an information theoretic proof that the bound $\lambda_0 \leq 5/3$ applies to \emph{all} (not limited to linear) feasible coding schemes. Consider any feasible coding scheme $\mathfrak{S}$. Suppose $\mathcal{K} = \{1,2\}, \mathcal{K}_B = \{1,2,4\}$, and $Y_0$ is uniformly distributed. A feasible scheme must be able to recover $Y_0$ from $Y_{\mathcal{K}}, B_{\mathcal{K}_B}$. From the decodability constraint (\ref{c:edec}), we have
\begin{eqnarray}
\kappa \lambda_0 \leq \kappa \lambda_0 (\mathfrak{S}) &=& H\left(Y_0 \right) \overset{(\ref{c:edec})}{=} I\left( Y_0; Y_1, Y_2, B_1, B_2, B_4 \right) \\
&=& I\left( Y_0; Y_1, B_1, B_2, B_4 \right) + I\left( Y_0; Y_2 \mid Y_1, B_1, B_2, B_4 \right) \\
&\overset{(\ref{ind})}{=}& I\left( Y_0; Y_1 \mid B_1, B_2, B_4 \right) + I\left( Y_0; Y_2 \mid Y_1, B_1, B_2, B_4 \right) \label{inf:1}\\
&\leq& I\left( Y_0, B_1, B_2; Y_1 \mid B_4 \right) + H\left(Y_2\right) \\
&\leq& \underbrace{ I\left( Y_0 ; Y_1 \mid B_4 \right) }_{\overset{(\ref{c:sec})}{\leq} 0} ~+~ I\left( B_1, B_2; Y_1 \mid Y_0, B_4 \right) + \kappa \label{inf:11}\\
\implies \kappa\lambda_0 - \kappa &\leq&  I\left( B_1, B_2; Y_1 \mid Y_0, B_4 \right). \label{inf:2}
\end{eqnarray}
(\ref{inf:1}) follows from the fact that $Y_0$ is independent of $B$. In (\ref{inf:11}), the first term is zero due to the security constraint (\ref{c:sec}), where $\mathcal{K}^c = \{1\}$ and $\mathcal{K}^c_B = \{4\}$.

Following the same procedure as above, and setting $\mathcal{K}_B$ as $\{1,3,4\}$ and $\{2,3,4\}$, respectively, we obtain
\begin{eqnarray}
\kappa\lambda_0 - \kappa &\leq&  I\left( B_1, B_3; Y_1 \mid Y_0, B_4 \right), \label{inf:3} \\
\kappa\lambda_0 - \kappa &\leq&  I\left( B_2, B_3; Y_1 \mid Y_0, B_4 \right). \label{inf:4}
\end{eqnarray}
Adding (\ref{inf:2}), (\ref{inf:3}), and (\ref{inf:4}), we have
\begin{eqnarray}
&& 3\kappa\lambda_0 - 3\kappa \notag\\
&\leq& I\left( B_1, B_2; Y_1 \mid Y_0, B_4 \right) + I\left( B_1, B_3; Y_1 \mid Y_0, B_4 \right) + I\left( B_2, B_3; Y_1 \mid Y_0, B_4 \right) \label{inf:5}\\
&\overset{ (\ref{ind})}{=}& I\left( B_1, B_2; Y_1 \mid Y_0, B_4 \right) + I\left( B_1, B_3; Y_1, Y_0, B_4 \right) + I\left( B_2, B_3; Y_1 \mid Y_0, B_4 \right)  \\
&\overset{(\ref{ind})}{\leq}& I\left( B_1, B_2; Y_1 \mid Y_0, B_4 \right) + I\left( B_1, B_3; Y_1 \mid Y_0, B_2, B_4 \right) + I\left( B_2, B_3; Y_1 \mid Y_0, B_4 \right) \\
&=& I\left( B_1, B_2; Y_1 \mid Y_0, B_4 \right) + I\left( B_3; Y_1 \mid Y_0, B_1, B_2, B_4 \right)  \notag\\
&&~+~ I\left( B_1 ; Y_1 \mid Y_0, B_2, B_4 \right) + I\left( B_2, B_3; Y_1 \mid Y_0, B_4 \right) \\
&\overset{(\ref{ind})}{\leq}& I\left( B_1, B_2, B_3; Y_1 \mid Y_0, B_4 \right) + I\left( B_1 ; Y_1 \mid Y_0, B_2, B_3, B_4 \right) + I\left( B_2, B_3; Y_1 \mid Y_0, B_4 \right) \\
&=& 2 I\left( B_1, B_2, B_3; Y_1 \mid Y_0, B_4 \right) \\
&\leq& 2 H(Y_1) ~\leq~ 2 \kappa \\
&\implies& 3\lambda_0 \leq 5,
\end{eqnarray}
thus producing the desired bound. Note that we repeatedly use the independence of $Y_0$ and $B$ and the uniformity of $B$ (so that $B_i$ are independent) to add conditioning terms and to combine the mutual information terms.

\section{Proof of Theorem \ref{thm:ab}}\label{sec:ab}
Consider any classical linear scheme for $\widetilde{\mbox{CS}}\widetilde{\mbox{SRA}}(N,K,N_B,K_B)$ as defined in Definition \ref{def:linear}, i.e., we are given $\lambda_0$, $\lambda_B$, $\kappa$, $q$, ${\bf A}_n$, ${\bf B}_n$, ${\bf Z}_n$, $\forall n \in [N]$ such that (\ref{c:edec}), (\ref{c:sec}),  (\ref{c:decb}) are satisfied. We show how this allows us to design a quantum coding scheme for $\widetilde{\mbox{QS}}\widetilde{\mbox{EA}}(N,K,N_B,K_B)$ as follows.

Consider the quantum encoding procedure first.
Suppose the quantum message $Q_0$ has $\kappa \lambda_0$ qudits where each qudit is $q$-dimensional. Suppose $Q_0$ is in an arbitrary state with density matrix $\omega_{Q_0}$. Without loss of generality, suppose $\omega_{Q_0}$ has a spectral decomposition $\omega_{Q_0} = \sum_{\bm{a}} p_{\bm{a}} \ket{\bm{a}} \bra{\bm{a}}$ and a purification $\ket{\varphi}_{RQ_0} = \sum_{\bm{a}} \sqrt{p_{\bm{a}}} \ket{\bm{a}} \ket{\bm{a}}$. The encoding proceeds as follows.
\begin{align}
&\ket{\varphi}_{R Q_0} \ket{\phi}_{A B} \\
&= \sum_{\bm{a} \in \mathbb{F}_q^{1\times \kappa\lambda_0}} \sqrt{p_{\bm{a}}} \ket{\bm{a}}_R \ket{\bm{a}}_{Q_0} \sum_{\bm{b} \in \mathbb{F}_q^{1\times N_B\kappa\lambda_B}} \frac{1}{\sqrt{q^{N_B\kappa\lambda_B}}} \ket{\bm{b}}_A \ket{\bm{b}}_B  \\
&\rightsquigarrow  \sum_{\bm{a}} \sqrt{p_{\bm{a}}}  \ket{\bm{a}}_R \sum_{\bm{b}, \bm{z} \in \mathbb{F}_q^{1\times L}} \frac{1}{\sqrt{q^{N_B\kappa\lambda_B+L}}} \ket{Y_1,Y_2,\dots,Y_N}_{Q_1\dots Q_N}  \ket{\bm{b}}_B,  \label{eq:enc1} 
\end{align}
where (\ref{eq:enc1}) is an isometric map from $Q_0A$ to  $Q_1\dots Q_N$, because (\ref{c:edec}) and (\ref{c:decb})  guarantee that 
from $(Y_n,n\in[N])=(\bm{a}{\bf A}_{n} +\bm{b}{\bf B}_{n} + \bm{z}{\bf Z}_{n}, n\in[N])$, one can recover $\bm{a}, \bm{b}$.
Observe that $R$ and $B$  go through an identity map and $RQ_1\dots Q_N B$ ends up in a pure state. Compared to Theorem 2 in \cite{Sun_Jafar_QuStorage} the main difference in the above encoding procedure is that here we have EA $AB$ and need to ensure that the encoding does not require access to $B$.

\begin{remark}
Note that the additional constraint (\ref{c:decb}) that we impose for $\widetilde{\mbox{CS}}\widetilde{\mbox{SRA}}$, guarantees that the above quantum encoding procedure is feasible for $\widetilde{\mbox{QS}}\widetilde{\mbox{EA}}$, i.e., the map to produce (\ref{eq:enc1}) from $Q_0A$ to $Q_1\dots Q_N$ is isometric. Recall that according to the problem formulation the encoding needs not be restricted to isometric maps. Restricted as it might seem, our approach (combined with space-sharing) provides optimal quantum schemes for all regimes where capacity is known (Case 1 and Case 2 in Theorem \ref{thm:region}) and is conjectured to be optimal in the remaining regime (Case 3) as well. 
\end{remark}

The quantum decoding process follows from  \cite[Theorem 2]{Sun_Jafar_QuStorage}. Note that the same classical constraints are satisfied, i.e., (\ref{c:edec}) for the decoding of $Y_0$ from each decoding set (i.e., any $K$ storage systems and any $K_B$ SR-systems) and (\ref{c:sec}) for the security of $Y_0$ from the complement of each decoding set (i.e., any $N-K$ storage systems and any $N_B-K_B$ SR-systems). Thus,  \cite[Theorem 2]{Sun_Jafar_QuStorage} guarantees  perfect recovery for any decoding set (the decoding sets are identical for $\widetilde{\mbox{CS}}\widetilde{\mbox{SRA}}(N,K,N_B,K_B)$ and $\widetilde{\mbox{QS}}\widetilde{\mbox{EA}}(N,K,N_B,K_B)$). For the sake of completeness let us briefly sketch the decoding procedure of \cite[Theorem 2]{Sun_Jafar_QuStorage} here.

Fix any $\mathcal{K}, \mathcal{K}_B$. Consider the state for $RQ_1\dots Q_N B = R Q_{\mathcal{K}} B_{\mathcal{K}_B} Q_{\mathcal{K}^c} B_{\mathcal{K}^c_B}$, expressed as follows.

\begin{eqnarray}
&& \sum_{\bm{a}} \sqrt{p_{\bm{a}}}  \ket{\bm{a}}_R \sum_{\bm{b}, \bm{z}} \frac{1}{\sqrt{q^{N_B\kappa\lambda_B+L}}} \ket{\bm{a} {\bf A}_{[N]} + \bm{b} {\bf B}_{[N]} + \bm{z} {\bf Z}_{[N]} }_{Q_1\dots Q_N}  \ket{\bm{b}}_B  \notag\\
&\equiv&  \sum_{\bm{a}} \ket{\bm{a}}_R \sum_{\bm{b}, \bm{z}} \ket{\bm{a} {\bf A}_{\mathcal{K}} + \bm{b} {\bf B}_{\mathcal{K}} + \bm{z} {\bf Z}_{\mathcal{K}} }_{Q_{\mathcal{K}}}  \ket{\bm{b}_{\mathcal{K}_B}}_{B_{\mathcal{K}_B}} \ket{\bm{a} {\bf A}_{\mathcal{K}^c} + \bm{b} {\bf B}_{\mathcal{K}^c} + \bm{z} {\bf Z}_{\mathcal{K}^c} }_{Q_{\mathcal{K}^c}}  \ket{\bm{b}_{\mathcal{K}^c_B}}_{B_{\mathcal{K}^c_B}}, \label{eq:dec1}
\end{eqnarray}
where for any matrix ${\bf A}$ and any set $\mathcal{K}$, ${\bf A}_{\mathcal{K}}$ denotes the sub-matrix of ${\bf A}$ with all columns whose indices are in $\mathcal{K} = \{k_1, \cdots, k_{|\mathcal{K}|}\}$,
\begin{eqnarray}
{\bf A}_{\mathcal{K}} \triangleq \left({\bf A}(:, k_1), \dots, {\bf A}(:, k_{|\mathcal{K}|}) \right).
\end{eqnarray}

The next key step is to perform a change of basis operation. Note that the decodability constraint (\ref{c:edec}) holds, i.e., from $(\bm{a} {\bf A}_{\mathcal{K}} + \bm{b} {\bf B}_{\mathcal{K}} + \bm{z} {\bf Z}_{\mathcal{K}}, \bm{b}_{\mathcal{K}_B}) \triangleq \bm{s}$, we may recover $\bm{a}$;  the security constraint (\ref{c:sec}) holds, i.e., $(\bm{a} {\bf A}_{\mathcal{K}^c} + \bm{b} {\bf B}_{\mathcal{K}^c} + \bm{z} {\bf Z}_{\mathcal{K}^c}, \bm{b}_{\mathcal{K}^c_B}) \triangleq \bm{n}$ is independent of $\bm{a}$. Then denoting $\bm{z}_1'$ as the common part of $\bm{s}$ and $\bm{n}$ (identified through the intersection of linear subspaces), $\bm{z}_2'$ as the part that is in $\bm{s}$ (a linear function of $\bm{s}$) but not in $\bm{n}$ (identified through the linear space that lies in $\bm{s}$, but has no overlap with $\bm{n}$), $\bm{z}_3'$ as the part that is in $\bm{n}$ but not in $\bm{s}$, we have
\begin{eqnarray}
&& \left(\bm{a} {\bf A}_{\mathcal{K}} + \bm{b} {\bf B}_{\mathcal{K}} + \bm{z} {\bf Z}_{\mathcal{K}}, \bm{b}_{\mathcal{K}_B} \right) \overset{\mbox{\scriptsize invertible}}{\longleftrightarrow} \left( {\bm a}, {\bm z}_1', {\bm z}_2', l({\bm z}_1', {\bm z}_2') \right), \\
&& \left(\bm{a} {\bf A}_{\mathcal{K}^c} + \bm{b} {\bf B}_{\mathcal{K}^c} + \bm{z} {\bf Z}_{\mathcal{K}^c}, \bm{b}_{\mathcal{K}^c_B} \right) = \left( l({\bm z}_1', \bm{z}_3') \right),
\end{eqnarray}
where $l$ denotes some linear combinations and the crucial observation is that $\bm{a}$ can be separated out from $\bm{s}$ while $\bm{a}$ cannot be obtained at all from $\bm{n}$. Furthermore, for any fixed $\bm{a}$, we have $(\bm{z}_1', \bm{z}_2', \bm{z}_3')$, which contains all $\bm{b}$ and $\bm{z}$ that appear in $Q_1\dots Q_N B$, is invertible to $(\bm{b}, \bm{z} {\bf Z}_{[N]})$. A  detailed treatment of the above change of basis operation can be found in \cite[Theorem 2]{Sun_Jafar_QuStorage}. 

We now proceed with the decoding procedure, continuing from (\ref{eq:dec1}).
\begin{eqnarray}
&\overset{(\ref{eq:dec1})}{\rightsquigarrow}& \sum_{\bm{a}} \ket{\bm{a}}_R \sum_{\bm{b}, \bm{z}}  \ket{{\bm a}, {\bm z}_1', {\bm z}_2', l({\bm z}_1', {\bm z}_2') } \ket{ l({\bm z}_1', \bm{z}_3') }_{Q_{\mathcal{K}^c} B_{\mathcal{K}^c_B}}  \\
&\equiv& \sum_{\bm{a}} \ket{\bm{a}}_R  \ket{\bm{a}}_{\widehat{Q}_0} \sum_{\bm{z}_1', \bm{z}_2', \bm{z}_3'}  \ket{ {\bm z}_1', {\bm z}_2', l({\bm z}_1', {\bm z}_2') } \ket{ l({\bm z}_1', \bm{z}_3') }_{Q_{\mathcal{K}^c} B_{\mathcal{K}^c_B}} \\
&=& \left(  \sum_{\bm{a}} \ket{\bm{a}}_R  \ket{\bm{a}}_{\widehat{Q}_0}  \right) \otimes \cdots .
\end{eqnarray}
Therefore, the decodability constraint (\ref{c:edec}) and the security constraint (\ref{c:sec}) together guarantee that we can unentangle $R \widehat{Q}_0$ from the rest of the codeword and the state of $R \widehat{Q}_0$ is identical to that of $R Q_0$. The rate tuple achieved for $\widetilde{\mbox{QS}}\widetilde{\mbox{EA}}(N,K,N_B,K_B)$ is $(\lambda_0, \lambda_B)$, same as that for $\widetilde{\mbox{CS}}\widetilde{\mbox{SRA}}(N,K,N_B,K_B)$.

\begin{remark}
Recall that according to (\ref{c:decb}), the code for $\widetilde{\mbox{CS}}\widetilde{\mbox{SRA}}$ requires the full recovery of $B$ from $Y_1, \dots, Y_N$. It is not difficult to see that such a constraint can sometimes exclude  optimal schemes, e.g., in the regime where $\lambda_B$ is large. However, as we show in the following sections, this code achieves all  (including conjectured) \emph{extreme} points of the capacity region $\mathcal{C}$. The corresponding insight for $\widetilde{\mbox{QS}}\widetilde{\mbox{EA}}$ is that while a pure $\rho$ might not always be optimal (e.g., for large $\lambda_B$) but a pure $\rho$ may still achieve all extreme points of the capacity region. The achievability of the remaining points is then implied by convexity of the capacity region, e.g., via space-sharing (see Appendix \ref{app:convex}).
\end{remark}

\section{Proof of Theorem \ref{thm:region}}\label{sec:region}

\subsection{Converse Proof of $\mathcal{C}$: Classical Storage}\label{sec:cregion}
Our problem formulation requires perfect recovery for an arbitrarily distributed $Y_0$, so in particular perfect recovery is required for a uniform $Y_0$, i.e., $H(Y_0) = \log_q|Y_0| = \kappa \lambda_0$. In the following, we assume $Y_0$ is uniform over $\mathcal{Y}_0$, a set of cardinality $q^{\kappa \lambda_0}$.

\subsubsection{Proof of $\mathcal{R}_{cut}$ Bound}
We prove that $ \lambda_0 \leq  \max(2K-N, 0) +  \lambda_B \cdot \max(2K_B - N_B, 0)$ for any feasible coding scheme $\mathfrak{S}$. 
Start from the entropic decodability condition (\ref{c:edec}) and set $\mathcal{K} = [K]$, $\mathcal{K}_B = [N_B]$.
\begin{eqnarray} 
\kappa \lambda_0 &\leq& \kappa \lambda_0(\mathfrak{S}) = H\left(Y_0 \right) \overset{(\ref{c:edec})}{=} I \left(Y_0 ; Y_{[K]}, B_{[K_B]} \right) \\
&\leq&  I\left( Y_0 ; Y_{[N-K]}, Y_{[N-K+1:K]} , B_{[N_B - K_B]}, B_{[N_B - K_B+1 : K_B]} \right) \label{eq:s11} \\
&\overset{(\ref{c:sec})}{=}& I\left(Y_0 ; Y_{[N-K+1:K]}, B_{[N_B - K_B+1 : K_B]}  \mid Y_{[N-K]}, B_{[N_B - K_B]} \right) \label{eq:s21} \\
&\leq&  H\left(Y_{[N-K+1:K]} \right) + H\left( B_{[N_B - K_B+1 : K_B]} \right) \label{eq:s31}\\
&\leq& \log_q \prod_{n \in [N-K+1:K]} |Y_n|  + \log_q \prod_{i \in [N_B-K_B+1:K_B]} |B_i| \\
&\leq& \kappa \cdot \max\left(2K-N, 0\right)  + \kappa \lambda_B(\mathfrak{S}) \cdot \max\left(2K_B-N_B, 0\right)   \\
\implies \lambda_0 &\leq& \max\left(2K-N, 0\right) + \lambda_B \cdot \max\left(2K_B-N_B, 0\right).
\end{eqnarray}
Step (\ref{eq:s21}) uses the security constraint (\ref{c:sec}) when $\mathcal{K}^c = [N-K], \mathcal{K}_B^c = [N_B-K_B]$.

\subsubsection{Proof of $\mathcal{R}_{\infty}$ Bound}\label{sec:cinf}
We prove that for any feasible coding scheme $\mathfrak{S}$, $\lambda_0 \leq \min(N, 2K) - \min(N-K, K) \frac{N_B}{K_B}$, which becomes
\begin{eqnarray}
\lambda_0 \leq N - (N-K) \frac{N_B}{K_B}~\mbox{when $K/N \geq 1/2$,} \label{c3}
\end{eqnarray}
and becomes
\begin{eqnarray}
\lambda_0 \leq K - 2K \frac{N_B}{K_B}~\mbox{when $K/N < 1/2$.} \label{c2}
\end{eqnarray}
(\ref{c3}) implies (\ref{c2}) because we may set $N = 2K$ in (\ref{c3}) so that it becomes (\ref{c2}). Reducing $N$ to $2K$ means that we only consider $2K$ out of the $N$ storage systems, which does not violate a converse argument. 
Therefore, it suffices to prove (\ref{c3}) only, considered next.
The proof is a generalization of the example in Section \ref{sec:infinity}, where the summation of the mutual information terms in (\ref{inf:5}) needs to be generalized (see Lemma \ref{lemma:sum}).

Consider $\mathcal{K} = [K] = [N-K] \cup [N-K+1:K], \mathcal{K}_B =  \mathcal{I} \cup [K_B+1:N_B]$ where $\mathcal{I} \subset [K_B], |\mathcal{I}| = 2K_B - N_B$. Note that $2K_B-N_B \leq K_B$ and $K/N \geq 1/2$. 
From the decodability constraint (\ref{c:edec}), we have
\begin{eqnarray}
\kappa \lambda_0 &\leq& \kappa \lambda_0(\mathfrak{S})  = H\left(Y_0 \right) \overset{(\ref{c:edec})}{=} I\left( Y_0; Y_{\mathcal{K}}, B_{\mathcal{K}_B} \right) \\
&=&  I\left( Y_0; Y_{[N-K]}, Y_{[N-K+1:K]}, B_{\mathcal{K}_B} \right) \\
&\leq& I\left( Y_0; Y_{[N-K]}, B_{\mathcal{K}_B} \right) + H\left( Y_{[N-K+1:K]} \right) \\
&\overset{(\ref{ind})}{=}& I\left( Y_0; Y_{[N-K]} \mid B_{\mathcal{I}}, B_{[K_B+1:N_B]} \right) + \kappa (2K-N)  \\
&\leq& I\left( Y_0, B_{\mathcal{I}}; Y_{[N-K]} \mid B_{[K_B+1:N_B]} \right) + \kappa (2K-N)  \\
&=& \underbrace{ I\left( Y_0 ; Y_{[N-K]} \mid B_{[K_B+1:N_B]} \right) }_{\overset{(\ref{c:sec})}{\leq} 0} ~+~ I\left( B_{\mathcal{I}}; Y_{[N-K]} \mid Y_0, B_{[K_B+1:N_B]} \right) \notag\\
&&+~\kappa (2K-N)  \\
&\implies& \kappa\lambda_0 - \kappa (2K-N)  \leq  I\left( B_{\mathcal{I}}; Y_{[N-K]} \mid Y_0, B_{[K_B+1:N_B]} \right). \label{cinf:1}
\end{eqnarray}
Next we focus on the RHS of (\ref{cinf:1}). Consider the following $K_B$ choices of $\mathcal{I}$.
\begin{eqnarray}
&&\mathcal{I}_1 = \{1,2,\cdots, 2K_B - N_B\}, \mathcal{I}_2 = \{2K_B - N_B+1, \cdots, 2(2K_B-N_B)\}_{mod}, \cdots, \notag\\
&& \mathcal{I}_{K_B} = \{(K_B-1)(2K_B - N_B) + 1, \cdots, K_B(2K_B - N_B)\}_{mod}. \label{cinf:iset}
\end{eqnarray}
The subscript $mod$ means that each element $i$ in the set is defined as $i  \mod K_B$. That is, each $\mathcal{I}_j, j \in [K_B]$ has $K_B$ elements in the set $[K_B]$ and the elements are taken consecutively (in a cyclic manner). In total, $\sum_{j \in [K_B]} |\mathcal{I}_j| = K_B (2K_B - N_B)$ so that overall the union of all $\mathcal{I}_j$ covers the set $[K_B]$ for a total of $2K_B - N_B$ times. For example, consider $K_B = 5$, $2K_B - N_B = 3$.
\begin{eqnarray}
\mathcal{I}_1 = \{1,2,3\}, \mathcal{I}_2 = \{4,5,1\}, \mathcal{I}_3 = \{2,3,4\}, \mathcal{I}_4 = \{5,1,2\}, \mathcal{I}_5 = \{3,4,5\}. \label{cinf:ex}
\end{eqnarray}
To proceed with the derivation of the converse, our goal is to treat the sum of the RHS of (\ref{cinf:1}) for all $\mathcal{I}_1, \dots, \mathcal{I}_{K_B}$. This result is stated in the following lemma.

\begin{lemma}\label{lemma:sum}
For the sets $\mathcal{I}_1, \dots, \mathcal{I}_{K_B}$ defined in (\ref{cinf:iset}) and any disjoint sets $\mathcal{J}_1, \mathcal{J}_2 \subset [K_B]$, i.e., $\mathcal{J}_1 \cap \mathcal{J}_2 = \emptyset$, we have
\begin{eqnarray}
&& I\left( B_{\mathcal{J}_1}; Y_{[N-K]} \mid Y_0, B_{[K_B+1:N_B]} \right) \leq I\left( B_{\mathcal{J}_1}; Y_{[N-K]} \mid B_{\mathcal{J}_2}, Y_0, B_{[K_B+1:N_B]} \right), \label{sum:1}\\
&& \sum_{j \in [K_B]} I\left( B_{\mathcal{I}_j}; Y_{[N-K]} \mid Y_0, B_{[K_B+1:N_B]} \right) \leq \left( 2K_B - N_B\right) I\left( B_{[K_B]}; Y_{[N-K]} \mid Y_0, B_{[K_B+1:N_B]} \right). \notag\\
&&\label{sum:2}
\end{eqnarray}
\end{lemma}

{\it Proof:} (\ref{sum:1}) states that we can add arbitrary conditional terms $B_\mathcal{J}$. It is a simple consequence of the independence of $Y_0$ and $B$ and the uniformity of $B$, proved as follows.
\begin{eqnarray}
I\left( B_{\mathcal{J}_1}; Y_{[N-K]} \mid Y_0, B_{[K_B+1:N_B]} \right) &\overset{(\ref{ind})}{=}&  I\left( B_{\mathcal{J}_1}; Y_{[N-K]}, Y_0, B_{[K_B+1:N_B]} \right) \\
&\leq& I\left( B_{\mathcal{J}_1}; Y_{[N-K]},  B_{\mathcal{J}_2}, Y_0, B_{[K_B+1:N_B]} \right) \\
&\overset{(\ref{ind})}{=}& I\left( B_{\mathcal{J}_1}; Y_{[N-K]} \mid B_{\mathcal{J}_2}, Y_0, B_{[K_B+1:N_B]} \right).
\end{eqnarray}

We are now ready to prove (\ref{sum:2}) by combining the mutual information terms with the chain rule and repeatedly using (\ref{sum:1}).
\begin{align}
& \sum_{j \in [K_B]} I\left( B_{\mathcal{I}_j}; Y_{[N-K]} \mid Y_0, B_{[K_B+1:N_B]} \right) \\
&= I\left( B_{\mathcal{I}_1}; Y_{[N-K]} \mid Y_0, B_{[K_B+1:N_B]} \right) +  I\left( B_{\mathcal{I}_2}; Y_{[N-K]} \mid Y_0, B_{[K_B+1:N_B]} \right) + \cdots \\
&\overset{(\ref{sum:1})}{\leq} I\left( B_{\mathcal{I}_1 \cup \mathcal{I}_2}; Y_{[N-K]} \mid Y_0, B_{[K_B+1:N_B]} \right) + \cdots ~~\mbox{if}~\mathcal{I}_1 \cup \mathcal{I}_2 \subset [K_B]\\
&\overset{(\ref{sum:1})}{\leq} I\left( B_{[K_B]}; Y_{[N-K]} \mid Y_0, B_{[K_B+1:N_B]} \right) + I\left( B_{\mathcal{I}_1 \cap \mathcal{I}_2}; Y_{[N-K]} \mid Y_0, B_{[K_B+1:N_B]} \right) + \cdots \notag\\
&~~~~~~\mbox{else if}~ [K_B] \subset \mathcal{I}_1 \cup \mathcal{I}_2 \\
&\overset{(\ref{sum:1})}{\leq} \cdots \\
&\overset{(\ref{sum:1})}{\leq} \left( 2K_B - N_B\right) I\left( B_{[K_B]}; Y_{[N-K]} \mid Y_0, B_{[K_B+1:N_B]} \right).
\end{align}
In the above proof, as adding conditioning preserves the direction of inequality (proved in (\ref{sum:1})), we keep taking union of the $\mathcal{I}_1, \mathcal{I}_2, \dots$ sets and split out a copy of $I\left( B_{[K_B]}; Y_{[N-K]} \mid Y_0, B_{[K_B+1:N_B]} \right)$ when the union covers $[K_B]$; and keep going until we have collected $2K_B - N_B$ copies. This process is easily explained by an example. Consider the sets in (\ref{cinf:ex}).
\begin{eqnarray}
&& I\left( B_1, B_2, B_3 ; Y_{[N-K]} \mid Y_0, B_{[K_B+1:N_B]} \right) +  I\left( B_4, B_5, B_1 ; Y_{[N-K]} \mid Y_0, B_{[K_B+1:N_B]} \right) \notag\\
&&+~  I\left( B_2, B_3, B_4 ; Y_{[N-K]} \mid Y_0, B_{[K_B+1:N_B]} \right) +  I\left( B_5, B_1, B_2 ; Y_{[N-K]} \mid Y_0, B_{[K_B+1:N_B]} \right) \notag\\
&&+~ I\left( B_3, B_4, B_5 ; Y_{[N-K]} \mid Y_0, B_{[K_B+1:N_B]} \right) \\
&\leq& I\left( B_1, B_2, B_3, B_4, B_5 ; Y_{[N-K]} \mid Y_0, B_{[K_B+1:N_B]} \right) +  I\left( B_1 ; Y_{[N-K]} \mid Y_0, B_{[K_B+1:N_B]} \right) \notag\\
&&+~  I\left( B_2, B_3, B_4 ; Y_{[N-K]} \mid Y_0, B_{[K_B+1:N_B]} \right) +  I\left( B_5, B_1, B_2 ; Y_{[N-K]} \mid Y_0, B_{[K_B+1:N_B]} \right) \notag\\
&&+~ I\left( B_3, B_4, B_5 ; Y_{[N-K]} \mid Y_0, B_{[K_B+1:N_B]} \right)\\
&\leq& I\left( B_1, B_2, B_3, B_4, B_5 ; Y_{[N-K]} \mid Y_0, B_{[K_B+1:N_B]} \right) +  I\left( B_1, B_2, B_3, B_4 ; Y_{[N-K]} \mid Y_0, B_{[K_B+1:N_B]} \right) \notag\\
&&+~  I\left( B_5, B_1, B_2 ; Y_{[N-K]} \mid Y_0, B_{[K_B+1:N_B]} \right) + I\left( B_3, B_4, B_5 ; Y_{[N-K]} \mid Y_0, B_{[K_B+1:N_B]} \right)\\
&\leq& 2I\left( B_1, B_2, B_3, B_4, B_5 ; Y_{[N-K]} \mid Y_0, B_{[K_B+1:N_B]} \right) \notag\\
&&+~  I\left( B_1, B_2 ; Y_{[N-K]} \mid Y_0, B_{[K_B+1:N_B]} \right) + I\left( B_3, B_4, B_5 ; Y_{[N-K]} \mid Y_0, B_{[K_B+1:N_B]} \right)\\
&\leq& 3I\left( B_1, B_2, B_3, B_4, B_5 ; Y_{[N-K]} \mid Y_0, B_{[K_B+1:N_B]} \right).
\end{eqnarray}
\hfill\qed

To complete the overall proof, we add (\ref{cinf:1}) for all $\mathcal{I} = \mathcal{I}_j, j \in [K_B]$.
\begin{eqnarray}
 \left( \kappa\lambda_0 - \kappa (2K-N)  \right) K_B &\leq& \sum_{j \in [K_B]}  I\left( B_{\mathcal{I}_j}; Y_{[N-K]} \mid Y_0, B_{[K_B+1:N_B]} \right) \\
 &\overset{(\ref{sum:2})}{\leq}& \left( 2K_B - N_B\right) I\left( B_{[K_B]}; Y_{[N-K]} \mid Y_0, B_{[K_B+1:N_B]} \right)\\
 &\leq& (2K_B - N_B) H(Y_{[N-K]}) \leq \kappa (N-K)(2K_B-N_B)  \\
 \implies \lambda_0 &\leq& \frac{(2K-N)K_B + (N-K) (2K_B - N_B)}{K_B} \\
 &=& N - (N-K) \frac{N_B}{K_B}.
\end{eqnarray}

\subsection{Achievability Proof of $\mathcal{C}$: Classical Storage}
\label{sec:c}
Recall that the achievable rate region is convex (i.e., any convex combination of rate tuples can be achieved through space sharing, refer to Appendix \ref{app:convex}). Also recall the fact that if a rate tuple $(\lambda_0, \lambda_B)$ is achievable, then $(\lambda_0-\alpha, \lambda_B + \beta)$ is achievable for any $\lambda_0\geq \alpha\geq 0$, $\beta \geq 0$. Thus, it suffices to consider only the extreme points. 

First consider the extreme point $(\lambda_0, \lambda_B) = \left(\max(2K-N, 0), 0 \right)$ that appears in all three regimes.
To see that $\lambda_0 = \max(2K-N, 0)$ is achievable when $\lambda_B = 0$, i.e., SR-systems are empty and will not be used, we note that 
the problem is reduced to the uniform capacity of MDS storage graphs \cite{Sun_Jafar_QuStorage}; according to Theorem 3 of \cite{Sun_Jafar_QuStorage}, $\lambda_0 = \max(2K-N, 0)$ is {\em linearly} achievable.

Next consider the remaining extreme points that are unique to the considered regime.

\subsubsection{$K/N \geq 1/2, K_B/N_B > 1/2$: Achievability of $(\lambda_0, \lambda_B)  = \left(N - \frac{(N-K) N_B}{K_B}, \frac{N-K}{K_B} \right)$}
This proof is a straightforward generalization of the example in Section \ref{ex:less}. We give a linear scheme that achieves the desired rate tuple. The parameters of the linear scheme are given as follows.
\begin{enumerate}
\item Field size $q$ is any prime power such that $q \geq 2N K_B$.
\item Scaling factor $\kappa = K_B$.
\item Precoding matrices are chosen such that
\begin{eqnarray}
\left( {\bf A}_1^{K_B \lambda_0 \times K_B}, \dots, {\bf A}_N^{K_B \lambda_0 \times K_B}; {\bf B}_1^{N_BK_B\lambda_B \times K_B}, \dots, {\bf B}_N^{N_BK_B\lambda_B \times K_B} \right) = {\bf H}_{NK_B \times NK_B},
\end{eqnarray}
where $K_B \lambda_0 + N_BK_B\lambda_B = N K_B$, and
${\bf H}$ is a Cauchy matrix with $\frac{1}{\alpha_i - \beta_j}$ being the element in the $i^{th}$ row and $j^{th}$ column; $\alpha_i, \beta_j, i, j \in [NK_B]$ are distinct elements in $\mathbb{F}_q$ (note that the field size $q$ is large enough for such distinct elements to exist).
\end{enumerate}

Local noise $Z$ is not used. The storage systems are set as
\begin{eqnarray}
\left( Y_1, \dots, Y_N \right)_{1 \times NK_B} = \left(Y_0, B_1, \dots, B_{N_B} \right)_{1\times NK_B} \times {\bf H}, \label{eq:hh}
\end{eqnarray}
where $Y_0$ contains $\kappa \lambda_0 = NK_B - (N-K)N_B$ elements in $\mathbb{F}_q$. The SR-system $B_i, i \in [N_B]$ contains $\kappa \lambda_B = (N-K)$ elements in $\mathbb{F}_q$, and $Y_n, n\in[N]$ contains $\kappa = K_B$ elements in $\mathbb{F}_q$.

We show that the above linear scheme works by verifying the constraints (\ref{c:edec}), (\ref{c:sec}),  (\ref{c:decb}) sequentially. 
Consider any $\mathcal{K} \in \binom{[N]}{K}$ and any $\mathcal{K}_B \in \binom{[N_B]}{K_B}$.
\begin{enumerate}
\item (\ref{c:edec}): We need to show that $Y_0$ is determined by $Y_{\mathcal{K}}, B_{\mathcal{K}_B}$. To see this, subtracting the contribution of $B_{\mathcal{K}_B}$ from $Y_{\mathcal{K}}$ gives us $(Y_0, B_{\mathcal{K}^c_B}) \times {\bf H}_{sub}^{KK_B \times KK_B}$ where ${\bf H}_{sub}$ is square because $$K_B \lambda_0 + (N_B-K_B)K_B\lambda_B = K K_B.$$ As ${\bf H}$ is Cauchy, $Y_0$ is obtained.

\item (\ref{c:sec}): We need to show that $Y_0$ is independent of $Y_{\mathcal{K}^c}, B_{\mathcal{K}^c_B}$. Proceed as follows.
\begin{eqnarray}
I\left( Y_0; Y_{\mathcal{K}^c}, B_{\mathcal{K}^c_B} \right) &\overset{(\ref{ind})}{=}& I\left( Y_0; Y_{\mathcal{K}^c} \mid B_{\mathcal{K}^c_B} \right) \\
&=& H \left( Y_{\mathcal{K}^c} \mid B_{\mathcal{K}^c_B} \right) - H\left( Y_{\mathcal{K}^c} \mid Y_0, B_{\mathcal{K}^c_B} \right) \\
&\leq& (N-K) K_B - H\left(  B_{\mathcal{K}_B}  {\bf H}_{sub}^{K^2_B\lambda_B \times (N-K) K_B} \right)  \\
&=& 0, 
\end{eqnarray}
where the last step follows from the fact that $$K^2_B\lambda_B = (N-K) K_B.$$

\item (\ref{c:decb}): ${\bf H}$ has full rank, so $B$ is determined by $Y_1, \dots, Y_N$ (see (\ref{eq:hh})).
\end{enumerate}

\subsubsection{$K/N < 1/2, K_B/N_B > 1/2$: Achievability of $(\lambda_0, \lambda_B) = \left(\frac{K(2K_B-N_B)}{2K_B}, \frac{K}{2K_B} \right)$}\label{sec:ach1}
The linear scheme of this section generalizes that of the example in Section \ref{ex:very1}. Specifically, the repetition code ($f_1, f_2$ is repeated in $Y_1, Y_2, Y_3$) in the example is generalized to an $(N, K)$ MDS code which reduces to the $(N,1)$ repetition code when $K=1$.
The parameters of the linear scheme that achieves the desired rate tuple are given as follows.
\begin{enumerate}
\item Field size $q$ is any prime power such that $q \geq 2N K_B > 2KK_B+KN_B$. 
\item Scaling factor $\kappa = 2K_B$.
\item Uniform local randomness $Z = (Z_1, \dots, Z_{(N-K)K_B}) \in \mathbb{F}_q^{1\times (N-K)K_B}$.
\item Precoding matrices are specified through the following steps. Note that $Y_0$ contains $\kappa \lambda_0 = K(2K_B-N_B)$ elements in $\mathbb{F}_q$ and $B_i, i \in [N_B]$ contains $\kappa \lambda_B = K$ elements in $\mathbb{F}_q$. 
First, choose a $2KK_B \times KN_B$ Cauchy matrix ${\bf F} = \left[\frac{1}{\alpha_i - \beta_j}\right]_{ij}$ where the element in the $i^{th}$ row and $j^{th}$ column is $\frac{1}{\alpha_i - \beta_j}$ and $\alpha_i, \beta_j, i \in [2KK_B], j \in [KN_B]$ are distinct elements in $\mathbb{F}_q$. Then define
\begin{eqnarray}
\left(\underbrace{\bm{f}_0}_{1\times K(N_B-K_B)}, \underbrace{\bm{f}_1}_{1\times K_B}, \dots, \underbrace{\bm{f}_K}_{1\times K_B}\right)_{1\times KN_B} \triangleq \left(Y_0, B_1, \dots, B_{N_B} \right)_{1 \times 2KK_B} \times {\bf F}_{2KK_B \times KN_B}. \label{ach1:f}
\end{eqnarray}
Second, pass $\bm{f}_1, \dots, \bm{f}_K$ through an $(N,K)$ MDS code to generate $\overline{\bm{f}}_1, \dots, \overline{\bm{f}}_N$ such that any $K$ distinct $\overline{\bm{f}}_i$ may recover $\bm{f}_1, \dots, \bm{f}_K$. To this end, choose a $K \times N$ Vandermonde matrix ${\bf V} = [\zeta_j^{i-1}]_{ij}$ where the element in the $i^{th}$ row and $j^{th}$ column is $\zeta_j^{i-1}, \forall i \in [K], j \in [N]$ and $\zeta_j$ are distinct elements in $\mathbb{F}_q$. Then define
\begin{eqnarray}
\left( \underbrace{\overline{\bm{f}}^T_1}_{K_B \times 1}, \dots, \underbrace{\overline{\bm{f}}^T_N}_{K_B \times 1} \right)_{K_B \times N} \triangleq \left( \bm{f}_1^T, \dots, \bm{f}_K^T \right)_{K_B \times K} \times {\bf V}_{K \times N}. \label{ach1:ff}
\end{eqnarray}
Third, choose an $NK_B \times NK_B$ Cauchy matrix ${\bf H} = \left[\frac{1}{\gamma_i - \delta_j}\right]_{ij}$ where the element in the $i^{th}$ row and $j^{th}$ column is $\frac{1}{\gamma_i - \delta_j}$ and $\gamma_i, \delta_j, i, j \in [NK_B]$ are distinct elements in $\mathbb{F}_q$. Then define
\begin{eqnarray}
\left( \underbrace{\bm{h}_1}_{1\times K_B}, \dots, \underbrace{\bm{h}_N}_{1\times K_B} \right)_{1\times NK_B} \triangleq \left(\underbrace{\bm{f}_0}_{1\times K(N_B-K_B)}, \underbrace{Y_0}_{1\times K(2K_B-N_B)}, \underbrace{Z}_{1\times (N-K)K_B} \right)_{1 \times NK_B} \times {\bf H}_{NK_B \times NK_B}. \notag\\
\label{ach1:h}
\end{eqnarray}
We are now ready to specify the storage systems. Note that $Y_n, n \in [N]$ contains $\kappa = 2K_B$ elements in $\mathbb{F}_q$.
\begin{eqnarray}
Y_n &=& \left( \underbrace{\overline{\bm{f}}_n}_{1\times K_B}, \underbrace{\bm{h}_n}_{1\times K_B} \right)_{1\times 2K_B}, \forall n \in [N]. \label{ach1:yn}
\end{eqnarray}
The above encoding procedure is linear, thus the existence of precoding matrices ${\bf A}_n, {\bf B}_n, {\bf Z}_n$ is implied.
\end{enumerate}

We show that the above linear scheme works by verifying the constraints (\ref{c:edec}), (\ref{c:sec}),  (\ref{c:decb}) sequentially. Consider any $\mathcal{K} = \binom{[N]}{K}, \mathcal{K}_B \in \binom{[N_B]}{K_B}$.
\begin{enumerate}
\item (\ref{c:edec}):
We need to show that $Y_0$ is determined by $Y_{\mathcal{K}}, B_{\mathcal{K}_B}$. Proceed as follows.
First, $Y_{\mathcal{K}}$ contains $\overline{\bm{f}}_{\mathcal{K}}$, which is invertible to $\bm{f}_{[K]}$ as ${\bf V}$ is a Vandermonde matrix.
Second, we may subtract the contribution of $B_{\mathcal{K}_B}$ from $\bm{f}_{[K]}$ and obtain
\begin{eqnarray}
&& \left( \underbrace{Y_0}_{1\times K(2K_B-N_B)}, \underbrace{ B_{\mathcal{K}^c_B}}_{1\times K(N_B-K_B)} \right)_{1\times KK_B} \times {\bf F}_{sub}^{K K_B \times KK_B}. 
\end{eqnarray}
${\bf F}_{sub}$ has full rank as it is a sub-matrix of the Cauchy matrix ${\bf F}$, then $Y_0$ is obtained with no error.

\item (\ref{c:sec}):
We need to show that $Y_0$ is independent of $Y_{\mathcal{K}^c}, B_{\mathcal{K}^c_B}$. To this end, we prove the following mutual information is zero.
\begin{eqnarray}
&& I\left(Y_0; Y_{\mathcal{K}^c}, B_{\mathcal{K}^c_B} \right) \notag\\
&\overset{(\ref{ind}), (\ref{ach1:yn})}{=}& I\left( Y_0; \overline{\bm{f}}_{\mathcal{K}^c}, \bm{h}_{\mathcal{K}^c} \mid B_{\mathcal{K}^c_B} \right) \\
&\leq& I\left( Y_0; \overline{\bm{f}}_{\mathcal{K}^c}, \bm{f}_{[K]}, \bm{h}_{\mathcal{K}^c} \mid B_{\mathcal{K}^c_B} \right) \\
&\overset{(\ref{ach1:ff})}{\leq}& I\left( Y_0; \bm{f}_{[K]}, \bm{h}_{\mathcal{K}^c} \mid B_{\mathcal{K}^c_B} \right) \label{eq:a11}\\
&=&  H\left( \bm{f}_{[K]}, \bm{h}_{\mathcal{K}^c} \mid B_{\mathcal{K}^c_B}  \right)  -  H \left( \bm{f}_{[K]}, \bm{h}_{\mathcal{K}^c} \mid Y_0, B_{\mathcal{K}^c_B}  \right) \\
&\leq& NK_B-  H \left( \bm{f}_{[K]} \mid Y_0, B_{\mathcal{K}^c_B} \right) -  H \left( \bm{h}_{\mathcal{K}^c} \mid \bm{f}_{[K]}, Y_0, B \right) \label{eq:a12} \\
&\overset{(\ref{ach1:f}), (\ref{ach1:h})}{=}& NK_B -  H \left( B_{\mathcal{K}} \times {\bf F}^{KK_B\times KK_B}_{sub} \mid Y_0, B_{\mathcal{K}^c_B} \right) \notag\\
&&~-  H \left( Z \times {\bf H}_{sub}^{(N-K)K_B \times (N-K) K_B} \mid Y_0, B \right)\\
&=& NK_B - KK_B - (N-K) K_B  = 0. 
\end{eqnarray}
(\ref{eq:a11}) follows from the fact that $\overline{\bm{f}}_{\mathcal{K}^c}$ is a deterministic function of $\bm{f}_{[K]}$ (refer to (\ref{ach1:ff})).
To obtain the first term of (\ref{eq:a12}), note that $\bm{f}_{[K]}, \bm{h}_{\mathcal{K}^c} $ contains $KK_B + (N-K)K_B = NK_B$ elements.

\item  (\ref{c:decb}): We need to show that $B$ is determined by $Y_{[N]}$. Proceed as follows. First, as Cauchy ${\bf H}_{NK_B\times NK_B}$ has full rank, we may from $\bm{h}_{[N]}$ (the last $K_B$ elements of $Y_{[N]}$), recover $\bm{f}_0, Y_0, Z$ with no error. Second, $Y_{[N]}$ contains $\overline{\bm{f}}_{[N]}$, from which we may obtain $\bm{f}_{[K]}$ as ${\bf V}$ is Vandermonde.
Third, we may eliminate the contribution of $Y_0$ from $\bm{f}_{[K]}$ and combine with $\bm{f}_0$ to obtain $(B_{[N_B]})_{1\times KN_B} \times {\bf F}_{sub}^{KN_B \times KN_B}$, from which $B_{[N_B]}$ can be recovered with no error as ${\bf F}$ is Cauchy. Thus SR $B = B_{[N_B]}$ is obtained.
\end{enumerate}

\subsubsection{$K/N < 1/2, K_B/N_B > 1/2$: Achievability of $(\lambda_0, \lambda_B) = \left(2K\left(1-\frac{N_B}{2K_B}\right), \frac{N-2K}{N_B}+\frac{K}{K_B} \right)$}
The linear scheme of this section generalizes that of the example in Section \ref{ex:very2}. Specifically, the repetition code (of $\bm{h}$ in $Y$) and the $(3,2)$ MDS code (of $\bm{b}^x$ in $Y$) in the example are generalized to a tuple of $(N,K)$ and $(N, N-K)$ MDS codes respectively that satisfy the property that the product of the parity check matrix of the $(N,K)$ MDS code and the generator matrix of the $(N,N-K)$ MDS code has full rank (set as generalized Reed-Solomon codes, see Lemma \ref{lemma:rank}). 
The parameters of the linear scheme that achieves the desired rate tuple are given as follows. Local randomness $Z$ is not used.
\begin{enumerate}
\item Field size $q$ is any prime power such that $q \geq \max(2KN_BK_B,N)$.
\item Scaling factor $\kappa = N_BK_B$.
\item Precoding matrices are specified through the following steps.  $Y_0$ contains $\kappa \lambda_0 = KN_B(2K_B-N_B)$ elements in $\mathbb{F}_q$ and $B_t, t \in [N_B]$ contains $\kappa \lambda_B = (N-K)K_B+K(N_B-K_B)$ elements in $\mathbb{F}_q$. First, we divide $B_t$ into $2$ parts, labelled by the superscript $x, y$.
\begin{eqnarray}
B_t = \left( \underbrace{ \bm{b}^{1,x}_t}_{1\times K_B} , \dots, \underbrace{ \bm{b}^{N-K,x}_t}_{1\times K_B}, \underbrace{\bm{b}_t^y}_{1\times K(N_B-K_B)} \right), t \in [N_B].
\end{eqnarray}
Second, define $\bm{\alpha} \triangleq (\alpha_1, \dots, \alpha_N) \in \mathbb{F}_q^{1\times N}$, $\bm{v} \triangleq (v_1, \dots, v_N) \in \mathbb{F}_q^{1\times N}$ where $\forall n \in [N], \alpha_n$ are distinct elements in $\mathbb{F}_q$ and $v_n \neq 0$; choose an $N \times N$ matrix $\mbox{GRS}_{N\times N}(\bm{v}, \bm{\alpha}) = \left[v_j \alpha_j^{i-1}\right]_{ij}$ where the element in the $i^{th}$ row and $j^{th}$ column is $v_j \alpha_j^{i-1}, i, j \in [N]$. Note that $\mbox{GRS}$ is the generator matrix of a generalized Reed-Solomon code. Define $\mbox{GRS}_{N\times N}(\bm{v}, \bm{\alpha}) \triangleq \left(  {\bf G}_{K \times N}; {\bf F}_{(N-K) \times N} \right)$, i.e., the first $K$ rows are defined as ${\bf G}$ and the last $N-K$ rows are defined as ${\bf F}$ and note that both ${\bf G}, {\bf F}$ are MDS, i.e., any maximal square sub-matrix has full rank.

Third, pass $\bm{b}^{1,x}_t, \dots, \bm{b}^{N-K,x}_t$ through an $(N,N-K)$ MDS code (with generator matrix ${\bf F}$) to generate $\bm{\overline{b}}^{1,x}_t, \dots, \bm{\overline{b}}^{N,x}_t$ such that any $(N-K)$ distinct $\bm{\overline{b}}^{x}_t$ may recover $\bm{b}^{1,x}_t, \dots, \bm{b}^{N-K,x}_t$, i.e., $\forall t \in [N_B]$,
\begin{eqnarray}
\left( \underbrace{\left(\bm{\overline{b}}^{1,x}_t\right)^T}_{K_B \times 1}, \dots, \underbrace{\left(\bm{\overline{b}}^{N,x}_t\right)^T}_{K_B \times 1} \right) _{K_B \times N} \triangleq \left( \left( \bm{b}^{1,x}_t \right)^T, \dots, \left( \bm{b}^{N-K,x}_t \right)^T \right)_{K_B \times (N-K)} \times {\bf F}_{(N-K) \times N}. \notag\\
\label{ach2:n-k}
\end{eqnarray}

Fourth, choose a $KN_BK_B \times KN_BK_B$ Cauchy matrix ${\bf H} = \left[\frac{1}{\beta_i - \gamma_j}\right]_{ij}$ where the element in the $i^{th}$ row and $j^{th}$ column is $\frac{1}{\beta_i - \gamma_j}$ and $\beta_i, \gamma_j, i, j \in [KN_BK_B]$ are distinct elements in $\mathbb{F}_q$. Then define
\begin{eqnarray}
\left( \underbrace{ \bm{h}^1_1}_{1 \times K_B}, \dots, \underbrace{\bm{h}^{K}_1}_{1 \times K_B}, \dots, \underbrace{\bm{h}^1_{N_B}}_{1 \times K_B}, \dots, \underbrace{\bm{h}^K_{N_B}}_{1 \times K_B} \right)_{1\times KN_BK_B} &\triangleq& \left(Y_0, \bm{b}_1^y, \dots, \bm{b}_{N_B}^y \right)_{1 \times KN_BK_B} \notag\\
&&~\times~ {\bf H}_{KN_BK_B \times KN_BK_B}.
\end{eqnarray}

Fifth, pass $\bm{h}^1_t, \dots, \bm{h}^K_t, t \in [N_B]$ through an $(N,K)$ MDS code (with generator matrix ${\bf G}$) to generate $\bm{\overline{h}}^1_t, \dots, \bm{\overline{h}}^N_t$ such that any $K$ distinct $\bm{\overline{h}}_t$ may recover $\bm{h}^1_t, \dots, \bm{h}^K_t$, i.e., $\forall t\in [N_B]$,
\begin{eqnarray}
\left( \underbrace{\left(\bm{\overline{h}}^{1}_t\right)^T}_{K_B \times 1}, \dots, \underbrace{\left(\bm{\overline{h}}^{N}_t\right)^T}_{K_B \times 1} \right) _{K_B \times N} \triangleq \left( \left( \bm{h}^{1}_t \right)^T, \dots, \left( \bm{h}^{K}_t \right)^T \right)_{K_B \times K} \times {\bf G}_{K \times N}. \label{ach2:k}
\end{eqnarray}

We are now ready to specify the storage systems. Note that $Y_n, n \in [N]$ contains $\kappa = N_BK_B$ elements in $\mathbb{F}_q$.
\begin{eqnarray}
Y_n &=& \left( \bm{\overline{h}}_1^n + \bm{\overline{b}}_1^{n,x}, \dots, \bm{\overline{h}}_{N_B}^n + \bm{\overline{b}}_{N_B}^{n,x} \right)_{1\times N_BK_B}, \forall n \in [N]. \label{ach2:y}
\end{eqnarray}
\end{enumerate}

We show that the above linear scheme works by verifying the constraints (\ref{c:edec}), (\ref{c:sec}),  (\ref{c:decb}) sequentially. Consider any $\mathcal{K} \in \binom{[N]}{K}, \mathcal{K}_B \in \binom{[N_B]}{K_B}$.
\begin{enumerate}
\item (\ref{c:edec}): 
We need to show that $Y_0$ is determined by $Y_{\mathcal{K}}, B_{\mathcal{K}_B}$. Proceed as follows.

$\bm{b}_{\mathcal{K}_B}^{1,x}, \dots, \bm{b}_{\mathcal{K}_B}^{N-K,x}$ are known from $B_{\mathcal{K}_B}$ and thus $\bm{\overline{b}}_{\mathcal{K}_B}^{1,x}, \dots, \bm{\overline{b}}_{\mathcal{K}_B}^{N,x}$ are known from $B_{\mathcal{K}_B}$ due to the $(N,N-K)$ MDS precoding in (\ref{ach2:n-k}). 
Then  subtract the contribution of $\bm{\overline{b}}_{\mathcal{K}_B}^{\mathcal{K},x}$ from $Y_{\mathcal{K}}$ and obtain  $\bm{\overline{h}}_{\mathcal{K}_B}^{\mathcal{K}}$, from which we can then obtain $\bm{h}_{\mathcal{K}_B}^{1}, \dots, \bm{h}_{\mathcal{K}_B}^{K}$ due to the $(N,K)$ MDS precoding in (\ref{ach2:k}). 
Note that
\begin{eqnarray}
\left( \bm{h}_{\mathcal{K}_B}^{1}, \dots, \bm{h}_{\mathcal{K}_B}^{K} \right)_{1\times KK^2_B} = \left( Y_0, \bm{b}_1^y, \dots, \bm{b}_{N_B}^y \right)_{1 \times KN_BK_B} \times {\bf H}_{sub}^{KN_BK_B \times KK_B^2}.
\end{eqnarray}
${\bf H}_{sub}$ denotes a sub-matrix of ${\bf H}$ where the superscript shows the dimension.
Note that $\bm{b}_{\mathcal{K}_B}^y$ are known from $B_{\mathcal{K}_B}$ and we may further subtract their contribution from $\bm{h}_{\mathcal{K}_B}^{1}, \dots, \bm{h}_{\mathcal{K}_B}^{K}$ and obtain
\begin{eqnarray}
\left( Y_0, \bm{b}_{\mathcal{K}_B^c}^y\right)_{1 \times KK^2_B} \times {\bf H}_{sub}^{KK_B^2 \times KK_B^2}.
\end{eqnarray}
${\bf H}_{sub}$ has full rank as it is a sub-matrix of the Cauchy matrix ${\bf H}$. Thus $Y_0$ is perfectly recovered.

\item (\ref{c:sec}): We need to show that $Y_0$ is independent of $Y_{\mathcal{K}^c}, B_{\mathcal{K}^c_B}$. Proceed as follows. 
\begin{eqnarray}
&& I\left(Y_0; Y_{\mathcal{K}^c}, B_{\mathcal{K}^c_B} \right) \notag\\
&\overset{(\ref{ind}), (\ref{ach2:y})}{=}& I\left( Y_0; \bm{\overline{h}}_{\mathcal{K}_B}^{\mathcal{K}^c} + \bm{\overline{b}}_{\mathcal{K}_B}^{\mathcal{K}^c,x}, \bm{\overline{h}}_{\mathcal{K}^c_B}^{\mathcal{K}^c} \mid B_{\mathcal{K}^c_B}  \right) \\
&\leq& I\left( Y_0; \bm{\overline{h}}_{\mathcal{K}_B}^{\mathcal{K}^c} + \bm{\overline{b}}_{\mathcal{K}_B}^{\mathcal{K}^c,x}, \bm{h}_{\mathcal{K}^c_B}^{[K]} \mid B_{\mathcal{K}^c_B} \right) \label{eq:ach21}\\
&=& H\left( \underbrace{ \bm{\overline{h}}_{\mathcal{K}_B}^{\mathcal{K}^c} + \bm{\overline{b}}_{\mathcal{K}_B}^{\mathcal{K}^c,x}, \bm{h}_{\mathcal{K}^c_B}^{[K]} }_{1\times \Delta} \bigg| B_{\mathcal{K}^c_B}  \right)  - H \left( \bm{\overline{h}}_{\mathcal{K}_B}^{\mathcal{K}^c} + \bm{\overline{b}}_{\mathcal{K}_B}^{\mathcal{K}^c,x}, \bm{h}_{\mathcal{K}^c_B}^{[K]} \mid Y_0, B_{\mathcal{K}^c_B}  \right) \\
&\leq& \Delta -  H \left( \bm{h}_{\mathcal{K}^c_B}^{[K]} \mid Y_0, B_{\mathcal{K}^c_B}  \right) -  H \left( \bm{\overline{h}}_{\mathcal{K}_B}^{\mathcal{K}^c} + \bm{\overline{b}}_{\mathcal{K}_B}^{\mathcal{K}^c,x} \mid \bm{h}_{\mathcal{K}^c_B}^{[K]}, Y_0, B_{\mathcal{K}^c_B}, \bm{b}^y_{[N_B]}  \right) \\
&\leq& \Delta -  H \left( \bm{b}_{\mathcal{K}_B}^{y}  \times {\bf H}_{sub}^{K(N_B-K_B)K_B\times K(N_B-K_B)K_B} \mid Y_0, B_{\mathcal{K}^c_B} \right) \notag\\
&&-~ H \left( \bm{\overline{b}}_{\mathcal{K}_B}^{\mathcal{K}^c,x} \mid Y_0, B_{\mathcal{K}^c_B}, \bm{b}^y_{[N_B]}  \right) \\
&=& \Delta -  K(N_B-K_B)K_B - H \left( \bm{b}_{\mathcal{K}_B}^{[N-K],x} \mid Y_0, B_{\mathcal{K}^c_B}, \bm{b}^y_{[N_B]}  \right) \label{eq:ach22} \\
&=& \Delta -  K(N_B-K_B)K_B - (N-K)K^2_B\\
&=& 0.
\end{eqnarray}
Note that the notation $\bm{h}_{\mathcal{X}}^{\mathcal{Y}} = \left\{ \bm{h}_{i}^j, i \in \mathcal{X}, j \in \mathcal{Y} \right\}$ and $\Delta = (N-K)K^2_B+K(N_B-K_B)K_B$.
(\ref{eq:ach21}) follows from the MDS encoding (\ref{ach2:k}) so that $\bm{\overline{h}}_{\mathcal{K}^c_B}^{\mathcal{K}^c}$ is a function of $\bm{h}_{\mathcal{K}^c_B}^{[K]}$ and $N-K > K$.
In (\ref{eq:ach22}), the third term follows from the MDS encoding (\ref{ach2:n-k}) so that $\bm{\overline{b}}_{\mathcal{K}_B}^{\mathcal{K}^c,x}$ is invertible to $\bm{b}_{\mathcal{K}_B}^{[N-K],x}$.

\item (\ref{c:decb}): We need to show that $B$ is determined by $Y_{[N]}$. 

First, it is well known that the dual of a generalized Reed-Solomon code is another generalized Reed-Solomon code \cite{MacWilliams_Sloane}. Then the null space of the row space of ${\bf G}_{K\times N}$ is the row space of an $(N-K) \times N$ matrix ${\bf G}^\perp_{(N-K) \times N}$ where ${\bf G}^\perp = \mbox{GRS}_{(N-K) \times N}(\bm{u}, \bm{\alpha})$, $\bm{u} = (u_1, \dots, u_N)$, and $u_n = \left( v_n \prod_{i \neq n} (\alpha_n - \alpha_i) \right)^{-1}, n \in [N]$, i.e.,
\begin{eqnarray}
{\bf G}^\perp_{(N-K) \times N} \times {\bf G}^T_{N\times K} = {\bf 0}_{(N-K) \times K}.
\end{eqnarray} 
Then we use ${\bf G}^\perp_{(N-K) \times N}$ to zero-force the $\bm{h}_t, t \in [N_B]$ terms in $Y_n$.
\begin{eqnarray}
{\bf G}^\perp_{(N-K) \times N} 
\underbrace{\left[\begin{array}{c}
Y_1\\
\vdots\\
Y_N
\end{array}
\right]}_{N \times N_BK_B} &=& 
{\bf G}^\perp_{(N-K) \times N} {\bf F}^T_{N\times (N-K)}
\underbrace{ \left[\begin{array}{ccc}
\bm{b}^{1,x}_1 & \cdots & \bm{b}^{1,x}_{N_B}\\
\vdots & \ddots & \vdots \\
\bm{b}^{N-K,x}_1 & \cdots & \bm{b}^{N-K,x}_{N_B}
\end{array}
\right]}_{(N-K) \times N_BK_B},
\end{eqnarray}
which from Lemma \ref{lemma:rank} (i.e., ${\bf G}^\perp{\bf F}^T$ has full rank), is invertible to $\bm{b}_{[N_B]}^{[N-K], x}$, i.e., we  recover all $\bm{b}^x$.

Second, given all $\bm{b}^x$, we  subtract their contribution from $Y_n, n \in [N]$ and obtain $\bm{\overline{h}}_{[N_B]}^{[N]}$, from which we  further obtain $\bm{h}_{[N_B]}^{[K]}$.

Finally, from $\bm{h}_{[N_B]}^{[K]}$, we obtain $B$ (i.e., all $\bm{b}^x, \bm{b}^y$).
\end{enumerate}

\begin{lemma}\label{lemma:rank}
\begin{eqnarray}
\mbox{rank}\left( {\bf G}^\perp_{(N-K) \times N} {\bf F}^T_{N\times (N-K)} \right) = N-K. \label{eq:rank}
\end{eqnarray}
\end{lemma}

{\it Proof:} To set up the proof by contradiction, let us assume $\left( {\bf G}^\perp{\bf F}^T \right)_{(N-K) \times (N-K)}$ does not have full rank, i.e., there exists a vector $\bm{\gamma} \in \mathbb{F}_q^{(N-K) \times 1}$ such that ${\bf G}^\perp{\bf F}^T \bm{\gamma} = {\bf 0}_{(N-K) \times 1}$. Then defining $\bm{\delta} \triangleq {\bf F}^T \bm{\gamma}$, we have ${\bf G}^\perp \bm{\delta} = {\bf 0}$, i.e., $\bm{\delta}$ lies in the null space of the row space of ${\bf G}^\perp$, which is the row space of ${\bf G}$ as ${\bf G}^\perp {\bf G}^T = {\bf 0}$. Now we arrive at the property that $\bm{\delta}$ lies in the row space of ${\bf G}$. However, $\bm{\delta} = {\bf F}^T \bm{\gamma}$ also lies in the row space of ${\bf F}$, i.e., $\bm{\delta}$ lies in the row space of both ${\bf F}$ and ${\bf G}$, which leads to the conclusion that $\bm{\delta} = {\bf 0}$ because the row spaces of ${\bf F}$ and ${\bf G}$ intersect only at the trivial zero vector as matrix $\left( {\bf G}; {\bf F}\right) = \mbox{GRS}_{N\times N}(\bm{v}, \bm{\alpha})$ has full rank. Finally, $\bm{\delta}$ being the zero vector indicates that $\bm{\gamma}$ must also be the zero vector as ${\bf F}^T$ has independent columns and $\bm{\delta} = {\bf F}^T \bm{\gamma}$. Finally, $\bm{\gamma}$ being the zero vector shows that ${\bf G}^\perp{\bf F}^T$ has full rank.

\hfill\qed

\subsection{$\mathcal{C}_Q$: Capacity region for $\widetilde{\mbox{QS}}\widetilde{\mbox{EA}}$}\label{sec:qregion}
Based on Theorem \ref{thm:ab} and the convexity of $\mathcal{C}_Q$ proved in Appendix \ref{app:convex}, the inner bound (achievable rate region) for $\mathcal{C}_{Q}$ is implied by the inner bound for the capacity region $\mathcal{C}$ of $\widetilde{\mbox{CS}}\widetilde{\mbox{SRA}}$, which is presented in Section \ref{sec:c}. 
Thus, here it suffices to prove only the outer bound (converse) for $\mathcal{C}_Q$. 

We have two bounds to prove, i.e., $\mathcal{R}_{cut}$ and $\mathcal{R}_{\infty}$.  In the following, unless explicitly stated  otherwise, the quantum information measure terms are with respect to the state $\rho_{RQ_1\dots Q_N B_1 \dots B_{N_B}}$. Recall from (\ref{q:dec}) that our problem formulation requires perfect recovery for {\em every} pure state $R Q_0$, so in particular perfect recovery is required for a maximally entangled state $R Q_0$ where $Q_0$ is maximally mixed, i.e., $H(R, Q_0)_{\ket{\varphi}_{RQ_0}} = 0$ and $H(R) = H(Q_0)_{\ket{\varphi}_{RQ_0}} = \log_q |Q_0|$. In the converse proof, we assume throughout that $R Q_0$ is maximally entangled. 

\subsubsection{Proof of $\mathcal{R}_{cut}$ Bound} 
We need to prove that every achievable tuple $(\lambda_0,\lambda_B)$ must satisfy the bound $ \lambda_0 \leq  \max(2K-N, 0) + \lambda_B\cdot\max(2K_B - N_B, 0) $. Let us proceed as follows. Consider any achievable tuple $(\lambda_0,\lambda_B)$. Since this tuple is achievable, there must exist a feasible coding scheme $\mathfrak{S}$ such that 
\begin{align}
\lambda_0(\mathfrak{S})&\geq \lambda_0,\label{eq:l0bound}\\
\lambda_B(\mathfrak{S})&\leq \lambda_B.\label{eq:lbbound}
\end{align}
Consider this coding scheme $\mathfrak{S}$. Besides $\lambda_0(\mathfrak{S}), \lambda_B(\mathfrak{S})$, since there is no potential for ambiguity for the remaining parameters of the coding scheme, we will suppress the label $\mathfrak{S}$ and write, e.g., $\kappa,q$ instead of $\kappa(\mathfrak{S}), q(\mathfrak{S})$ to simplify the notation.

The perfect recovery of $R Q_0$ implies the entropic condition \cite{Schumacher_Nielsen, Sun_Jafar_QuStorage},
\begin{eqnarray}
\forall \mathcal{K} \in \binom{[N]}{K}, \forall \mathcal{K}_B \in \binom{[N_B]}{K_B}, ~~ I\left(R; Q_{\mathcal{K}}, B_{\mathcal{K}_B} \right) = 2 H(R) = 2\kappa \lambda_0(\mathfrak{S}). \label{q:edec}
\end{eqnarray}
Then from weak monotonicity, we have
\begin{eqnarray}
&& I\left(R; Q_{\mathcal{K}}, B_{\mathcal{K}_B} \right) + I \left(R; Q_{\mathcal{K}^c}, B_{\mathcal{K}^c_B} \right) \leq 2 H(R) = 2 \kappa \lambda_0(\mathfrak{S})\\
&\overset{(\ref{q:edec})}{\implies}& I\left(R; Q_{\mathcal{K}^c}, B_{\mathcal{K}^c_B} \right) = 0, ~\forall \mathcal{K}, \forall \mathcal{K}_B. \label{q:esec}
\end{eqnarray}
From (\ref{q:esec}), setting $\mathcal{K} = [N-K+1:N]$ and $\mathcal{K}_B = [N_B-K_B+1:N_B]$, we have
\begin{eqnarray}
I\left(R; Q_{[N-K]}, B_{[N_B - K_B]} \right) = 0. \label{eq:sec1}
\end{eqnarray}
Then from (\ref{q:edec}), setting $\mathcal{K} = [K], \mathcal{K}_B = [K_B]$, we have
\begin{align}
&2\kappa \lambda_0(\mathfrak{S}) \notag\\
&= I \left(R ; Q_{[K]}, B_{[K_B]} \right) \\
&\leq  I \left(R ; Q_{[N-K]}, Q_{[N-K+1:K]} , B_{[N_B - K_B]}, B_{[N_B - K_B+1:K_B]} \right) \label{eq:s1} \\
&\overset{(\ref{eq:sec1})}{=} I \left( R ; Q_{[N-K+1:K]}, B_{[N_B - K_B+1:K_B]} \mid Q_{[N-K]}, B_{[N_B - K_B]} \right) \\
&\leq 2 H \left( Q_{[N-K+1:K]} \right) + 2 H \left( B_{[N_B - K_B+1:K_B]} \right) \\
&\leq 2 \log_q \prod_{n \in [N-K+1:K]} |Q_n| + 2 \log_q \prod_{i \in [N_B-K_B+1:K_B]} |B_i| \\
&\leq 2\kappa\cdot\max\left(2K-N, 0\right)  + 2\kappa \lambda_B(\mathfrak{S}) \cdot\max\left(2K_B-N_B, 0 \right) \\
\implies \lambda_0(\mathfrak{S}) &\leq \max\left(2K-N, 0\right) + \lambda_B(\mathfrak{S}) \cdot\max\left(2K_B-N_B, 0\right)\\
\stackrel{\eqref{eq:l0bound}\eqref{eq:lbbound}}{\implies} \lambda_0 &\leq \max\left(2K-N, 0\right) + \lambda_B\cdot\max\left(2K_B-N_B, 0\right).
\end{align}
(\ref{eq:s1}) follows from the fact that quantum conditional mutual information is non-negative.

\subsubsection{Proof of $\mathcal{R}_{\infty}$ Bound}
Second, consider $\mathcal{R}_\infty$. Interestingly, the proof of the quantum setting parallels that of the classical setting in Section \ref{sec:cinf}. The differences are highlighted here while similarities are briefly described. Following exactly the same reasoning as that of the classical setting, the $K/N < 1/2$ case follows from the $K/N \geq 1/2$ case  and it suffices to prove 
\begin{eqnarray}
\lambda_0 \leq N - (N-K) \frac{N_B}{K_B}~\mbox{when $K/N \geq 1/2$.} 
\end{eqnarray}

Consider $\mathcal{K} = [K] = [N-K] \cup [N-K+1:K], \mathcal{K}_B =  \mathcal{I} \cup [K_B+1:N_B]$ where $\mathcal{I} \subset [K_B], |\mathcal{I}| = 2K_B - N_B$. Note that $2K_B-N_B \leq K_B$ and $K/N \geq 1/2$. 
Consider any feasible scheme $\mathfrak{S}$. From the quantum entropic decodability constraint (\ref{q:edec}), we have 
\begin{eqnarray}
2 \kappa \lambda_0 &\leq& 2 \kappa \lambda_0(\mathfrak{S})  \overset{(\ref{q:edec})}{=}  I\left( R; Q_{[N-K]}, Q_{[N-K+1:K]}, B_{\mathcal{K}_B} \right) \\
&\leq& I\left( R; Q_{[N-K]}, B_{\mathcal{K}_B} \right) + 2 H\left( Q_{[N-K+1:K]} \right) \label{qinf:1} \\
&\overset{(\ref{product})}{\leq}& I\left( R, B_{\mathcal{I}}; Q_{[N-K]} \mid B_{[K_B+1:N_B]} \right) + 2\kappa (2K-N)  \\
&=& \underbrace{ I\left( R ; Q_{[N-K]} \mid B_{[K_B+1:N_B]} \right) }_{\overset{(\ref{q:esec})}{\leq} 0} ~+~ I\left( B_{\mathcal{I}}; Q_{[N-K]} \mid R, B_{[K_B+1:N_B]} \right) \notag\\
&&+~2\kappa (2K-N)  \label{qinf:2}\\
&\implies& 2\left( \kappa\lambda_0 - \kappa (2K-N) \right) \leq  I\left( B_{\mathcal{I}}; Q_{[N-K]} \mid R, B_{[K_B+1:N_B]} \right). \label{qinf:3}
\end{eqnarray}
(\ref{qinf:1}) follows from the following quantum mutual information bound 
$$I\left( R; Q_{[N-K+1:K]} \mid Q_{[N-K]}, B_{\mathcal{K}_B} \right) \leq 2 H\left( Q_{[N-K+1:K]} \right),$$
where we have a factor of $2$ before the entropy term $H\left( Q_{[N-K+1:K]} \right)$ because of the weak monotonicity of quantum entropy (in contrast to the strong monotonicity of classical entropy). The first term of (\ref{qinf:2}) is zero because of the implicit quantum security constraint, proved in (\ref{q:esec}), which says that due to weak monotonicity (monogamy of entanglement), a decoding set preserves all entanglement with the reference system, leaving its complement in a product state with the reference system.

Next we add (\ref{qinf:3}) for all $\mathcal{I} = \mathcal{I}_j, j \in [K_B]$ where $\mathcal{I}_j$ is defined in (\ref{cinf:iset}).
\begin{eqnarray}
 2 \left( \kappa\lambda_0 - \kappa (2K-N)  \right) K_B &\leq& \sum_{j \in [K_B]}  I\left( B_{\mathcal{I}_j}; Q_{[N-K]} \mid R, B_{[K_B+1:N_B]} \right) \\
 &\overset{(\ref{sum:2})}{\leq}& \left( 2K_B - N_B\right) I\left( B_{[K_B]}; Q_{[N-K]} \mid R, B_{[K_B+1:N_B]} \right)   \label{qinf:4}\\
 &\leq& 2 \kappa (N-K)(2K_B-N_B) \label{qinf:5}\\
 \implies \lambda_0 &\leq& N - (N-K) \frac{N_B}{K_B}.
\end{eqnarray}
In (\ref{qinf:4}), we  apply (\ref{sum:2}) in Lemma \ref{lemma:sum}, proved in the classical setting because the proof of (\ref{sum:2}) only uses following two properties that hold in both classical and quantum settings.
\begin{enumerate}
\item $I(Y_0; B) = 0$ and $H(B_1, \dots, B_{N_B}) = H(B_1) + \cdots + H(B_{N_B})$, while the corresponding quantum version also holds, i.e., $I(R; B) = 0$ as $R$ and $B$ are in a product state (see (\ref{product})) and the $B_i, i \in [N_B]$ entanglement is in a product state, i.e., $H(B_1, \dots, B_{N_B}) = H(B_1) + \cdots + H(B_{N_B})$ (see (\ref{ent})).
\item Strong sub-additivity of entropy, which holds for both quantum and classical entropy \cite{Lieb_Ruskai}. 
\end{enumerate}
We may verify each step of the proof of (\ref{sum:2}) and will find that replacing $Y_0$ by $R$ and replacing $Y_n$ by $Q_n$, all the steps remain valid. 
Similar to above, the only difference lies in how to bound the mutual information term in (\ref{qinf:4}), where for the quantum setting, we have a factor of $2$. The proof of the quantum converse is thus complete.

\section{Conclusion}
The problem of erasure-prone quantum storage with erasure-prone entanglement assistance, $\widetilde{\mbox{QS}}\widetilde{\mbox{EA}}$, is explored by connecting it to an analogous problem of erasure-prone classical storage with erasure-prone shared-randomness assistance, $\widetilde{\mbox{CS}}\widetilde{\mbox{SRA}}$. The capacity regions for both $\widetilde{\mbox{QS}}\widetilde{\mbox{EA}}$ and $\widetilde{\mbox{CS}}\widetilde{\mbox{SRA}}$ are characterized and shown to be identical in all parameter regimes, with the exception of one regime where the capacity remains open. It is conjectured that the capacity regions, $\mathcal{C}_Q$ and $\mathcal{C}$, respectively, are identical even in this regime, and that they match an inner bound (a feasible coding scheme) that is presented in Theorem \ref{thm:region}. Let us conclude by listing some promising directions for future work.
\begin{enumerate}
\item The validity of the aforementioned conjecture is an important open question. Resolving this question may require a deeper understanding of the various converse arguments in order to find the right combination of bounds to cover the regime that remains open. 

\item The connection between $\widetilde{\mbox{QS}}\widetilde{\mbox{EA}}$ and $\widetilde{\mbox{CS}}\widetilde{\mbox{SRA}}$, while clearly established in Theorem \ref{thm:ab} from an achievability perspective, remains intriguing in terms of converse bounds. Since the capacity regions for the two settings seem to match, it is natural to ask if there is also a  mechanism to translate an information theoretic converse bound from one setting to the other.

\item Figure \ref{fig:flowchart} shows that there is much room to reduce the size requirements for our coding schemes. Observe that the field $\mathbb{F}_q$ in the specialized coding scheme of Figure \ref{fig:flowchart} can be any field (even $\mathbb{F}_2$) but our generalizable construction for the same setting in Section  \ref{ex:very2} assumes $q\geq 13$. The maximum storage possible under strict constraints on the sizes of storage and EA systems remains an important open question for future work.

\item Our formulation of $\widetilde{\mbox{QS}}\widetilde{\mbox{EA}}$  does not allow any \emph{precoding} (i.e., before the message appears at the encoder) across EA systems to counter the EA erasures. However precoding across EA systems may be possible in some settings, e.g., if the EA systems are co-located. The capacity for similar alternative formulations is also of interest.

\item Besides what is shown in Figure \ref{fig:channel}, the $\widetilde{\mbox{QS}}\widetilde{\mbox{EA}}$ formulation does not allow any additional classical communication. On the other hand, it does allow joint processing of all unerased storage and EA systems at the decoder. Other formulations of interest may  restrict access to a subset of unerased systems at the decoder while allowing additional classical communication among surviving systems \cite{chen2024capacities}.
\end{enumerate}
Indeed, unreliable entanglement assistance is an important and  underexplored topic with broad applicability across a wide variety of settings.

\appendix
\section{Convexity of $\mathcal{C}_Q, \mathcal{C}$ and Space Sharing}\label{app:convex}
We prove that $\mathcal{C}_Q$, the capacity region for $\widetilde{\mbox{QS}}\widetilde{\mbox{EA}}$ is convex and note that the convexity of $\mathcal{C}$ for $\widetilde{\mbox{CS}}\widetilde{\mbox{SRA}}$ follows from similar reasoning.

We show that $\mathcal{C}_Q$ is convex, i.e., if $(\lambda_0(1), \lambda_B(1)) \in \mathcal{C}_Q$ and $(\lambda_0(2), \lambda_B(2)) \in \mathcal{C}_Q$, then $(\mu \lambda_0(1) + (1-\mu) \lambda_0(2), \mu \lambda_B(1) + (1-\mu) \lambda_B(2)) \in \mathcal{C}_Q$ for $0 \leq \mu \leq 1$. The proof is based on a  \emph{space-sharing} argument, essentially a direct construction of a coding scheme as a convex combination of other coding schemes, analogous to the time-sharing argument that is quite common in information theory.

As $(\lambda_0(1), \lambda_B(1)) \in \mathcal{C}_Q$, there exists a feasible coding scheme $\mathfrak{S}_1$ with parameters $$\left(\lambda_0(\mathfrak{S}_1), \lambda_B(\mathfrak{S}_1),\kappa(\mathfrak{S}_1),q(\mathfrak{S}_1),\mbox{ENC}(\mathfrak{S}_1), \mbox{DEC}^{\mathcal{K},\mathcal{K}_B} (\mathfrak{S}_1) \right),$$
where $\lambda_0(\mathfrak{S}_1) \geq \lambda_0(1), \lambda_B(\mathfrak{S}_1) \leq \lambda_B(1)$. Similarly, $(\lambda_0(2), \lambda_B(2)) \in \mathcal{C}_Q$ indicates that there is a feasible coding scheme $\mathfrak{S}_2$ with parameters $$\left(\lambda_0(\mathfrak{S}_2), \lambda_B(\mathfrak{S}_2),\kappa(\mathfrak{S}_2),q(\mathfrak{S}_2),\mbox{ENC}(\mathfrak{S}_2), \mbox{DEC}^{\mathcal{K},\mathcal{K}_B} (\mathfrak{S}_2) \right),$$
where $\lambda_0(\mathfrak{S}_2) \geq \lambda_0(2), \lambda_B(\mathfrak{S}_2) \leq \lambda_B(2)$. Next, based on $\mathfrak{S}_1$, $\mathfrak{S}_2$, we construct a coding scheme $\mathfrak{S}$ as follows. As rationals are dense over reals, for any $\mu \in [0,1]$ and any $\epsilon \in \mathbb{R}_{\geq 0}$, we can find a rational $\mu_\epsilon = u/v$ with $u, v \in \mathbb{Z}_{\geq 0}$ and $u \leq v$ such that $|\mu - \mu_\epsilon| \leq \epsilon$.

\subsection{$q(\mathfrak{S}_1) = q(\mathfrak{S}_2)$} \label{sec:sameq}
First, let us start with the simpler case where $q(\mathfrak{S}_1) = q(\mathfrak{S}_2) \triangleq q$. The parameters of $\mathfrak{S}$ are
\begin{eqnarray}
&& \left(\lambda_0(\mathfrak{S}), \lambda_B(\mathfrak{S}),\kappa(\mathfrak{S}),q(\mathfrak{S}) = q,\mbox{ENC}(\mathfrak{S}), \mbox{DEC}^{\mathcal{K},\mathcal{K}_B} (\mathfrak{S}) \right), \\
&& \lambda_0(\mathfrak{S}) = \frac{u}{v} \lambda_0(\mathfrak{S}_1) + \frac{v-u}{v} \lambda_0(\mathfrak{S}_2), \\
&&  \lambda_B(\mathfrak{S}) = \frac{u}{v} \lambda_B(\mathfrak{S}_1) + \frac{v-u}{v} \lambda_B(\mathfrak{S}_2), \\
&& \kappa(\mathfrak{S}) = v \cdot \kappa(\mathfrak{S}_1) \cdot \kappa(\mathfrak{S}_2), \\
&& \mbox{ENC}(\mathfrak{S}) = \left(\mbox{ENC}(\mathfrak{S}_1)\right)^{{\otimes u\kappa(\mathfrak{S}_2)}} \otimes\left(\mbox{ENC}(\mathfrak{S}_2)\right)^{\otimes(v-u)\kappa(\mathfrak{S}_1)},\label{share:enc}\\
&& \mbox{DEC}^{\mathcal{K},\mathcal{K}_B} (\mathfrak{S}) = \left( \mbox{DEC}^{\mathcal{K},\mathcal{K}_B} (\mathfrak{S}_1)\right)^{\otimes{u\kappa(\mathfrak{S}_2)}} \otimes \left(  \mbox{DEC}^{\mathcal{K},\mathcal{K}_B} (\mathfrak{S}_2)\right)^{\otimes{(v-u)\kappa(\mathfrak{S}_1)}}. \label{share:dec}
\end{eqnarray}
The quantum message $Q_0$ has $\kappa(\mathfrak{S}) \lambda_0(\mathfrak{S})$ qudits, and the EA $A_1$ $\dots$ $A_{N_B}$ $B_1$ $\dots$$ B_{N_B}$,  has $\kappa(\mathfrak{S}) \lambda_B(\mathfrak{S})$ qudits, 
\begin{align}
\kappa(\mathfrak{S}) \lambda_0(\mathfrak{S}) &=\kappa(\mathfrak{S}_1) \cdot  \kappa(\mathfrak{S}_2)\cdot\left( u  \lambda_0(\mathfrak{S}_1) + (v-u) \lambda_0(\mathfrak{S}_2)\right),\\
\kappa(\mathfrak{S}) \lambda_B(\mathfrak{S}) &= \kappa(\mathfrak{S}_1) \cdot \kappa(\mathfrak{S}_2)\cdot\left(u \lambda_B(\mathfrak{S}_1) + (v-u)  \lambda_B(\mathfrak{S}_2)\right).
\end{align}
 Then $\mbox{ENC}(\mathfrak{S})$ and $\mbox{DEC}^{\mathcal{K},\mathcal{K}_B} (\mathfrak{S})$ in (\ref{share:enc}) and (\ref{share:dec}) refer to the following encoding and decoding operations. $Q_0 = Q_0(1,1) \dots$ $Q_0(1,u\kappa(\mathfrak{S}_2)) Q_0(2,1) \dots Q_0(2,(v-u)\kappa(\mathfrak{S}_1))$ is divided into $u \kappa(\mathfrak{S}_2)+(v-u)\kappa(\mathfrak{S}_1)$ blocks, where $Q_0(1,i), i \in [u\kappa(\mathfrak{S}_2)]$ has $\kappa(\mathfrak{S}_1) \lambda_0(\mathfrak{S}_1)$ qudits and $Q_0(2,j), j \in [(v-u)\kappa(\mathfrak{S}_1)]$ has $\kappa(\mathfrak{S}_2) \lambda_0(\mathfrak{S}_2)$ qudits.
Similarly, $A_m, B_m, m \in [N_B]$ is divided into $u\kappa(\mathfrak{S}_2) + (v-u)\kappa(\mathfrak{S}_1)$ blocks where $A_m(1,i), B_m(1,i), i \in [u\kappa(\mathfrak{S}_2)]$ each has $\kappa(\mathfrak{S}_1) \lambda_B(\mathfrak{S}_1)$ qudits and $A_m(2,j), B_m(2,j), j \in [(v-u)\kappa(\mathfrak{S}_1)]$ each has $\kappa(\mathfrak{S}_2) \lambda_B(\mathfrak{S}_2)$ qudits.

Next, for each $i \in [u\kappa(\mathfrak{S}_2)]$, $Q_0(1,i), A_m(1,i), B_m(1,i)$ is encoded through the map $\mbox{ENC}(\mathfrak{S}_1)$ and decoded through the map $\mbox{DEC}^{\mathcal{K},\mathcal{K}_B} (\mathfrak{S}_1)$ so that the final output $\widehat{Q}_0(1,i)$ preserves the same entanglements with the rest of the universe that were originally associated with $Q_0(1,i)$. Similarly, for each $j \in [(v-u)\kappa(\mathfrak{S}_1)]$, $Q_0(2,j), A_m(2,j), B_m(2,j)$ is encoded through the map $\mbox{ENC}(\mathfrak{S}_2)$ and decoded through the map $\mbox{DEC}^{\mathcal{K},\mathcal{K}_B} (\mathfrak{S}_2)$ so that the final output $\widehat{Q}_0(2,j)$ preserves the same entanglements with the rest of the universe that were originally associated with $Q_0(2,j)$. Combining all outputs $\widehat{Q}_0 \triangleq \widehat{Q}_0(1,1) \dots \widehat{Q}_0(1,u\kappa(\mathfrak{S}_2)) \widehat{Q}_0(2,1) \dots \widehat{Q}_0(2,(v-u)\kappa(\mathfrak{S}_1))$, the quantum system $R \widehat{Q}_0$ is recovered whose state is identical to $\ket{\varphi}_{RQ_0}$.

Note that 
\begin{eqnarray}
&& \lambda_0(\mathfrak{S}) = \frac{u}{v} \lambda_0(\mathfrak{S}_1) + \frac{v-u}{v} \lambda_0(\mathfrak{S}_2) \geq  \frac{u}{v} \lambda_0(1) + \frac{v-u}{v} \lambda_0(2),\\
&&  \lambda_B(\mathfrak{S}) = \frac{u}{v} \lambda_B(\mathfrak{S}_1) + \frac{v-u}{v} \lambda_B(\mathfrak{S}_2) \leq \frac{u}{v} \lambda_B(1) + \frac{v-u}{v} \lambda_B(2).
\end{eqnarray}
Thus, the rate tuple $(\lambda_0, \lambda_B) = ( \mu_\epsilon \lambda_0(1) + (1-\mu_\epsilon) \lambda_0(2), \mu_\epsilon \lambda_B(1) + (1-\mu_\epsilon) \lambda_B(2) )$ is achievable. Letting $\epsilon \rightarrow 0$, the rate tuple $(\lambda_0, \lambda_B) = ( \mu \lambda_0(1) + (1-\mu) \lambda_0(2), \mu \lambda_B(1) + (1-\mu) \lambda_B(2) )$ is in the closure of the set of achievable rate tuples and the proof of convexity is complete.

\subsection{General $q(\mathfrak{S}_1), q(\mathfrak{S}_2)$}
Second, we consider the general case where $q(\mathfrak{S}_1), q(\mathfrak{S}_2)$ might be different. Here the key observation is that the closure of the set of achievable rate tuples is the same for all $q \geq 2$, i.e., for any $q \geq 2$, we may transform any feasible coding scheme $\mathfrak{S}$ with parameter $q(\mathfrak{S})$ to another scheme $\mathfrak{S}'$ such that $q(\mathfrak{S}') = q$ while achieving the same rate tuple. This result is stated in the following lemma.

\begin{lemma}\label{lemma:q}
For any $\epsilon$, $0<\epsilon < \lambda_0$, any $q \in \mathbb{Z}_{\geq 2}$ and any feasible coding scheme $\mathfrak{S}$ with parameters $$\left(\lambda_0(\mathfrak{S}), \lambda_B(\mathfrak{S}),\kappa(\mathfrak{S}),q(\mathfrak{S}),\mbox{ENC}(\mathfrak{S}), \mbox{DEC}^{\mathcal{K},\mathcal{K}_B} (\mathfrak{S}) \right),$$
that achieves rate tuple $(\lambda_0, \lambda_B)$, i.e., $\lambda_0(\mathfrak{S}) \geq \lambda_0, \lambda_B(\mathfrak{S}) \leq \lambda_B$, there exists another feasible coding scheme $\mathfrak{S}'$ with parameters $$\left(\lambda_0(\mathfrak{S}'), \lambda_B(\mathfrak{S}'),\kappa(\mathfrak{S}'),q(\mathfrak{S}') = q,\mbox{ENC}(\mathfrak{S}'), \mbox{DEC}^{\mathcal{K},\mathcal{K}_B} (\mathfrak{S}') \right),$$
that achieves rate tuple $(\lambda_0 - \epsilon, \lambda_B + \epsilon)$, i.e., $\lambda_0(\mathfrak{S}') \geq \lambda_0 - \epsilon, \lambda_B(\mathfrak{S}') \leq \lambda_B + \epsilon$. Therefore, the closure of the set of achievable rate tuples does not depend on $q$.
\end{lemma}

{\it Proof:} Note that $\epsilon$, $\lambda_0(\mathfrak{S}), \lambda_B(\mathfrak{S}), \kappa(\mathfrak{S}), q(\mathfrak{S}), q$ are given and $\epsilon < \lambda_0 \leq \lambda_0(\mathfrak{S})$. Then we can find integers $T, T', \kappa' \in \mathbb{Z}_{\geq 1}$ such that
\begin{eqnarray}
T \kappa(\mathfrak{S}) \frac{\log_2 q(\mathfrak{S})}{ \log_2 q} \frac{\lambda_B(\mathfrak{S}) }{ \lambda_B(\mathfrak{S}) + \epsilon} \leq T' \kappa' \leq T \kappa(\mathfrak{S}) \frac{\log_2 q(\mathfrak{S})}{ \log_2 q} \frac{\lambda_0(\mathfrak{S})}{ \lambda_0(\mathfrak{S}) - \epsilon} \label{eq:integer}
\end{eqnarray}
and the parameters of $\mathfrak{S}'$ are set as
\begin{eqnarray}
&& \lambda_0(\mathfrak{S}') =\lambda_0 (\mathfrak{S}) \frac{\kappa(\mathfrak{S})}{\kappa'} \frac{T}{T'} \frac{\log_2 q(\mathfrak{S})}{ \log_2 q}, \label{eq:l0}\\
&& \lambda_B(\mathfrak{S}') = \lambda_B (\mathfrak{S}) \frac{\kappa(\mathfrak{S})}{\kappa'} \frac{T}{T'} \frac{\log_2 q(\mathfrak{S})}{ \log_2 q}, \label{eq:lb} \\
&& \kappa(\mathfrak{S}') = \kappa' T', \label{eq:lk}\\
&& \mbox{ENC}(\mathfrak{S}') = \left(\mbox{ENC}(\mathfrak{S})\right)^{\otimes T},\\
&& \mbox{DEC}^{\mathcal{K},\mathcal{K}_B} (\mathfrak{S}') = \left( \mbox{DEC}^{\mathcal{K},\mathcal{K}_B} (\mathfrak{S})\right)^{\otimes T}.
\end{eqnarray}

For the coding scheme $\mathfrak{S}'$, the quantum message $Q_0$ has dimension $$|Q_0| = q^{\kappa(\mathfrak{S}') \lambda_0(\mathfrak{S}')} \overset{(\ref{eq:l0}), (\ref{eq:lk})}{=} q(\mathfrak{S})^{\kappa(\mathfrak{S}) \lambda_0(\mathfrak{S}) \times T} $$
and for the entanglement $AB$, each $A_m, B_m, m \in [N_B]$ has dimension $$|A_m| = |B_m| = q^{\kappa(\mathfrak{S}') \lambda_B(\mathfrak{S}')} \overset{(\ref{eq:lb}), (\ref{eq:lk})}{=} q(\mathfrak{S})^{\kappa(\mathfrak{S}) \lambda_B(\mathfrak{S}) \times T}.$$
$Q_0, A_m, B_m$ is encoded and decoded through $T$ parallel concatenations of $\mbox{ENC}(\mathfrak{S})$ and $\mbox{DEC}^{\mathcal{K},\mathcal{K}_B} (\mathfrak{S})$; perfect recovery is thus guaranteed. Furthermore,
\begin{eqnarray}
&& \lambda_0(\mathfrak{S}') \overset{(\ref{eq:integer}),(\ref{eq:l0})}{\geq} \lambda_0(\mathfrak{S}) - \epsilon \geq  \lambda_0- \epsilon, \\
&& \lambda_B(\mathfrak{S}') \overset{(\ref{eq:integer}),(\ref{eq:lb})}{\leq} \lambda_B (\mathfrak{S}) + \epsilon \leq  \lambda_B + \epsilon, 
\end{eqnarray}
so that the rate tuple $(\lambda_0 - \epsilon, \lambda_B + \epsilon)$ is achieved.
\hfill\qed

Equipped with Lemma \ref{lemma:q}, we may transform the two schemes $\mathfrak{S}_1$, $\mathfrak{S}_2$ to $\mathfrak{S}'_1$, $\mathfrak{S}'_2$ so that $q(\mathfrak{S}'_1) = q(\mathfrak{S}'_2) = q$. Then we can apply the same space sharing scheme in Appendix \ref{sec:sameq} to construct $\mathfrak{S}'$ so that letting $\epsilon \rightarrow 0$, any convex combination of rate tuples is achieved. The proof is thus complete.

\section{$(N_B = K_B = 1, K/N < 1/2)$: Achievability of $(\lambda_0, \lambda_B) = (K/2, K/2)$}\label{sec:perfect}
Consider $\widetilde{\mbox{QS}}\widetilde{\mbox{EA}}$ with $N_B = K_B = 1$ so that the entanglement $B$ is not erasure-prone. The setting reduces to entanglement-assisted quantum MDS codes where the rate tuple $(\lambda_0, \lambda_B) = (K/2, K/2)$ is known to be achievable by a teleportation based scheme \cite{Grassl_Huber_Winter}. Here we specialize the classical code in Section \ref{sec:ach1} and translate through Theorem \ref{thm:ab} to provide an alternative quantum scheme that does not involve teleportation or measurement. 

Consider encoding first. Set $\kappa = 2$ and choose $q \geq 2N$ as a prime power.
The quantum message $Q_0$ has $\kappa \lambda_0 = K$ qudits and each qudit is $q$-dimensional. Suppose $Q_0$ is in an arbitrary state with density matrix $\omega_{Q_0}$.  Without loss of generality, suppose $\omega_{Q_0}$ has a spectral decomposition $\omega_{Q_0} = \sum_{\bm{a}\in\mathbb{F}_q^{1\times K}} p_{\bm{a}} \ket{\bm{a}} \bra{\bm{a}}$ and purification $RQ_0$ in the pure state $\ket{\varphi}_{RQ_0}=\sum_{\bm{a}\in\mathbb{F}_q^{1\times K}} \sqrt{p_{\bm{a}}} \ket{\bm{a}} \ket{\bm{a}}$. The EA $B$ comprises $N_B \kappa \lambda_B = K$ qudits. The encoding proceeds as follows. 
\begin{align}
& \ket{\varphi}_{RQ_0}\ket{\phi}_{AB}\\
&= \sum_{\bm{a} \in \mathbb{F}_q^{1\times K}} \sqrt{p_{\bm{a}}} \ket{\bm{a}}_R \ket{\bm{a}}_{Q_0} \sum_{\bm{b} \in \mathbb{F}_q^{1\times K}} \frac{1}{\sqrt{q^{K} }} \ket{\bm{b}}_A \ket{\bm{b}}_B  \\
&\rightsquigarrow  \sum_{\bm{a} \in \mathbb{F}_q^{1\times K}} \sqrt{p_{\bm{a}}} \ket{\bm{a}}_R \ket{\bm{a}}_{Q_0} \sum_{\bm{b} \in \mathbb{F}_q^{1\times K}} \frac{1}{\sqrt{q^{K} }} \ket{\bm{b}}_A \ket{\bm{b}}_B  \sum_{\bm{z} \in \mathbb{F}_q^{1\times (N-K)}} \frac{1}{\sqrt{q^{N-K} }} \ket{\bm{z}} \\
&=  \sum_{\bm{a}} \sqrt{p_{\bm{a}}} \ket{\bm{a}}_R  \sum_{\bm{b}, \bm{z}} \frac{1}{\sqrt{q^{N} }} \ket{\bm{a}}_{Q_0} \ket{\bm{b}}_{A} \ket{\bm{z}} \ket{\bm{b}}_B  \\
&\rightsquigarrow  \sum_{\bm{a}} \sqrt{p_{\bm{a}}} \ket{\bm{a}}_R  \sum_{\bm{b}, \bm{z}} \frac{1}{\sqrt{q^{N} }} \ket{f_1(\bm{a}+\bm{b}), h_1(\bm{a}, \bm{z})}_{Q_1} \dots  \ket{f_N(\bm{a}+\bm{b}), h_N(\bm{a}, \bm{z})}_{Q_N} \ket{\bm{b}}_B.  \label{perfect:1}
\end{align}
Evidently, the encoded state $\rho_{RQ_1\dots Q_NB}$ in \eqref{eq:rho} is a pure state in this case. In (\ref{perfect:1}), we define the functions $f_1,\dots, f_N$ as follows.
\begin{align}
& \left( f_1(\bm{a}+\bm{b}), \dots, f_N(\bm{a}+\bm{b}) \right) = \left(\bm{a}+\bm{b} \right){\bf V},&&\forall \bm{a},\bm{b}\in\mathbb{F}_q^{1\times K}.\\
\intertext{The LHS is a vector in $\mathbb{F}_q^{1\times N}$. On the RHS, $\bm{a}+\bm{b}\in\mathbb{F}_q^{1\times K}$, and ${\bf V}\in\mathbb{F}_q^{K\times N}$ is defined as follows.}
& {\bf V} = \left[ \alpha_j^{i-1} \right]_{ij}, i \in [K], j \in [N], \alpha_j ~\mbox{are distinct in}~\mathbb{F}_q. \\
\intertext{The functions $h_1,\dots, h_N$ in (\ref{perfect:1}) are defined as}
& \left( h_1(\bm{a}, \bm{z}), \dots, h_N(\bm{a}, \bm{z}) \right)_{1\times N} = \left( \bm{a}, \bm{z} \right)_{1 \times N} \times {\bf H}_{N \times N},&&\forall \bm{a}\in\mathbb{F}_q^{1\times K}, \bm{z}\in\mathbb{F}_q^{1\times (N-K)}.\\
\intertext{The LHS is a vector in $\mathbb{F}_q^{1\times N}$. On the RHS, $(\bm{a},\bm{z})\in\mathbb{F}_q^{1\times N}$, and ${\bf H}\in\mathbb{F}_q^{N\times N}$ is defined as follows.}
& {\bf H}_{N \times N} = \left[ \frac{1}{\beta_i - \gamma_j}\right]_{ij}, i \in [N], j \in [N], ~\beta_i, \gamma_j~\mbox{are distinct in}~\mathbb{F}_q.
\end{align}
Note that (\ref{perfect:1}) is an isometry because $(f_{[N]}(\bm{a}+\bm{b}), h_{[N]}(\bm{a}, \bm{z}))$ determines $\bm{a}, \bm{b}, \bm{z}$. Note that ${\bf V}$ is a Vandermonde matrix and ${\bf H}$ is a Cauchy matrix so that both have full rank.

Consider decoding next. For any $\mathcal{K} \in \binom{[N]}{K}$, the state for $R Q_1 \dots Q_N B = R Q_{\mathcal{K}} Q_{\mathcal{K}^c} B$ is expressed as follows.
\begin{align}
&\sum_{\bm{a}} \sqrt{p_{\bm{a}}} \ket{\bm{a}}_R  \sum_{\bm{b}, \bm{z}} \frac{1}{\sqrt{q^{N} }} \ket{f_{\mathcal{K}}(\bm{a}+\bm{b}), h_{\mathcal{K}}(\bm{a}, \bm{z})}_{Q_{\mathcal{K}}}  \ket{\bm{b}}_B  \ket{f_{\mathcal{K}^c}(\bm{a}+\bm{b}), h_{\mathcal{K}^c}(\bm{a}, \bm{z})}_{Q_{\mathcal{K}^c}} \\
&\rightsquigarrow  \sum_{\bm{a}} \sqrt{p_{\bm{a}}} \ket{\bm{a}}_R  \sum_{\bm{b}, \bm{z}} \frac{1}{\sqrt{q^{N} }} \ket{\bm{a} {\bf V}_{sub}^{K\times K}} \ket{h_{\mathcal{K}}(\bm{a}, \bm{z})}  \ket{\bm{b}} \ket{f_{\mathcal{K}^c}(\bm{a}+\bm{b}), h_{\mathcal{K}^c}(\bm{a}, \bm{z})}_{Q_{\mathcal{K}^c}}  \label{perfect:2}\\
&\rightsquigarrow  \sum_{\bm{a}} \sqrt{p_{\bm{a}}} \ket{\bm{a}}_R  \sum_{\bm{b}, \bm{z}} \frac{1}{\sqrt{q^{N} }} \ket{\bm{a}} \ket{h_{\mathcal{K}}(\bm{a}, \bm{z})}  \ket{\bm{b}} \ket{f_{\mathcal{K}^c}(\bm{a}+\bm{b}), h_{\mathcal{K}^c}(\bm{a}, \bm{z})}_{Q_{\mathcal{K}^c}}
\label{perfect:3} \\
&\rightsquigarrow  \sum_{\bm{a}} \sqrt{p_{\bm{a}}} \ket{\bm{a}}_R  \sum_{\bm{b}, \bm{z}} \frac{1}{\sqrt{q^{N} }} \ket{\bm{a}} \ket{\bm{z} {\bf H}_{sub}^{(N-K) \times K}}  \ket{\bm{a}+\bm{b}}\bigg| {\color{black} \bm{f}_{\mathcal{K}^c}( \underbrace{ \bm{a}+\bm{b}}_{\triangleq \bm{b}'} ), \underbrace{ \bm{h}_{\mathcal{K}^c}(\bm{a}, \bm{z})}_{\triangleq \bm{z}'} } \bigg\rangle_{Q_{\mathcal{K}^c}}  \label{perfect:4}\\
&\rightsquigarrow  \sum_{\bm{a}} \sqrt{p_{\bm{a}}} \ket{\bm{a}}_R  \sum_{\bm{b}', \bm{z}'} \frac{1}{\sqrt{q^{N} }} \ket{\bm{a}} \ket{\bm{z}' {\bf \overline{H}}_{sub}^{(N-K) \times K}}  \ket{\bm{b}'} \ket{f_{\mathcal{K}^c}(  \bm{b}' ),  \bm{z}' }_{Q_{\mathcal{K}^c}}      \label{perfect:5} \\
&= \left( \sum_{\bm{a}} \sqrt{p_{\bm{a}}} \ket{\bm{a}}_R \ket{\bm{a}}_{\widehat{Q}_0} \right)\otimes\left(\sum_{\bm{b}', \bm{z}'} \frac{1}{\sqrt{q^{N} }} \ket{\bm{z}' {\bf \overline{H}}_{sub}^{(N-K) \times K}}  \ket{\bm{b}'} \ket{f_{\mathcal{K}^c}(  \bm{b}' ),  \bm{z}' }_{Q_{\mathcal{K}^c}} \right).  \label{perfect:6} 
\end{align}
Thus, $R\hat{Q}_0$ is recovered in the same pure state as the original message and reference, $RQ_0$.
In (\ref{perfect:2}), $f_{\mathcal{K}}(\bm{a} + \bm{b}) = (\bm{a} + \bm{b}) \times {\bf V}(:, \mathcal{K})$. The matrix ${\bf V}(:, \mathcal{K})\in\mathbb{F}_q^{K\times K}$ is denoted as ${\bf V}_{sub}^{K\times K}$. As $\ket{b}$ is available (from the EA $B$), we  subtract the contribution of $\bm{b}$ and obtain $\ket{\bm{a} {\bf V}_{sub}^{K\times K}}\ket{b}$ through a unitary transformation of $\ket{f_{\mathcal{K}}(\bm{a} + \bm{b})}\ket{b}$. In the next step, (\ref{perfect:3}) is obtained because ${\bf V}_{sub}^{K\times K}$ (comprised of $K$ columns of the Vandermonde matrix ${\bf V}$) has full rank and thus a unitary map  transforms $\ket{\bm{a} {\bf V}_{sub}}$ to $\ket{\bm{a}}$. In (\ref{perfect:4}), $h_{\mathcal{K}}(\bm{a}, \bm{z}) = (\bm{a}, \bm{z}) \times {\bf H}(:, \mathcal{K}) = \bm{a} \times {\bf H}(1:K, \mathcal{K}) + \bm{z} \times {\bf H}(K+1:N, \mathcal{K})$, 
where $i:j$ denotes the index tuple $(i, i+1, \cdots, j)$ and the shorthand notation $:$ in ${\bf H}(:, \mathcal{K})$ denotes the tuple that contains all indices, i.e., all rows.
The matrix ${\bf H}(K+1:N, \mathcal{K})\in\mathbb{F}_q^{(N-K)\times K}$ is denoted as ${\bf H}_{sub}^{(N-K)\times K}$. To obtain (\ref{perfect:5}), note that 
\begin{eqnarray}
&& \bm{z}'  = h_{\mathcal{K}^c}(\bm{a}, \bm{z}) 
= \bm{a} \times {\bf H}(1:K, \mathcal{K}^c) + \bm{z} \times {\bf H}(K+1:N, \mathcal{K}^c), ~ \bm{z}'  \in \mathbb{F}_q^{1\times (N-K)} \\
&\implies& \bm{z} =   \left(  \bm{z}'  -  \bm{a} \times {\bf H}(1:K, \mathcal{K}^c) \right) \times {\bf H}^{-1}(K+1:N, \mathcal{K}^c)_{(N-K) \times (N-K)}, 
\end{eqnarray}
and we define the $(N-K)\times K$ matrix ${\bf \overline{H}}_{sub}^{(N-K)\times K} \triangleq  {\bf H}^{-1}(K+1:N, \mathcal{K}^c) \times {\bf H}_{sub}^{(N-K)\times K}$.
Note that ${\bf H}(K+1:N, \mathcal{K}^c)$ has full rank as it is a square sub-matrix of the Cauchy matrix ${\bf H}$. A unitary map thus transforms $\ket{\bm{a}} \ket{\bm{z} {\bf H}_{sub}^{(N-K) \times K}}$ into $\ket{\bm{a}} \ket{\bm{z}' {\bf \overline{H}}_{sub}^{(N-K) \times K}}$.
In (\ref{perfect:5}), we also replace the sum over $\bm{b}, \bm{z}$ with the sum over $\bm{b}', \bm{z}'$ because for any given $\bm{a}$, the mapping between $(\bm{b}, \bm{z})$ and $(\bm{b}', \bm{z}')$ is invertible so that when $(\bm{b}, \bm{z})$ takes all values in $\mathbb{F}_q^{1\times N}$, $(\bm{b}', \bm{z}')$ also takes all values in $\mathbb{F}_q^{1\times N}$. Thus, the decoding is successful. 

\bibliography{Thesis}
\end{document}